\documentclass[a4paper,11pt]{article}
\pdfoutput=1 
\date{}
\usepackage{jheppub}
\usepackage{amsmath,amssymb,amsfonts,amsxtra, mathrsfs,graphics,graphicx,amsthm,epsfig, youngtab,bm,longtable,float,tikz,empheq}
\usepackage[margin=0.5cm]{caption}
\usepackage{thmtools}

\declaretheorem[numbered=no]{definition}
\allowdisplaybreaks[1]
\usetikzlibrary{positioning}
\usetikzlibrary{automata}
\usetikzlibrary{arrows}
\usetikzlibrary{calc}
\usetikzlibrary{decorations.markings}
\usetikzlibrary{decorations.pathreplacing}
\usetikzlibrary{intersections}
\usetikzlibrary{positioning}
\usetikzlibrary{topaths}
\usetikzlibrary{shapes.geometric}
\usetikzlibrary{shapes.misc}
\tikzset{cf-group/.style = {
    shape = rounded rectangle, minimum size=1.0cm,
    rotate=90,
    rounded rectangle right arc = none,
    draw}}
\tikzset{cross/.style={path picture={ 
  \draw[black]
(path picture bounding box.south east) -- (path picture bounding box.north west) (path picture bounding box.south west) -- (path picture bounding box.north east);
}}}

\newcommand*\widefbox[1]{\fbox{\hspace{2em}#1\hspace{2em}}}

\newcommand{\be}{\begin{equation}}
\newcommand{\ee}{\end{equation}}
\newcommand{\ba}{\begin{array}}
\newcommand{\ea}{\end{array}} 
\newcommand{\bi}{\begin{itemize}}
\newcommand{\ei}{\end{itemize}}
\def\vec#1{\bm{#1}}
\def\bea#1\eea{\allowdisplaybreaMs \begin{align}#1\end{align}}
 \newcommand{\ben}{\begin{enumerate}}
\newcommand{\een}{\end{enumerate}}
\newcommand{\bean}{\begin{eqnarray*}}
\newcommand{\eean}{\end{eqnarray*}}
\newcommand{\eref}[1]{(\ref{#1})}

\newcommand{\nn}{\nonumber}

\newcommand{\tr}{\mathrm{Tr}}

\newcommand{\tq}{\widetilde{q}}

\newcommand{\BR}{\mathbb{R}}

\newcommand{\comment}[1]{}
\newcommand{\CF}{{\cal F}}

\newcommand{\CS}{{\cal S}}

\newcommand{\CT}{{\cal T}}
\newcommand{\CD}{{\cal D}}
\newcommand{\CM}{{\cal M}}
\newcommand{\CO}{{\cal O}}

\newcommand{\CN}{{\cal N}}

\newcommand{\CP}{{\cal P}}

\newcommand{\CZ}{{\cal Z}}
\newcommand{\CR}{{\cal R}}
\newcommand{\CI}{{\cal I}}
\newcommand{\CU}{{\cal U}}

\newcommand{\tm}{\widetilde{m}}

\newcommand{\wt}{\widetilde}
\newcommand{\wh}{\widehat}

\newcommand{\ch}{\cosh \pi}
\newcommand{\s}{\sigma}

\newcommand{\Secref}[1]{Section~\ref{#1}}

\newcommand{\appref}[1]{App.~\ref{#1}}

\newcommand{\figref}[1]{Fig.~\ref{#1}}
\renewcommand{\eqref}[1]{(\ref{#1})}

\title{Line Defects in Three Dimensional Mirror Symmetry beyond ADE quivers}
\author{Anindya Dey}
\affiliation{Department of Physics and Astronomy, Johns Hopkins University, 3400 North Charles Street,
Baltimore, MD 21218, USA}
\emailAdd{anindya.hepth@gmail.com}
\abstract{Understanding the map of line defects in a Quantum Field Theory under a given duality is 
generically a difficult problem. This paper is the second in a series which aims to address this 
question in the context of 3d $\CN=4$ mirror symmetry. A general prescription for constructing vortex defects and 
their mirror maps in quiver gauge theories beyond the $A$-type was presented by the author in 
an earlier paper \cite{Dey:2021jbf}, where specific examples involving $D$-type and affine $D$-type quivers were discussed. 
In this paper, we apply the aforementioned prescription to construct a family of vortex defects as coupled 3d-1d systems 
in quiver gauge theories beyond the $ADE$-type, and study their mirror maps. 
Specifically, we focus on a class of quiver gauge theories involving unitary gauge nodes with 
edge multiplicity greater than 1, i.e. two gauge nodes in these theories may be connected by multiple bifundamental hypermultiplets. 
Quiver gauge theories of this type arise as 3d mirrors of certain Argyres-Douglas theories compactified on a circle. 
Some of these quiver gauge theories are known to have a pair of 3d mirrors, which are themselves related by an IR duality, discussed 
recently in \cite{Dey:2021rxw}. For a concrete example where a pair of 3d mirrors do exist, we study how the vortex defects constructed 
using our prescription map to Wilson defects in each mirror theory.}

\begin{document}

\maketitle
\section{Introduction and summary of results}

\subsection{Background and the basic idea of the paper}
Line defects constitute a very important class of observables in Quantum Field Theories, and encode a wealth 
of information about the physics of the system. In particular, they are the charged operators under the one-form
symmetries. In supersymmetric theories, the BPS line defects are associated with a rich set of algebraic and geometric 
structures that have been studied in various space-time dimensions. 

A rather ubiquitous feature of Quantum Field Theories is the existence of UV/IR dualities -- a generic name for 
the phenomena where a set of theories with completely different descriptions at a certain energy scale flow to 
the same physical theory at some other energy scale. Among other things, these dualities provide an extremely useful tool 
to probe strongly-coupled physics from a field theory perspective. \\

Given the importance of these two widely studied subjects in our understanding of Quantum Field Theories, one is 
naturally led to the question as to how line defects transform across a given duality. It turns out that 
determining the map of line defects in a Quantum Field Theory under a given duality is generically a difficult problem. 
For example, the map of a gauge Wilson defect, which is labelled by a representation of the gauge group, 
to a dual defect may be extremely non-trivial, given the fact that the dual theory in general has a different gauge group. 
The current work is the second paper (after \cite{Dey:2021jbf}) in a series which addresses this question in the context of 
3d $\CN=4$ mirror symmetry \cite{Intriligator:1996ex} for generic quiver gauge theories. 
For this specific IR duality, half-BPS Wilson defects are expected to map to half-BPS vortex 
defects, although writing down the precise map of the operators in a generic dual pair is a challenging task. 
Even for the case of $A$-type quivers (i.e. linear quivers with unitary gauge groups), for which mirror symmetry 
has long been understood as $S$-duality in a simple Type IIB setting \cite{Hanany:1996ie}, the problem of mapping half-BPS defects 
was solved fairly recently \cite{Assel:2015oxa}.  The Type IIB construction of \cite{Hanany:1996ie} was suitably extended in \cite{Assel:2015oxa}
to incorporate the defects. 

In a recent paper \cite{Dey:2021jbf}, the author presented a general 
procedure for constructing vortex defects in quiver gauge theories beyond the $A$-type and studying their maps 
under 3d mirror symmetry to Wilson defects. The construction involves a certain generalization of the $S$-operation of 
Witten \cite{Witten:2003ya}, which was defined as an \textit{$S$-type operation} in \cite{Dey:2020hfe}. 
As discussed in \cite{Dey:2021jbf}, these $S$-type operations 
can be used to construct vortex defects and the associated mirror maps in a large class of non-linear quiver gauge theories, 
starting from the vortex defects in the $A$-type (linear) quivers. 
The procedure in question is entirely field theoretic, and does not rely on any String Theory realization 
of the quiver gauge theory. This procedure was used in \cite{Dey:2021jbf} to study the 
line defects and the associated mirror maps in quiver gauge theories of the $D$-type and the affine $D$-type.\\

In the present paper, we initiate the study of half-BPS line defects and their mirror maps in quiver gauge theories 
beyond the $ADE$ type. For the sake for specificity, we will focus on a class of quiver gauge theories involving 
unitary gauge nodes, such that a pair of nodes maybe connected by multiple bifundamental hypermultiplets.
We will refer to such theories as quivers with \textit{edge multiplicity} greater than one. Quiver gauge theories of this type are known to 
arise as mirrors of certain 3d SCFTs \cite{boalch2008irregular, Xie:2012hs,Xie:2013jc,Wang:2018gvb, Xie:2021ewm}, 
which are obtained by the circle-compactification of Argyres-Douglas theories
\cite{Argyres:1995jj, Argyres:1995xn,Gaiotto:2009hg} and flowing to the IR. 
We construct a certain family of vortex defects in these theories as coupled 3d-1d systems, and study their mirror maps.

It has been shown recently \cite{Dey:2021rxw} that there is generically a pair of 3d mirrors for a 
given quiver of the above type (i.e. with edge multiplicity greater than one) associated with an Argyres-Douglas theory
\footnote{In other words, the 3d SCFT obtained by the circle reduction of the Argyres-Douglas theory has two different Lagrangian 
descriptions, in addition to having a Lagrangian 3d mirror.}. 
One of the mirrors is a unitary quiver 
gauge theory of non-ADE type with matter in fundamental/bifundamental representation as well as certain hypermultiplets 
in the determinant/anti-determinant representation of the gauge group - we will refer to this theory as the ``unitary mirror" 
(or, $U$-mirror). The other mirror is a linear chain of unitary and special unitary gauge nodes with 
fundamental/bifundamental matter \cite{Closset:2020afy, Giacomelli:2020ryy} -- we will refer to this theory 
as the ``unitary-special unitary mirror" (or, $U-SU$-mirror). 
The two mirrors are related by an IR duality, which was studied in \cite{Dey:2021rxw}. 
We will discuss how the vortex defects that we construct in specific examples map to Wilson defects in the 
$U$-mirror as well as the $U-SU$-mirror.\\

The paper is organized as follows. In \Secref{VD-LQ}, we summarize the quiver notation and 
briefly review the vortex defects for the $A$-type quivers. 
We then introduce the 3d-1d coupled quivers that realize vortex defects for gauge nodes connected by edges with 
higher multiplicity, in \Secref{VD-edges}.
In \Secref{sec:Ab-Ex}, we construct these vortex defects in an Abelian quiver gauge theory using the $S$-type operations 
and derive their mirror maps. The Abelian quiver in question is a generalized version of the \textit{complete graph} quiver 
associated with the $(A_2, A_{3p-1})$ theory. Next, we construct vortex defects in a non-Abelian quiver gauge theory 
which is associated with the $D_9(SU(3))$ Argyres-Douglas theory. This theory admits a $U$-mirror as well as a 
$U-SU$-mirror. In \Secref{NAb-nodef}-\Secref{NAb-VD}, we study the mirror maps of the vortex defects in the $U$-mirror, and extend 
them to the $U-SU$-mirror in \Secref{NAb-VD-SU}. A brief summary of the $S$-type operations is given in \appref{sec:SOp-app}, while 
various details of the Witten index and sphere partition function computations can be found in \appref{sec:WI-app}, \appref{sec:PF-Ab-app},
and \appref{sec:PF-NAb-app}.

\subsection{Summary of results}

The main results of the paper are summarized as follows:

\subsubsection*{Coupled quivers with higher edge multiplicity}

Consider a generic 3d $\CN=4$ quiver $\CT$ with unitary gauge nodes and at least a single edge of multiplicity $p>1$. 
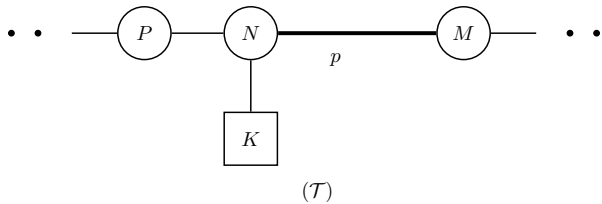
\begin{center}
 \scalebox{.7}{\begin{tikzpicture}[node distance=2cm, nnode/.style={circle,draw,thick, fill, inner sep=1 pt},cnode/.style={circle,draw,thick,minimum size=1.0 cm},snode/.style={rectangle,draw,thick,minimum size=1.0 cm},rnode/.style={red, circle,draw,thick,fill=red!30 ,minimum size=1.0cm}]
 \node[cnode] (20) at (-4,0){$P$} ;
 \node[nnode] (1) at (-6.5,0){} ;
\node[nnode] (2) at (-6,0){} ;
\node[](3) at (-5.5,0){};
\node[cnode] (4) at (-2,0) {$N$};
\node[cnode] (5) at (2,0) {$M$};
\node[snode] (6) at (-2,-2) {$K$};
\draw[thick] (3) -- (20);
\draw[thick] (4) -- (20);
\draw[thick] (4) -- (6);
\node[](8) at (3.5,0){};
\node[nnode] (9) at (4,0){} ;
\node[nnode] (10) at (4.5,0){} ;
\draw[thick] (5) -- (8);
\draw[line width=0.75mm, black] (4) to (5);
\node[text width=1cm](17) at (0, -0.5){$p$};
\node[text width=0.1cm](18) at (-1,-3){$(\CT)$};
\end{tikzpicture}}
\end{center}
The thick edge connecting the $U(N)$ and $U(M)$ gauge nodes labelled by an integer $p$ denotes $p$ hypermultiplets in the 
bifundamental representation of $U(N) \times U(M)$. We introduce a class of vortex defects for the $U(N)$ gauge node in 
$\CT$ which are realized by the following pair of coupled 3d-1d quivers $(\CT[V_{Q,R,p'}^{(I)\,-}])$ and $(\CT[V_{Q,R,p'}^{(II)\,+}])$, 
where the 1d quiver is a (2,2) gauged SQM. 
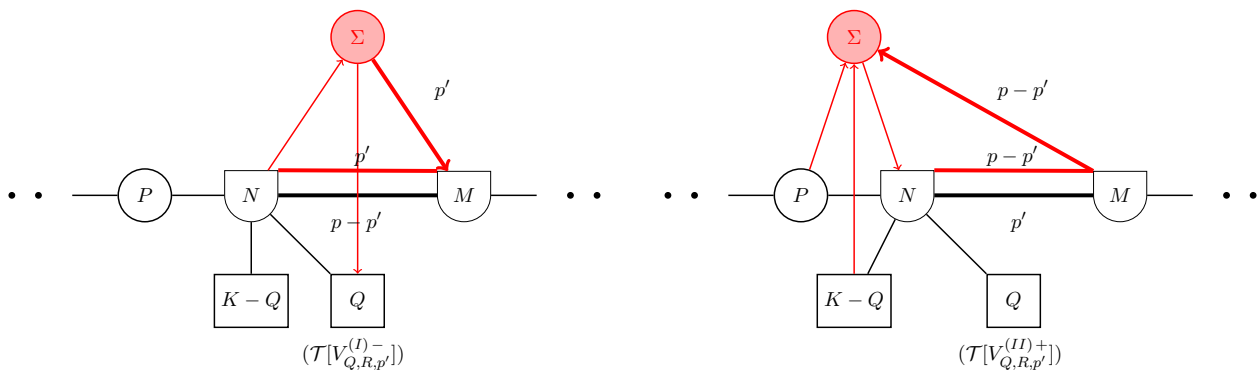
\begin{center}
\begin{tabular}{ccc}
 \scalebox{.7}{\begin{tikzpicture}[node distance=2cm, nnode/.style={circle,draw,thick, fill, inner sep=1 pt},cnode/.style={circle,draw,thick,minimum size=1.0 cm},snode/.style={rectangle,draw,thick,minimum size=1.0 cm},rnode/.style={red, circle,draw,thick,fill=red!30 ,minimum size=1.0cm}]
 \node[cnode] (20) at (-4,0){$P$} ;
 \node[nnode] (1) at (-6.5,0){} ;
\node[nnode] (2) at (-6,0){} ;
\node[](3) at (-5.5,0){};
\node[cf-group] (4) at (-2,0) {\rotatebox{-90}{$N$}};
\node[cf-group] (5) at (2,0) {\rotatebox{-90}{$M$}};
\node[snode] (6) at (-2,-2) {$K-Q$};
\node[snode] (11) at (0,-2) {$Q$};
\draw[thick] (3) -- (20);
\draw[thick] (4) -- (20);
\draw[thick] (4) -- (6);
\node[rnode] (7) at (0,3) {$\Sigma$};
\node[](8) at (3.5,0){};
\node[nnode] (9) at (4,0){} ;
\node[nnode] (10) at (4.5,0){} ;
\draw[thick] (5) -- (8);
\draw[thick] (4) -- (11);
\draw[line width=0.75mm, red] (4.south east) to (5.north east);
\draw[line width=0.75mm, black] (4.south west) to (5.north west);
\draw[red, thick, ->] (4) -- (7);
\draw[line width=0.75mm, red, ->] (7) -- (5);
\draw[red, thick, ->] (7) -- (11);
\node[text width=0.1cm](15) at (0,0.75){$p'$};
\node[text width=0.1cm](16) at (1.5, 2){$p'$};
\node[text width=1cm](17) at (0, -0.5){$p-p'$};
\node[text width=0.1cm](18) at (-1,-3){$(\CT[V_{Q,R,p'}^{(I)\,-}])$};
\end{tikzpicture}}
& \qquad
&  \scalebox{.7}{\begin{tikzpicture}[node distance=2cm, nnode/.style={circle,draw,thick, fill, inner sep=1 pt},cnode/.style={circle,draw,thick,minimum size=1.0 cm},snode/.style={rectangle,draw,thick,minimum size=1.0 cm},rnode/.style={red, circle,draw,thick,fill=red!30 ,minimum size=1.0cm}]
 \node[cnode] (20) at (-4,0){$P$} ;
 \node[nnode] (1) at (-6.5,0){} ;
\node[nnode] (2) at (-6,0){} ;
\node[](3) at (-5.5,0){};
\node[cf-group] (4) at (-2,0) {\rotatebox{-90}{$N$}};
\node[cf-group] (5) at (2,0) {\rotatebox{-90}{$M$}};
\node[snode] (6) at (-3,-2) {$K-Q$};
\node[snode] (11) at (0,-2) {$Q$};
\draw[thick] (3) -- (20);
\draw[thick] (4) -- (20);
\draw[thick] (4) -- (6);
\node[rnode] (7) at (-3,3) {$\Sigma$};
\node[](8) at (3.5,0){};
\node[nnode] (9) at (4,0){} ;
\node[nnode] (10) at (4.5,0){} ;
\draw[thick] (5) -- (8);
\draw[thick] (4) -- (11);
\draw[line width=0.75mm, red] (4.south east) to (5.north east);
\draw[line width=0.75mm, black] (4.south west) to (5.north west);
\draw[red, thick, ->] (20) -- (7);
\draw[red, thick, ->] (6) -- (7);
\draw[red, thick, ->] (7) -- (4);
\draw[line width=0.75mm, red, ->] (5.north east) -- (7);
\node[text width= 1 cm](15) at (0,0.75){$p-p'$};
\node[text width=1 cm](16) at (0.2, 2){$p-p'$};
\node[text width=0.1cm](17) at (0, -0.5){$p'$};
\node[text width=0.1cm](18) at (-1,-3){$(\CT[V_{Q,R,p'}^{(II)\,+}])$};
\end{tikzpicture}}
\end{tabular}
\end{center}

The coupled quivers above have some additional ingredients, compared to the more familiar quivers in \cite{Assel:2015oxa}. 
Consider the quiver $(\CT[V_{Q,R,p'}^{(I)\,-}])$ for example. We show in red the 1d (2,2) multiplets as well as the 
$p'$ 3d bifundamental hypermultiplets which couple to 
the 1d chiral multiplets by a cubic superpotential of the form \eref{3d-1dSup-2}.  
A red, thin directed line denotes a single 1d chiral multiplet, while a red, thick directed line labelled by an integer 
$p'$ denotes $p'$ chiral multiplets. The red node $\Sigma$ denotes a 1d linear quiver of the form given in \figref{1d-gen-2}. 

The coupled quiver is realized by identifying the $[U(N) \times U(Q) \times U(M) \times U(p')]_{1d}$ flavor symmetry of the 
SQM to some $[U(N) \times U(Q) \times U(M) \times U(p')]_{3d}$  flavor/gauge symmetry of the 3d theory, via the 
cubic superpotential. The identification of the $U(N) \times U(Q) \times U(M)$ factor works in the standard fashion 
\cite{Assel:2015oxa}, while $U(p')_{1d}$ is identified with the $U(p')_{3d} \subset U(p)_{3d}$ flavor symmetry associated with $p'$ of the 
3d bifundamental hypers. Finally, the superscripts $\mp$ in the quivers $(\CT[V_{Q,R,p'}^{(I)\,-}])$ and $(\CT[V_{Q,R,p'}^{(II)\,+}])$ 
imply that the 1d FI parameters should be chosen to be negative definite and positive definite respectively.

For a detailed description of the quiver notation and the associated superpotential, we refer the reader to 
\Secref{VD-LQ}-\Secref{VD-edges}.

\subsubsection*{Map of defects under mirror symmetry}

We construct vortex defects in specific examples of quivers with edges of higher multiplicity. The generalities of the 
construction, which involves implementing a sequence of the $S$-type operations on vortex 
defects in $A$-type quivers, is summarized in \appref{sec:SOp-app}. 
We determine the map of these vortex defects to Wilson defects under mirror symmetry. For simplifying our presentation, 
we will restrict ourselves to examples of 3d quivers $\CT$ where $M=N=1$ (see the figures above), 
and the other gauge nodes being generically non-Abelian. \\

The first example we study is an Abelian quiver pair of the form:
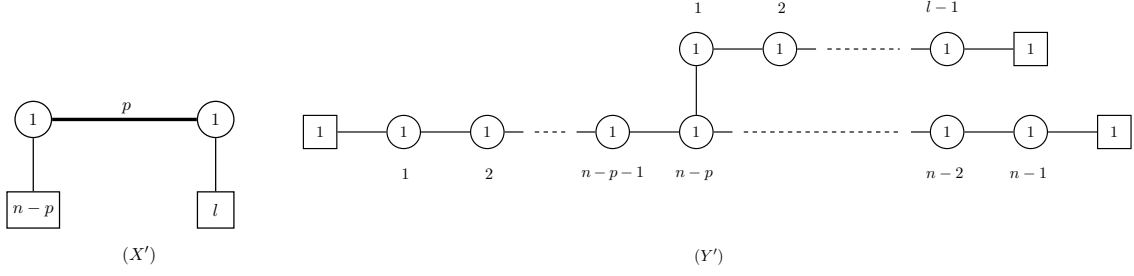
\begin{center}
\scalebox{0.6}{\begin{tikzpicture}[node distance=2cm,cnode/.style={circle,draw,thick,minimum size=8mm},snode/.style={rectangle,draw,thick,minimum size=8mm},pnode/.style={rectangle,red,draw,thick,minimum size=8mm}]
\node[cnode] (1) at (-2,0) {$1$};
\node[cnode] (2) at (2,0) {$1$};
\node[snode] (3) at (-2,-2) {$n-p$};
\node[snode] (4) at (2,-2) {$l$};
\draw[thick] (1) -- (3);
\draw[thick] (2) -- (4);
\draw[line width=0.75mm, black] (1) to (2);
\node[text width=0.1cm](15) at (0,0.25){$p$};
\node[text width=0.1cm](16) at (0,-3){$(X')$};
\end{tikzpicture}}
\qquad 
\scalebox{0.55}{\begin{tikzpicture}[node distance=2cm,cnode/.style={circle,draw,thick,minimum size=8mm},snode/.style={rectangle,draw,thick,minimum size=8mm},pnode/.style={rectangle,red,draw,thick,minimum size=8mm}]
\node[snode] (1) at (-2,0) {$1$};
\node[cnode] (2) at (0,0) {$1$};
\node[cnode] (3) at (2,0) {$1$};
\node[] (4) at (3,0) {};
\node[] (5) at (4,0) {};
\node[cnode] (6) at (5,0) {$1$};
\node[cnode] (7) at (7,0) {$1$};
\node[] (8) at (8,0) {};
\node[] (9) at (12,0) {};
\node[cnode] (10) at (13,0) {$1$};
\node[cnode] (11) at (15,0) {$1$};
\node[snode] (12) at (17,0) {$1$};
\node[cnode] (13) at (7,2) {$1$};
\node[cnode] (14) at (9,2) {$1$};
\node[] (15) at (10,2) {};
\node[] (16) at (12,2) {};
\node[cnode] (17) at (13,2) {$1$};
\node[snode] (30) at (15,2) {$1$};
\draw[thick] (1) -- (2);
\draw[thick] (2) -- (3);
\draw[thick] (3) -- (4);
\draw[thick,dashed] (4) -- (5);
\draw[thick] (5) -- (6);
\draw[thick] (6) -- (7);
\draw[thick] (7) -- (8);
\draw[thick,dashed] (8) -- (9);
\draw[thick] (9) -- (10);
\draw[thick] (10) -- (11);
\draw[thick] (11) -- (12);
\draw[thick] (7) -- (13);
\draw[thick] (13) -- (14);
\draw[thick] (14) -- (15);
\draw[thick,dashed] (15) -- (16);
\draw[thick] (16) -- (17);
\draw[thick] (17) -- (30);
\node[text width=0.1cm](20) at (0,-1) {$1$};
\node[text width=0.1cm](21) at (2,-1) {$2$};
\node[text width= 1.5 cm](22) at (5,-1) {$n-p-1$};
\node[text width=1 cm](23) at (7,-1) {$n-p$};
\node[text width=1 cm](24) at (13,-1) {$n-2$};
\node[text width=1 cm](25) at (15,-1) {$n-1$};
\node[text width=0.1 cm](26) at (7,3) {$1$};
\node[text width=0.1 cm](27) at (9,3) {$2$};
\node[text width=1 cm](28) at (13,3) {$l-1$};
\node[text width=0.1cm](30) at (7,-3){$(Y')$};
\end{tikzpicture}}
\end{center}
For the special case of $n=2p$ and $l=p$ , the quiver $X'$ reproduces the complete graph quiver with three vertices and 
edge multiplicity $p$ -- the 3d mirror of the $(A_2, A_{3p-1})$ AD theory reduced on a circle. The mirror theory $Y'$ was 
first found in \cite{Dey:2020hfe}, and discussed further in \cite{Dey:2021rxw}. 

We find a concrete example of a vortex defect of the form described above and its mirror dual in this case, as shown below.

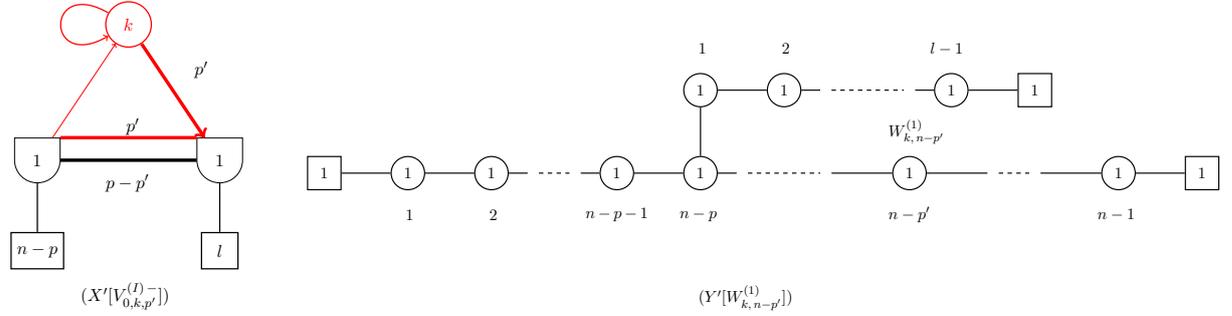
\begin{center}
\begin{tabular}{ccc}
 \scalebox{.6}{\begin{tikzpicture}[node distance=2cm,cnode/.style={circle,draw,thick,minimum size=8mm},snode/.style={rectangle,draw,thick,minimum size=8mm},pnode/.style={rectangle,red,draw,thick,minimum size=8mm}, nnode/.style={circle, red, draw,thick,minimum size=1.0cm}, lnode/.style={shape = rounded rectangle, minimum size= 1cm, rotate=90, rounded rectangle right arc = none, draw}]
\node[lnode] (1) at (-2,0) {\rotatebox{-90}{1}};
\node[lnode] (2) at (2,0) {\rotatebox{-90}{1}};
\node[snode] (3) at (-2,-2) {$n-p$};
\node[snode] (4) at (2,-2) {$l$};
\draw[thick] (1) -- (3);
\draw[thick] (2) -- (4);
\node[nnode] (5) at (0,3) {$k$};
\draw[red, thick, ->] (5) to [out=150,in=210,looseness=8] (5);
\draw[line width=0.75mm, red] (1.south east) to (2.north east);
\draw[line width=0.75mm, black] (1.south west) to (2.north west);
\draw[red, ->] (1) -- (5);
\draw[line width=0.75mm, red, ->] (5) -- (2);
\node[text width=0.1cm](15) at (0,0.75){$p'$};
\node[text width=0.1cm](16) at (1.5, 2){$p'$};
\node[text width=1cm](17) at (0, -0.5){$p-p'$};
\node[text width=0.1cm](18) at (-1,-3){$(X'[V^{(I)\,-}_{0,k,p'}])$};
\end{tikzpicture}}
& \qquad
&\scalebox{0.55}{\begin{tikzpicture}[node distance=2cm,cnode/.style={circle,draw,thick,minimum size=8mm},snode/.style={rectangle,draw,thick,minimum size=8mm},pnode/.style={rectangle,red,draw,thick,minimum size=8mm}]
\node[snode] (1) at (-2,0) {$1$};
\node[cnode] (2) at (0,0) {$1$};
\node[cnode] (3) at (2,0) {$1$};
\node[] (4) at (3,0) {};
\node[] (5) at (4,0) {};
\node[cnode] (6) at (5,0) {$1$};
\node[cnode] (7) at (7,0) {$1$};
\node[] (8) at (8,0) {};
\node[] (9) at (10,0) {};
\node[cnode] (10) at (12,0) {$1$};
\node[] (40) at (14,0) {};
\node[] (41) at (15,0) {};
\node[cnode] (11) at (17,0) {$1$};
\node[snode] (12) at (19,0) {$1$};
\node[cnode] (13) at (7,2) {$1$};
\node[cnode] (14) at (9,2) {$1$};
\node[] (15) at (10,2) {};
\node[] (16) at (12,2) {};
\node[cnode] (17) at (13,2) {$1$};
\node[snode] (30) at (15,2) {$1$};
\draw[thick] (1) -- (2);
\draw[thick] (2) -- (3);
\draw[thick] (3) -- (4);
\draw[thick,dashed] (4) -- (5);
\draw[thick] (5) -- (6);
\draw[thick] (6) -- (7);
\draw[thick] (7) -- (8);
\draw[thick,dashed] (8) -- (9);
\draw[thick] (9) -- (10);
\draw[thick] (10) -- (40);
\draw[thick, dashed] (40) -- (41);
\draw[thick] (41) -- (11);
\draw[thick] (11) -- (12);
\draw[thick] (7) -- (13);
\draw[thick] (13) -- (14);
\draw[thick] (14) -- (15);
\draw[thick,dashed] (15) -- (16);
\draw[thick] (16) -- (17);
\draw[thick] (17) -- (30);
\node[text width=0.1cm](20) at (0,-1) {$1$};
\node[text width=0.1cm](21) at (2,-1) {$2$};
\node[text width= 1.5 cm](22) at (5,-1) {$n-p-1$};
\node[text width=1 cm](23) at (7,-1) {$n-p$};
\node[text width=1 cm](24) at (12,-1) {$n-p'$};
\node[text width=1 cm](31) at (12,1) {$W^{(1)}_{k,\, n-p'}$};
\node[text width=1 cm](25) at (17,-1) {$n-1$};
\node[text width=0.1 cm](26) at (7,3) {$1$};
\node[text width=0.1 cm](27) at (9,3) {$2$};
\node[text width=1 cm](28) at (13,3) {$l-1$};
\node[text width=0.1cm](30) at (7,-3){$(Y'[W^{(1)}_{k,\, n-p'}])$};
\end{tikzpicture}}
\end{tabular}
\end{center}

The dual defect in this case is a Wilson defect of charge $k$ for the $U(1)$ gauge shown in the figure on the 
right. The details of the computation for the above result and the related notation can be found in \Secref{sec:Ab-Ex}.\\

The second example we study is a non-Abelian quiver pair of the following form (the numbers above the gauge nodes 
are labels with no physical significance).

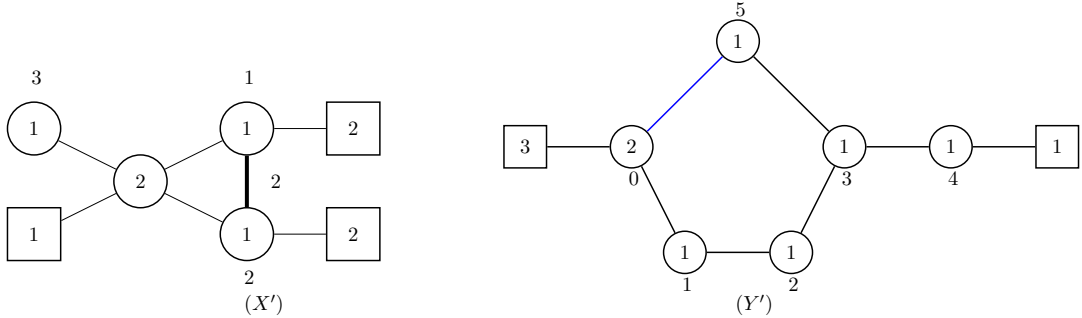
\begin{center}
\begin{tabular}{ccc}
\scalebox{.7}{\begin{tikzpicture}[
cnode/.style={circle,draw,thick, minimum size=1.0cm},snode/.style={rectangle,draw,thick,minimum size=1cm}]
\node[cnode] (9) at (0,1){1};
\node[snode] (10) at (0,-1){1};
\node[cnode] (11) at (2, 0){2};
\node[cnode] (12) at (4, 1){1};
\node[cnode] (13) at (4, -1){1};
\node[snode] (14) at (6, 1){$2$};
\node[snode] (15) at (6, -1){$2$};
\draw[-] (9) -- (11);
\draw[-] (10) -- (11);
\draw[-] (12) -- (11);
\draw[-] (13) -- (11);
\draw[-] (12) -- (14);
\draw[-] (13) -- (15);
\draw[line width=0.75mm, black] (12) to (13);
\node[text width=0.1cm](20) at (4.5,0){$2$};
\node[text width=0.1cm](21)[above=0.2 cm of 9]{3};
\node[text width=0.1cm](23)[above=0.2 cm of 12]{1};
\node[text width=0.1cm](24)[below=0.05 cm of 13]{2};
\node[text width=0.1cm](31)[below=0.5 cm of 13]{$(X')$};
\end{tikzpicture}}
&\qquad  \qquad
&\scalebox{.7}{\begin{tikzpicture}[node distance=2cm,cnode/.style={circle,draw,thick,minimum size=8mm},snode/.style={rectangle,draw,thick,minimum size=8mm},pnode/.style={rectangle,red,draw,thick,minimum size=8mm}]
\node[cnode] (1) at (-3,0) {$2$};
\node[snode] (2) at (-5,0) {$3$};
\node[cnode] (3) at (-1,2) {$1$};
\node[cnode] (4) at (-2,-2) {$1$};
\node[cnode] (5) at (0,-2) {$1$};
\node[cnode] (6) at (1,0) {$1$};
\node[cnode] (7) at (3,0) {$1$};
\node[snode] (8) at (5,0) {$1$};
\draw[thick] (1) -- (2);
\draw[thick, blue] (1) -- (3);
\draw[thick] (1) -- (4);
\draw[thick] (4) -- (5);
\draw[thick] (5) -- (6);
\draw[thick] (6) -- (7);
\draw[thick] (7) -- (8);
\draw[thick] (3) -- (6);
\node[text width=0.1cm](30) at (-1,-3){$(Y')$};
\node[text width=0.1cm](40) at (-3, -0.6){0};
\node[text width=0.1cm](41) at (-2, -2.6){1};
\node[text width=0.1cm](42) at (0, -2.6){2};
\node[text width=0.1cm](43) at (1, -0.6){3};
\node[text width=0.1cm](44) at (3, -0.6){4};
\node[text width=0.1cm](45) at (-1, 2.6){5};
\end{tikzpicture}}
\end{tabular}
\end{center}
The theory $X'$ is a non-ADE-type quiver gauge theory with unitary gauge groups and hypermultiplets 
in fundamental/bifundamental representation of the gauge group. In particular, the gauge nodes 
$U(1)_1$ and $U(1)_2$ are connected by an edge of multiplicity 2. The dual theory $Y'$ is also 
a non-ADE-type quiver gauge theory built out of unitary gauge nodes, 
and fundamental/bifundamental matter. In addition, $Y'$ has a single hypermultiplet which transforms 
in the determinant representation of the $U(2)$ gauge group and has charge 1 under one of the 
adjacent $U(1)$ gauge nodes, as denoted by the blue line in the figure.

The theory $X'$ is the 3d mirror associated with the circle reduction of the AD theory $D_9(SU(3))$. 
The quiver $Y'$ is the $U$-mirror of $X'$ (as described above), which was first found in \cite{Dey:2020hfe} 
and discussed further in \cite{Dey:2021rxw}. \\

For the dual pair $(X',Y')$, we first construct a vortex defect in $X'$ that does not involve the bifundamental 
hypers of the edge with multiplicity 2. The coupled quiver for the vortex defect and its mirror dual are given as 
follows. 

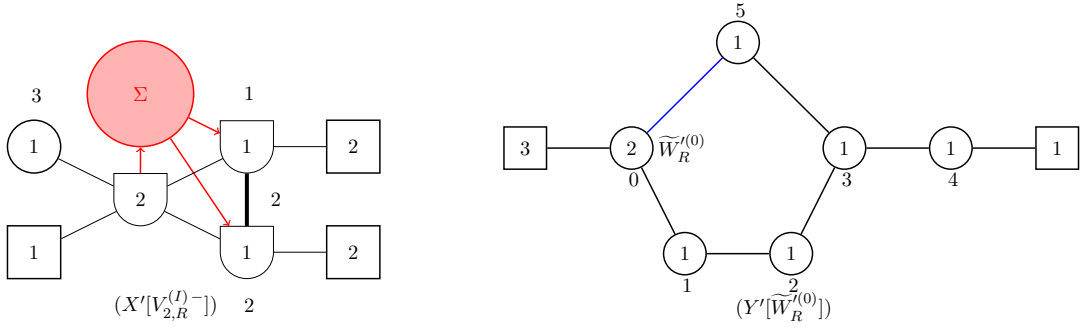
\begin{center}
\begin{tabular}{ccc}
\scalebox{.7}{\begin{tikzpicture}[node distance=2cm, nnode/.style={circle,draw,thick, red, fill=red!30, minimum size=2.0 cm},cnode/.style={circle,draw,thick,minimum size=1.0 cm},snode/.style={rectangle,draw,thick,minimum size=1.0 cm}]
\node[cnode] (1) at (0,1) {1} ;
\node[cf-group] (2) at (2,0) {\rotatebox{-90}{2}};
\node[snode] (3) at (0,-1) {1};
\node[cf-group] (5) at (4, 1) {\rotatebox{-90}{1}};
\node[cf-group] (6) at (4, -1) {\rotatebox{-90}{1}};
\node[nnode] (7) at (2,2) {$\Sigma$};
\node[snode] (8) at (6,1) {2};
\node[snode] (9) at (6,-1) {2};
\draw[red, thick, ->] (2)--(7);
\draw[red, thick, ->] (7)--(5);
\draw[red, thick, ->] (7)--(6);
\draw[-] (1) -- (2);
\draw[-] (2) -- (3);
\draw[-] (2) -- (5);
\draw[-] (2) -- (6);
\draw[-] (5) -- (8);
\draw[-] (6) -- (9);
\draw[line width=0.75mm, black] (5) to (6);
\node[text width=1cm](9) at (2, -2) {$(X'[V^{(I)\,-}_{2,R}])$};
\node[text width=0.1cm](21)[above=0.2 cm of 1]{3};
\node[text width=0.1cm](23) at (4,2){1};
\node[text width=0.1cm](24) at (4,-2){2};
\node[text width=0.1cm](20) at (4.5,0){$2$};
\end{tikzpicture}}
& \qquad  \qquad
&\scalebox{.7}{\begin{tikzpicture}[node distance=2cm,cnode/.style={circle,draw,thick,minimum size=8mm},snode/.style={rectangle,draw,thick,minimum size=8mm},pnode/.style={rectangle,red,draw,thick,minimum size=8mm}]
\node[cnode] (1) at (-3,0) {$2$};
\node[snode] (2) at (-5,0) {$3$};
\node[cnode] (3) at (-1,2) {$1$};
\node[cnode] (4) at (-2,-2) {$1$};
\node[cnode] (5) at (0,-2) {$1$};
\node[cnode] (6) at (1,0) {$1$};
\node[cnode] (7) at (3,0) {$1$};
\node[snode] (8) at (5,0) {$1$};
\draw[thick] (1) -- (2);
\draw[thick, blue] (1) -- (3);
\draw[thick] (1) -- (4);
\draw[thick] (4) -- (5);
\draw[thick] (5) -- (6);
\draw[thick] (6) -- (7);
\draw[thick] (7) -- (8);
\draw[thick] (3) -- (6);
\node[text width=1cm](36) at (-2, 0) {$\wt{W}'^{(0)}_{R}$};
\node[text width=0.1cm](40) at (-3, -0.6){0};
\node[text width=0.1cm](41) at (-2, -2.6){1};
\node[text width=0.1cm](42) at (0, -2.6){2};
\node[text width=0.1cm](43) at (1, -0.6){3};
\node[text width=0.1cm](44) at (3, -0.6){4};
\node[text width=0.1cm](45) at (-1, 2.6){5};
\node[text width=0.1cm](30) at (-1,-3){$(Y'[\wt{W}'^{(0)}_{R}])$};
\end{tikzpicture}}
\end{tabular}
\end{center}

The vortex defect in this case is labelled by a representation $R$ of $U(2)$ which is encoded in the 1d quiver $\Sigma$ 
(see \Secref{VD-LQ} for details). The dual is a Wilson defect for the $U(2)$ gauge node in the theory $Y'$ labelled by the same 
representation $R$.\\

Next, we construct a vortex defect in $X'$ that involves the $U(1)_1 \times U(1)_2$ bifundamental hypermultiplets. 
The defect and its dual are shown below. The dual involves a Wilson defect of charge $k$ and $-k$ for the gauge 
nodes labelled $U(1)_1$ and $U(1)_3$ respectively.

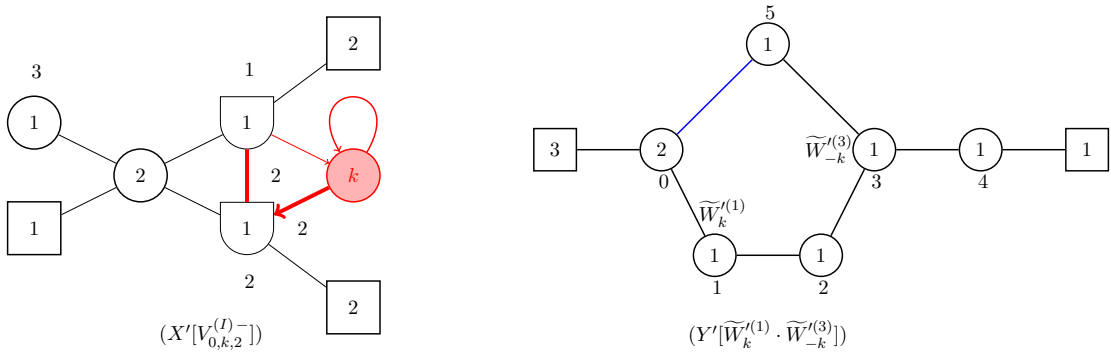
\begin{center}
\begin{tabular}{ccc}
\scalebox{.7}{\begin{tikzpicture}[nnode/.style={circle,draw,thick, red, fill=red!30, minimum size=1.0 cm},
cnode/.style={circle,draw,thick, minimum size=1.0cm},snode/.style={rectangle,draw,thick,minimum size=1cm}]
\node[cnode] (9) at (0,1){1};
\node[snode] (10) at (0,-1){1};
\node[cnode] (11) at (2, 0){2};
\node[cf-group] (12) at (4, 1){\rotatebox{-90}{1}};
\node[cf-group] (13) at (4, -1){\rotatebox{-90}{1}};
\node[snode] (14) at (6, 2.5){$2$};
\node[snode] (15) at (6, -2.5){$2$};
\node[nnode] (16) at (6,0){$k$};
\draw[red, thick, ->] (16) to [out=60,in=120,looseness=8] (16);
\draw[-] (9) -- (11);
\draw[-] (10) -- (11);
\draw[-] (12) -- (11);
\draw[-] (13) -- (11);
\draw[-] (12) -- (14);
\draw[-] (13) -- (15);
\draw[red,->] (12) -- (16);
\draw[red, line width=0.75mm, ->] (16) -- (13);
\draw[line width=0.75mm, red] (12) to (13);
\node[text width=0.1cm](20) at (4.5,0){$2$};
\node[text width=0.1cm](21)[above=0.2 cm of 9]{3};
\node[text width=0.1cm](23) at (4,2){1};
\node[text width=0.1cm](24) at (4,-2){2};
\node[text width=0.1cm](25) at (5,-1){2};
\node[text width=1.3 cm](31) at (3,-3){$(X'[V_{0,k,2}^{(I)\,-}])$};
\end{tikzpicture}}
& \qquad  \qquad
& \scalebox{.7}{\begin{tikzpicture}[node distance=2cm,cnode/.style={circle,draw,thick,minimum size=8mm},snode/.style={rectangle,draw,thick,minimum size=8mm},pnode/.style={rectangle,red,draw,thick,minimum size=8mm}]
\node[cnode] (1) at (-3,0) {$2$};
\node[snode] (2) at (-5,0) {$3$};
\node[cnode] (3) at (-1,2) {$1$};
\node[cnode] (4) at (-2,-2) {$1$};
\node[cnode] (5) at (0,-2) {$1$};
\node[cnode] (6) at (1,0) {$1$};
\node[cnode] (7) at (3,0) {$1$};
\node[snode] (8) at (5,0) {$1$};
\draw[thick] (1) -- (2);
\draw[thick, blue] (1) -- (3);
\draw[thick] (1) -- (4);
\draw[thick] (4) -- (5);
\draw[thick] (5) -- (6);
\draw[thick] (6) -- (7);
\draw[thick] (7) -- (8);
\draw[thick] (3) -- (6);
\node[text width=1cm](36) at (-1.8, -1.2) {$\wt{W}'^{(1)}_k$};
\node[text width=1cm](37) at (0.2, 0) {$\wt{W}'^{(3)}_{-k}$};
\node[text width=0.1cm](40) at (-3, -0.6){0};
\node[text width=0.1cm](41) at (-2, -2.6){1};
\node[text width=0.1cm](42) at (0, -2.6){2};
\node[text width=0.1cm](43) at (1, -0.6){3};
\node[text width=0.1cm](44) at (3, -0.6){4};
\node[text width=0.1cm](45) at (-1, 2.6){5};
\node[text width=3 cm](30) at (-1,-3.5){$(Y'[\wt{W}'^{(1)}_k \cdot \wt{W}'^{(3)}_{-k}])$};
\end{tikzpicture}}
\end{tabular}
\end{center}

The details of the computation and the related notation are given in \Secref{NAb-nodef}-\Secref{NAb-VD}.

\subsubsection*{Hopping duality for the vortex defects}

A vortex defect can be realized by multiple 3d-1d quivers. At the level of supersymmetric observables, like 
the sphere partition function for example, the matrix models associated to such quivers can be mapped to 
one another by a simple change of variables. This is known as \textit{hopping duality} \cite{Assel:2015oxa}. 
The generic coupled systems $(\CT[V_{Q,R,p'}^{(I)\,-}])$ and $(\CT[V_{Q,R,p'}^{(II)\,+}])$, described above,
are therefore hopping duals. For the examples of vortex defects discussed above, the associated hopping 
dualities are listed below.

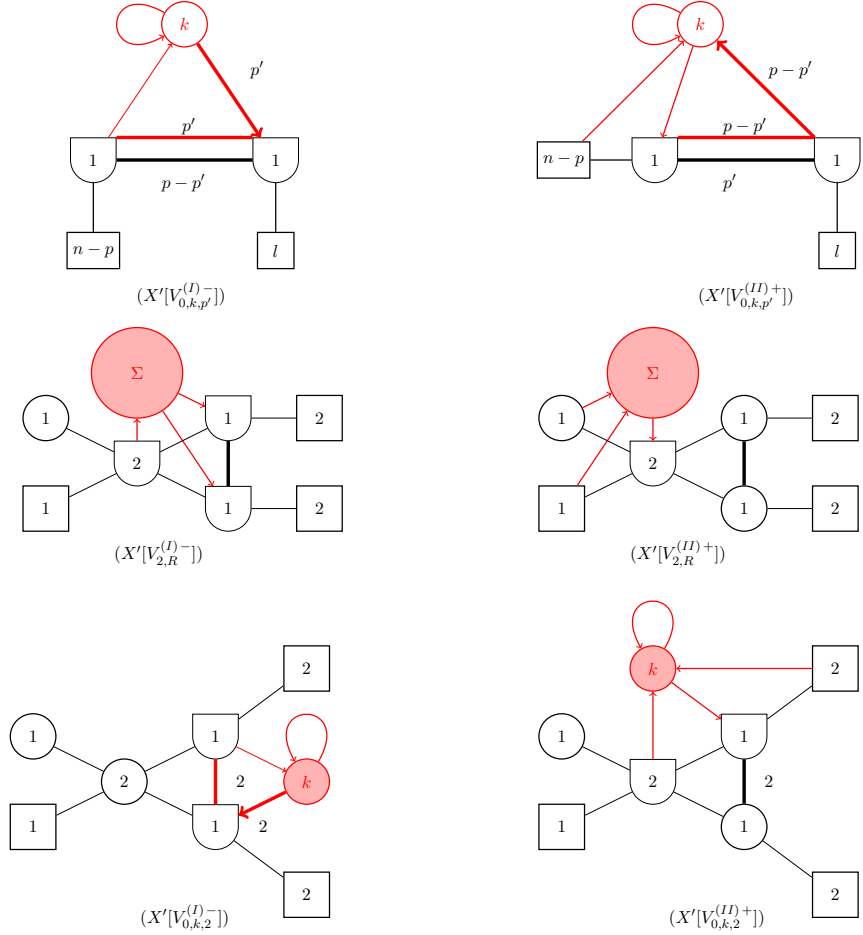
\begin{center}
\begin{tabular}{ccc}
 \scalebox{.6}{\begin{tikzpicture}[node distance=2cm,cnode/.style={circle,draw,thick,minimum size=8mm},snode/.style={rectangle,draw,thick,minimum size=8mm},pnode/.style={rectangle,red,draw,thick,minimum size=8mm}, nnode/.style={circle, red, draw,thick,minimum size=1.0cm}, lnode/.style={shape = rounded rectangle, minimum size= 1cm, rotate=90, rounded rectangle right arc = none, draw}]
\node[lnode] (1) at (-2,0) {\rotatebox{-90}{1}};
\node[lnode] (2) at (2,0) {\rotatebox{-90}{1}};
\node[snode] (3) at (-2,-2) {$n-p$};
\node[snode] (4) at (2,-2) {$l$};
\draw[thick] (1) -- (3);
\draw[thick] (2) -- (4);
\node[nnode] (5) at (0,3) {$k$};
\draw[red, thick, ->] (5) to [out=150,in=210,looseness=8] (5);
\draw[line width=0.75mm, red] (1.south east) to (2.north east);
\draw[line width=0.75mm, black] (1.south west) to (2.north west);
\draw[red, ->] (1) -- (5);
\draw[line width=0.75mm, red, ->] (5) -- (2);
\node[text width=0.1cm](15) at (0,0.75){$p'$};
\node[text width=0.1cm](16) at (1.5, 2){$p'$};
\node[text width=1cm](17) at (0, -0.5){$p-p'$};
\node[text width=0.1cm](18) at (-1,-3){$(X'[V^{(I)\,-}_{0,k,p'}])$};
\end{tikzpicture}}
& \qquad \qquad
& \scalebox{.6}{\begin{tikzpicture}[node distance=2cm,cnode/.style={circle,draw,thick,minimum size=8mm},snode/.style={rectangle,draw,thick,minimum size=8mm},pnode/.style={rectangle,red,draw,thick,minimum size=8mm}, nnode/.style={circle, red, draw,thick,minimum size=1.0cm}, lnode/.style={shape = rounded rectangle, minimum size= 1cm, rotate=90, rounded rectangle right arc = none, draw}]
\node[lnode] (1) at (-2,0) {\rotatebox{-90}{1}};
\node[lnode] (2) at (2,0) {\rotatebox{-90}{1}};
\node[snode] (3) at (-4,0) {$n-p$};
\node[snode] (4) at (2,-2) {$l$};
\draw[thick] (1) -- (3);
\draw[thick] (2) -- (4);
\node[nnode] (5) at (-1,3) {$k$};
\draw[red, thick, ->] (5) to [out=150,in=210,looseness=8] (5);
\draw[line width=0.75mm, red] (1.south east) to (2.north east);
\draw[line width=0.75mm, black] (1.south west) to (2.north west);
\draw[red, thick, ->] (5) -- (1);
\draw[line width=0.75mm, red, ->] (2) -- (5);
\draw[red, thick, ->] (3) -- (5);
\node[text width=1cm](15) at (0,0.75){$p-p'$};
\node[text width=1cm](16) at (1, 2){$p-p'$};
\node[text width=1cm](17) at (0, -0.5){$p'$};
\node[text width=0.1cm](18) at (-1,-3){$(X'[V^{(II)\,+}_{0,k,p'}])$};
\end{tikzpicture}}\\
\scalebox{.6}{\begin{tikzpicture}[node distance=2cm, nnode/.style={circle,draw,thick, red, fill=red!30, minimum size=2.0 cm},cnode/.style={circle,draw,thick,minimum size=1.0 cm},snode/.style={rectangle,draw,thick,minimum size=1.0 cm}]
\node[cnode] (1) at (0,1) {1} ;
\node[cf-group] (2) at (2,0) {\rotatebox{-90}{2}};
\node[snode] (3) at (0,-1) {1};
\node[cf-group] (5) at (4, 1) {\rotatebox{-90}{1}};
\node[cf-group] (6) at (4, -1) {\rotatebox{-90}{1}};
\node[nnode] (7) at (2,2) {$\Sigma$};
\node[snode] (8) at (6,1) {2};
\node[snode] (9) at (6,-1) {2};
\draw[red, thick, ->] (2)--(7);
\draw[red, thick, ->] (7)--(5);
\draw[red, thick, ->] (7)--(6);
\draw[-] (1) -- (2);
\draw[-] (2) -- (3);
\draw[-] (2) -- (5);
\draw[-] (2) -- (6);
\draw[-] (5) -- (8);
\draw[-] (6) -- (9);
\draw[line width=0.75mm, black] (5) to (6);
\node[text width=1cm](9) at (2, -2) {$(X'[V^{(I)\,-}_{2,R}])$};
\end{tikzpicture}}
&\qquad  \qquad \qquad
&\scalebox{.6}{\begin{tikzpicture}[node distance=2cm, nnode/.style={circle,draw,thick, red, fill=red!30, minimum size=2.0 cm},cnode/.style={circle,draw,thick,minimum size=1.0 cm},snode/.style={rectangle,draw,thick,minimum size=1.0 cm}]
\node[cnode] (1) at (0,1) {1} ;
\node[cf-group] (2) at (2,0) {\rotatebox{-90}{2}};
\node[snode] (3) at (0,-1) {1};
\node[cnode] (5) at (4, 1) {1};
\node[cnode] (6) at (4, -1) {1};
\node[nnode] (7) at (2,2) {$\Sigma$};
\node[snode] (8) at (6,1) {2};
\node[snode] (9) at (6,-1) {2};
\draw[red, thick, ->] (7)--(2);
\draw[red, thick, ->] (3)--(7);
\draw[red, thick, ->] (1)--(7);
\draw[-] (1) -- (2);
\draw[-] (2) -- (3);
\draw[-] (2) -- (5);
\draw[-] (2) -- (6);
\draw[-] (5) -- (8);
\draw[-] (6) -- (9);
\draw[line width=0.75mm, black] (5) to (6);
\node[text width=1cm](9) at (2, -2) {$(X'[V^{(II)\,+}_{2,R}])$};
\end{tikzpicture}}\\
\scalebox{.6}{\begin{tikzpicture}[nnode/.style={circle,draw,thick, red, fill=red!30, minimum size=1.0 cm},
cnode/.style={circle,draw,thick, minimum size=1.0cm},snode/.style={rectangle,draw,thick,minimum size=1cm}]
\node[cnode] (9) at (0,1){1};
\node[snode] (10) at (0,-1){1};
\node[cnode] (11) at (2, 0){2};
\node[cf-group] (12) at (4, 1){\rotatebox{-90}{1}};
\node[cf-group] (13) at (4, -1){\rotatebox{-90}{1}};
\node[snode] (14) at (6, 2.5){$2$};
\node[snode] (15) at (6, -2.5){$2$};
\node[nnode] (16) at (6,0){$k$};
\draw[red, thick, ->] (16) to [out=60,in=120,looseness=8] (16);
\draw[-] (9) -- (11);
\draw[-] (10) -- (11);
\draw[-] (12) -- (11);
\draw[-] (13) -- (11);
\draw[-] (12) -- (14);
\draw[-] (13) -- (15);
\draw[red,->] (12) -- (16);
\draw[red, line width=0.75mm, ->] (16) -- (13);
\draw[line width=0.75mm, red] (12) to (13);
\node[text width=0.1cm](20) at (4.5,0){$2$};
\node[text width=0.1cm](25) at (5,-1){2};
\node[text width=1.3 cm](31) at (3,-3){$(X'[V_{0,k,2}^{(I)\,-}])$};
\end{tikzpicture}}
& \qquad \qquad \qquad
& \scalebox{.6}{\begin{tikzpicture}[nnode/.style={circle,draw,thick, red, fill=red!30, minimum size=1.0 cm},
cnode/.style={circle,draw,thick, minimum size=1.0cm},snode/.style={rectangle,draw,thick,minimum size=1cm}]
\node[cnode] (9) at (0,1){1};
\node[snode] (10) at (0,-1){1};
\node[cf-group] (11) at (2, 0){\rotatebox{-90}{2}};
\node[cf-group] (12) at (4, 1){\rotatebox{-90}{1}};
\node[cnode] (13) at (4, -1){1};
\node[snode] (14) at (6, 2.5){$2$};
\node[snode] (15) at (6, -2.5){$2$};
\node[nnode] (16) at (2, 2.5){$k$};
\draw[red, thick, ->] (16) to [out=60,in=120,looseness=8] (16);
\draw[-] (9) -- (11);
\draw[-] (10) -- (11);
\draw[-] (12) -- (11);
\draw[-] (13) -- (11);
\draw[-] (12) -- (14);
\draw[-] (13) -- (15);
\draw[red,thick, ->] (16) -- (12);
\draw[line width=0.75mm] (12) to (13);
\draw[red, thick, ->] (11) -- (16);
\draw[red, thick, ->] (14) -- (16);
\node[text width=0.1cm](20) at (4.5,0){$2$};
\node[text width=1.3 cm](31) at (3,-3){$(X'[V_{0,k,2}^{(II)\,+}])$};
\end{tikzpicture}}
\end{tabular}
\end{center}

\subsubsection*{Map of defects under IR duality}

The non-Abelian quiver $X'$, discussed above, has a second Lagrangian mirror, which we refer to 
as the $U-SU$ mirror $Y''$.

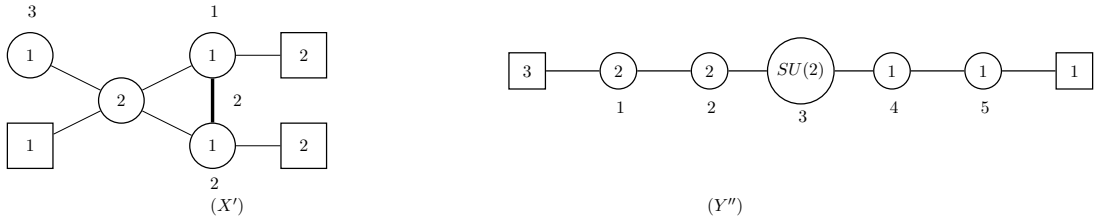
\begin{center}
\begin{tabular}{ccc}
\scalebox{.6}{\begin{tikzpicture}[
cnode/.style={circle,draw,thick, minimum size=1.0cm},snode/.style={rectangle,draw,thick,minimum size=1cm}]
\node[cnode] (9) at (0,1){1};
\node[snode] (10) at (0,-1){1};
\node[cnode] (11) at (2, 0){2};
\node[cnode] (12) at (4, 1){1};
\node[cnode] (13) at (4, -1){1};
\node[snode] (14) at (6, 1){$2$};
\node[snode] (15) at (6, -1){$2$};
\draw[-] (9) -- (11);
\draw[-] (10) -- (11);
\draw[-] (12) -- (11);
\draw[-] (13) -- (11);
\draw[-] (12) -- (14);
\draw[-] (13) -- (15);
\draw[line width=0.75mm, black] (12) to (13);
\node[text width=0.1cm](20) at (4.5,0){$2$};
\node[text width=0.1cm](21)[above=0.2 cm of 9]{3};
\node[text width=0.1cm](23)[above=0.2 cm of 12]{1};
\node[text width=0.1cm](24)[below=0.05 cm of 13]{2};
\node[text width=0.1cm](31)[below=0.5 cm of 13]{$(X')$};
\end{tikzpicture}}
& \qquad \qquad \qquad
& \scalebox{.6}{\begin{tikzpicture}[node distance=2cm,cnode/.style={circle,draw,thick,minimum size=8mm},snode/.style={rectangle,draw,thick,minimum size=8mm},pnode/.style={rectangle,red,draw,thick,minimum size=8mm}]
\node[cnode] (1) at (-3,0) {$2$};
\node[snode] (2) at (-5,0) {$3$};
\node[cnode] (3) at (-1,0) {$2$};
\node[cnode] (4) at (1,0) {$SU(2)$};
\node[cnode] (5) at (3,0) {$1$};
\node[cnode] (6) at (5,0) {$1$};
\node[snode] (7) at (7,0) {$1$};
\draw[thick] (1) -- (2);
\draw[thick] (1) -- (3);
\draw[thick] (3) -- (4);
\draw[thick] (4) -- (5);
\draw[thick] (5) -- (6);
\draw[thick] (6) -- (7);
\node[text width=0.1cm](50) at (-3, -0.8) {1};
\node[text width=0.1cm](50) at (-1, -0.8) {2};
\node[text width=0.1cm](50) at (1, -1) {3};
\node[text width=0.1cm](50) at (3, -0.8) {4};
\node[text width=0.1cm](50) at (5, -0.8) {5};
\node[text width=0.1cm](30) at (-1,-3){$(Y'')$};
\end{tikzpicture}}
\end{tabular}
\end{center}

In \Secref{sec:NAb-Ex}, we work out the mirror maps for the vortex defects in $X'$, mentioned above, to Wilson
defects in the $U-SU$ mirror $Y''$. The dual Wilson defect, in each case, is of the following form.

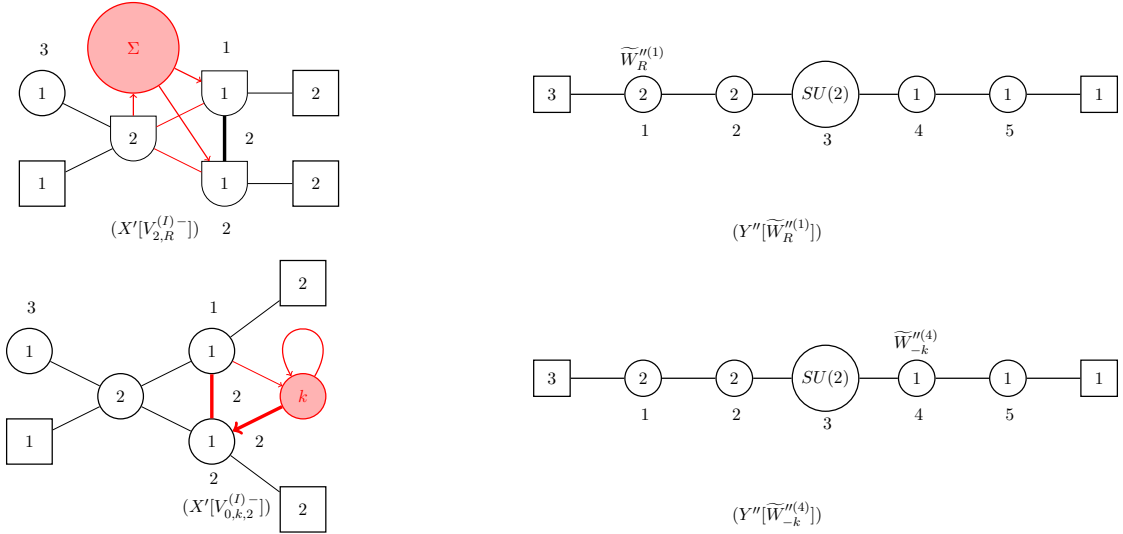
\begin{center}
\begin{tabular}{ccc}
\scalebox{.6}{\begin{tikzpicture}[node distance=2cm, nnode/.style={circle,draw,thick, red, fill=red!30, minimum size=2.0 cm},cnode/.style={circle,draw,thick,minimum size=1.0 cm},snode/.style={rectangle,draw,thick,minimum size=1.0 cm}]
\node[cnode] (1) at (0,1) {1} ;
\node[cf-group] (2) at (2,0) {\rotatebox{-90}{2}};
\node[snode] (3) at (0,-1) {1};
\node[cf-group] (5) at (4, 1) {\rotatebox{-90}{1}};
\node[cf-group] (6) at (4, -1) {\rotatebox{-90}{1}};
\node[nnode] (7) at (2,2) {$\Sigma$};
\node[snode] (8) at (6,1) {2};
\node[snode] (9) at (6,-1) {2};
\draw[red, thick, ->] (2)--(7);
\draw[red, thick, ->] (7)--(5);
\draw[red, thick, ->] (7)--(6);
\draw[-] (1) -- (2);
\draw[-] (2) -- (3);
\draw[-, red] (2) -- (5);
\draw[-, red] (2) -- (6);
\draw[-] (5) -- (8);
\draw[-] (6) -- (9);
\draw[line width=0.75mm, black] (5) to (6);
\node[text width=1cm](9) at (2, -2) {$(X'[V^{(I)\,-}_{2,R}])$};
\node[text width=0.1cm](21)[above=0.2 cm of 1]{3};
\node[text width=0.1cm](23) at (4,2){1};
\node[text width=0.1cm](24) at (4,-2){2};
\node[text width=0.1cm](20) at (4.5,0){$2$};
\end{tikzpicture}}
& \qquad \qquad \qquad
& \scalebox{.6}{\begin{tikzpicture}[node distance=2cm,cnode/.style={circle,draw,thick,minimum size=8mm},snode/.style={rectangle,draw,thick,minimum size=8mm},pnode/.style={rectangle,red,draw,thick,minimum size=8mm}]
\node[cnode] (1) at (-3,0) {$2$};
\node[snode] (2) at (-5,0) {$3$};
\node[cnode] (3) at (-1,0) {$2$};
\node[cnode] (4) at (1,0) {$SU(2)$};
\node[cnode] (5) at (3,0) {$1$};
\node[cnode] (6) at (5,0) {$1$};
\node[snode] (7) at (7,0) {$1$};
\draw[thick] (1) -- (2);
\draw[thick] (1) -- (3);
\draw[thick] (3) -- (4);
\draw[thick] (4) -- (5);
\draw[thick] (5) -- (6);
\draw[thick] (6) -- (7);
\node[text width=1cm](36) at (-3, 0.8) {$\wt{W}''^{(1)}_{R}$};
\node[text width=0.1cm](50) at (-3, -0.8) {1};
\node[text width=0.1cm](50) at (-1, -0.8) {2};
\node[text width=0.1cm](50) at (1, -1) {3};
\node[text width=0.1cm](50) at (3, -0.8) {4};
\node[text width=0.1cm](50) at (5, -0.8) {5};
\node[text width=0.1cm](30) at (-1,-3){$(Y''[\wt{W}''^{(1)}_{R}])$};
\end{tikzpicture}}\\
\scalebox{.6}{\begin{tikzpicture}[nnode/.style={circle,draw,thick, red, fill=red!30, minimum size=1.0 cm},
cnode/.style={circle,draw,thick, minimum size=1.0cm},snode/.style={rectangle,draw,thick,minimum size=1cm}]
\node[cnode] (9) at (0,1){1};
\node[snode] (10) at (0,-1){1};
\node[cnode] (11) at (2, 0){2};
\node[cnode] (12) at (4, 1){1};
\node[cnode] (13) at (4, -1){1};
\node[snode] (14) at (6, 2.5){$2$};
\node[snode] (15) at (6, -2.5){$2$};
\node[nnode] (16) at (6,0){$k$};
\draw[red, thick, ->] (16) to [out=60,in=120,looseness=8] (16);
\draw[-] (9) -- (11);
\draw[-] (10) -- (11);
\draw[-] (12) -- (11);
\draw[-] (13) -- (11);
\draw[-] (12) -- (14);
\draw[-] (13) -- (15);
\draw[red,->] (12) -- (16);
\draw[red, line width=0.75mm, ->] (16) -- (13);
\draw[line width=0.75mm, red] (12) to (13);
\node[text width=0.1cm](20) at (4.5,0){$2$};
\node[text width=0.1cm](21)[above=0.2 cm of 9]{3};
\node[text width=0.1cm](23)[above=0.2 cm of 12]{1};
\node[text width=0.1cm](24)[below=0.05 cm of 13]{2};
\node[text width=0.1cm](25) at (5,-1){2};
\node[text width=1.3 cm](31)[below=0.5 cm of 13]{$(X'[V^{(I)\,-}_{0,k,2}])$};
\end{tikzpicture}}
& \qquad \qquad \qquad
& \scalebox{.6}{\begin{tikzpicture}[node distance=2cm,cnode/.style={circle,draw,thick,minimum size=8mm},snode/.style={rectangle,draw,thick,minimum size=8mm},pnode/.style={rectangle,red,draw,thick,minimum size=8mm}]
\node[cnode] (1) at (-3,0) {$2$};
\node[snode] (2) at (-5,0) {$3$};
\node[cnode] (3) at (-1,0) {$2$};
\node[cnode] (4) at (1,0) {$SU(2)$};
\node[cnode] (5) at (3,0) {$1$};
\node[cnode] (6) at (5,0) {$1$};
\node[snode] (7) at (7,0) {$1$};
\draw[thick] (1) -- (2);
\draw[thick] (1) -- (3);
\draw[thick] (3) -- (4);
\draw[thick] (4) -- (5);
\draw[thick] (5) -- (6);
\draw[thick] (6) -- (7);
\node[text width=1cm](36) at (3, 0.8) {$\wt{W}''^{(4)}_{-k}$};
\node[text width=0.1cm](50) at (-3, -0.8) {1};
\node[text width=0.1cm](50) at (-1, -0.8) {2};
\node[text width=0.1cm](50) at (1, -1) {3};
\node[text width=0.1cm](50) at (3, -0.8) {4};
\node[text width=0.1cm](50) at (5, -0.8) {5};
\node[text width=0.1cm](30) at (-1,-3){$(Y''[\wt{W}''^{(4)}_{-k}])$};
\end{tikzpicture}}
\end{tabular}
\end{center}


\section{Coupled quivers for half-BPS vortex defects}

\subsection{Linear Quivers with unitary gauge groups}\label{VD-LQ}

\begin{figure}[htbp]
\begin{center}
\scalebox{0.6}{\begin{tikzpicture}[node distance=2cm,
cnode/.style={circle,draw,thick, minimum size=1.0cm},snode/.style={rectangle,draw,thick,minimum size=1cm}, pnode/.style={circle,double,draw,thick, minimum size=1.0cm}]
\node[cnode] (1) at (-2,0) {$N_1$};
\node[cnode] (2) at (-2,2) {$N_2$};
\node[cnode] (3) at (0,2) {$SU(N_3)$};
\node[cnode] (4) at (0,0) {$SU(N_4)$};
\node[snode] (5) at (-4,0) {$M_1$};
\node[snode] (6) at (-4,2) {$M_2$};
\node[snode] (7) at (0,4) {$M_3$};
\node[snode] (8) at (2,0) {$M_4$};
\draw[line width=0.75mm, black] (1) -- (2);
\draw[-] (2) -- (3);
\draw[-] (3) -- (4);
\draw[-] (1) -- (4);
\draw[-] (1) -- (5);
\draw[-] (2) -- (6);
\draw[-] (3) -- (7);
\draw[-] (4) -- (8);
\draw[-] (2) -- (4);
\draw[-] (1) -- (3);
\node[text width=0.1](40) at (-1.8, 1){$p$};
\node[cnode] (9) at (6,4) {$N$};
\node[text width=3cm](10) at (8.5, 4){$U(N)$ vector multiplet};
\node[cnode] (11) at (6,2.5) {$SU(N)$};
\node[text width=3cm](12) at (8.5, 2.5){$SU(N)$ vector multiplet};
\node[snode] (13) at (6,1) {$M$};
\node[cnode] (14) at (4.5,1) {$G$};
\draw[-] (13)--(14);
\node[text width=3cm](15) at (8.5, 1){$M$ hypers in fund. of $G$};
\node[cnode] (16) at (6, -0.5) {$G'$};
\node[cnode] (17) at (4.5, -0.5) {$G$};
\draw[-] (16)--(17);
\node[text width=3cm](18) at (8.5, - 0.5){Single $G \times G'$ bifund. hyper};
\node[cnode] (31) at (6,-2) {$G'$};
\node[cnode] (32) at (4.5,-2) {$G$};
\draw[line width=0.75mm, black] (31)--(32);
\node[text width=0.2cm](33) at (5.2, -1.5){$p$};
\node[text width=3cm](34) at (8.5, -2){$p$ $G \times G'$ bifund. hypers};
\end{tikzpicture}}
\end{center}
\caption{\footnotesize{A quiver gauge theory with $G=U(N_1) \times U(N_2) \times SU(N_3) \times SU(N_4)$ and hypermultiplets in bifundamental and fundamental representations, with the conventions listed on the RHS.}}
\label{fig: QuivConv}
\end{figure}
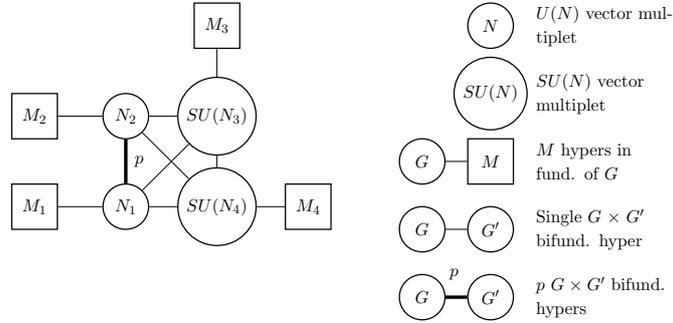

Let us begin by summarizing the 3d $\CN=4$ quiver gauge theory notation that we will be relevant for the rest of the paper,
as given in \figref{fig: QuivConv}.
 
The quivers we consider will also involve hypermultiplets transforming as powers of the determinant representation 
(or anti-determinant representation) of certain unitary factors in the quiver gauge group, which we will denote by the following notation:
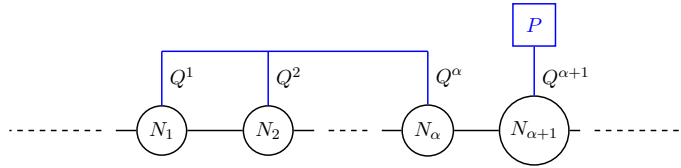
\begin{center}
\scalebox{0.7}{\begin{tikzpicture}[node distance=2cm,cnode/.style={circle,draw,thick,minimum size=8mm},snode/.style={rectangle,draw,thick,minimum size=8mm},pnode/.style={rectangle,red,draw,thick,minimum size=8mm}]
\node[] (10) at (-3,0) {};
\node[] (1) at (-1,0) {};
\node[cnode] (2) at (0,0) {$N_1$};
\node[cnode] (3) at (2,0) {$N_2$};
\node[] (4) at (3,0) {};
\node[] (5) at (4,0) {};
\node[cnode] (6) at (5,0) {$N_{\alpha}$};
\node[cnode] (7) at (7,0) {$N_{\alpha+1}$};
\node[] (8) at (8,0) {};
\node[] (9) at (10,0) {};
\node[snode, blue] (12) at (7,2) {$P$};
\draw[thick] (1) -- (2);
\draw[thick] (2) -- (3);
\draw[thick] (3) -- (4);
\draw[thick,dashed] (4) -- (5);
\draw[thick] (5) -- (6);
\draw[thick] (6) -- (7);
\draw[thick] (7) -- (8);
\draw[thick,blue] (7) -- (12);
\draw[thick,dashed] (8) -- (9);
\draw[thick,dashed] (1) -- (10);
\draw[thick, blue] (2) -- (0,1.5);
\draw[thick, blue] (0,1.5) -- (5,1.5);
\draw[thick, blue] (2,1.5) -- (3);
\draw[thick, blue] (5,1.5) -- (6);
\node[text width=.2cm](15) at (0.25,1){$Q^1$};
\node[text width=.2cm](16) at (2.25,1){$Q^2$};
\node[text width=.2cm](17) at (5.25,1){$Q^\alpha$};
\node[text width=.2cm](18) at (7.25,1){$Q^{\alpha+1}$};
\end{tikzpicture}}
\end{center}
The blue line connecting the gauge nodes $U(N_1)$, $U(N_2)$ and $U(N_\alpha)$
represents a \textit{single} hypermultiplet transforming in powers of the (anti)determinant 
representations of the said unitary groups. The precise powers are given by the charges $\{Q_i=\pm k_i\,N_i\}$,
where $k_i$ is a positive integer, such that the hypermultiplet transforms in the $k_i$-th power of the determinant or anti-determinant 
representation of $U(N_i)$, depending on whether $Q_i \gtrless 0$. Multiple copies of such a hypermultiplet will be represented by a 
thick blue line with an integer denoting the multiplicity. Similarly, $P$ copies of a hypermultiplet transforming under 
the (anti)determinant representation of a single gauge node (and uncharged under the other gauge nodes) by a blue line connected 
to a blue flavor node.\\

\begin{figure}[htbp]
\begin{center}
\begin{tabular}{ccc}
\scalebox{0.8}{\begin{tikzpicture}[node distance=2cm, nnode/.style={circle,draw,thick, fill, inner sep=1 pt},cnode/.style={circle,draw,thick,minimum size=1.0 cm},snode/.style={rectangle,draw,thick,minimum size=1.0 cm},rnode/.style={red, circle,draw,thick,fill=red!30 ,minimum size=1.0cm}]
\node[nnode] (1) at (-1.5,0){} ;
\node[nnode] (2) at (-1,0){} ;
\node[](3) at (-0.5,0){};
\node[cnode] (4) at (1,0) {$P$};
\node[cf-group] (5) at (3,0) {\rotatebox{-90}{$N$}};
\node[cf-group] (6) at (5,0) {\rotatebox{-90}{$M$}};
\node[snode] (12) at (2,-1.5) {$K-Q$};
\node[snode] (13) at (4,-1.5) {$Q$};
\node[rnode] (7) at (4,2) {$\Sigma$};
\draw[thick,-] (3) -- (4);
\draw[thick,-] (4) -- (5);
\draw[thick, -] (5) to (6);
\draw[thick, -] (5) to (12);
\draw[thick, -] (5) to (13);
\draw[red, thick,->] (5) -- (7);
\draw[red, thick,->] (7) -- (6);
\draw[red, thick,->] (7) -- (13);
\node[] (9) at (6.5,0){} ;
\node[nnode] (10) at (7,0){} ;
\node[nnode] (11) at (7.5,0){} ;
\draw[-] (6) -- (9);
\node[text width=0.1cm](20)[below=2.2 cm of 5]{$(\CT[V_{Q,R}^r])$};
\end{tikzpicture}}
& \qquad \qquad
& \scalebox{0.8}{\begin{tikzpicture}[node distance=2cm, nnode/.style={circle,draw,thick, fill, inner sep=1 pt},cnode/.style={circle,draw,thick,minimum size=1.0 cm},snode/.style={rectangle,draw,thick,minimum size=1.0 cm},rnode/.style={red, circle,draw,thick,fill=red!30 ,minimum size=1.0cm}]
\node[nnode] (2) at (0.5,0){} ;
\node[nnode] (3) at (1,0){} ;
\node[](4) at (1.5,0){};
\node[cf-group] (5) at (3,0) {\rotatebox{-90}{$P$}};
\node[cf-group] (6) at (5,0) {\rotatebox{-90}{$N$}};
\node[cnode] (8) at (7,0) {$M$};
\node[snode] (12) at (4,-1.5) {$K-Q$};
\node[snode] (13) at (6,-1.5) {$Q$};
\node[rnode] (7) at (4,2) {$\Sigma$};
\draw[thick,-] (4) -- (5);
\draw[thick, -] (5) to (6);
\draw[thick, -] (8) to (6);
\draw[thick, -] (6) to (12);
\draw[thick, -] (6) to (13);
\draw[red, thick,->] (5) -- (7);
\draw[red, thick,->] (7) -- (6);
\draw[red, thick,->] (12) -- (7);
\node[] (9) at (8.5,0){} ;
\node[nnode] (10) at (9,0){} ;
\node[nnode] (11) at (9.5,0){} ;
\draw[-] (8) -- (9);
\node[text width=0.1cm](20)[below=2.2 cm of 6]{$(\CT[V_{Q,R}^l])$};
\end{tikzpicture}}
\end{tabular}
\caption{\footnotesize{A pair of generic 3d-1d systems realizing a vortex defect in the 3d linear quiver.}}
\label{3d-1d-gen}
\end{center}
\end{figure}
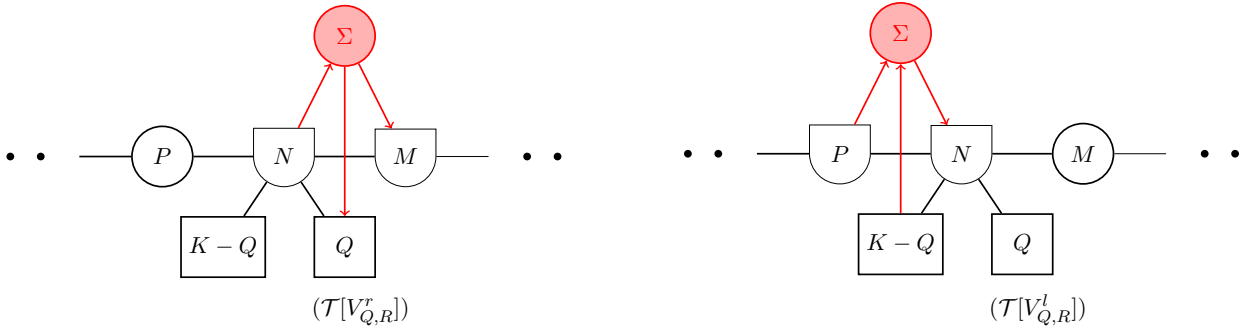

\begin{figure}[htbp]
\begin{center}
\scalebox{0.8}{\begin{tikzpicture}[
nnode/.style={circle,draw,thick, fill=blue,minimum size= 6mm},cnode/.style={circle,draw,thick,minimum size=4mm},snode/.style={rectangle,draw,thick,minimum size=6mm},rnode/.style={red, circle,draw,thick,fill=red!30 ,minimum size=4mm},rrnode/.style={red, circle,draw,thick,fill=red!30 ,minimum size=1.0cm}]
\node[rnode] (1) at (-4,0) {$n_1$};
\node[rnode] (2) at (-2,0) {$n_2$};
\node[] (3) at (0,0.2){};
\node[] (4) at (0,-0.2){};
\node[circle,draw,thick, fill, inner sep=1 pt] (5) at (0,0){} ;
\node[circle,draw,thick, fill, inner sep=1 pt] (6) at (0.5,0){} ;
\node[circle,draw,thick, fill, inner sep=1 pt] (7) at (1,0){} ;
\node[] (8) at (1,0.2){};
\node[] (9) at (1,-0.2){};
\node[rnode] (10) at (3,0) {$n_{P-1}$};
\node[rnode] (11) at (5,0) {$n_{P}$};
\node[snode] (12) at (7,1) {$N_R$};
\node[snode] (13) at (7,-1) {$N_L$};
\draw[red, thick, ->] (1) to [out=30,in=150] (2);
\draw[red, thick, ->] (2) to [out=210,in=330] (1);
\draw[red, thick, ->] (2) to [out=30,in=150] (3);
\draw[red, thick, ->] (4) to [out=210,in=330] (2);
\draw[red, thick, ->] (8) to [out=30,in=150] (10);
\draw[red, thick, ->] (10) to [out=210,in=330] (9);
\draw[red, thick, ->] (10) to [out=30,in=150] (11);
\draw[red, thick, ->] (11) to [out=210,in=330] (10);
\draw[red, thick, ->] (11) to (12.south west);
\draw[red, thick, ->] (13.north west) to (11);
\draw[red, thick, ->] (1) to [out=60,in=120,looseness=8] (1);
\draw[red, thick, ->] (2) to [out=60,in=120,looseness=8] (2);
\draw[red, thick, ->] (10) to [out=60,in=120,looseness=8] (10);
\draw[red, thick, ->] (11) to [out=60,in=120,looseness=8] (11);
\node[] (17) at (8,0) {$\Longleftrightarrow$};
\node[rrnode] (14) at (10,0) {$\Sigma$};
\node[snode] (15) at (12,1) {$N_R$};
\node[snode] (16) at (12,-1) {$N_L$};
\draw[red, thick, ->] (14) to (15.south west);
\draw[red, thick, ->] (16.north west) to (14);
\end{tikzpicture}}
\caption{\footnotesize{The generic form of a (2,2) SQM that realizes a vortex defect. Note that $N_L=N$, $N_R=M+Q$ for the right SQM, and $N_L=P+K-Q$, $N_R=N$ for the left SQM.}}
\label{1d-gen}
\end{center}
\end{figure}
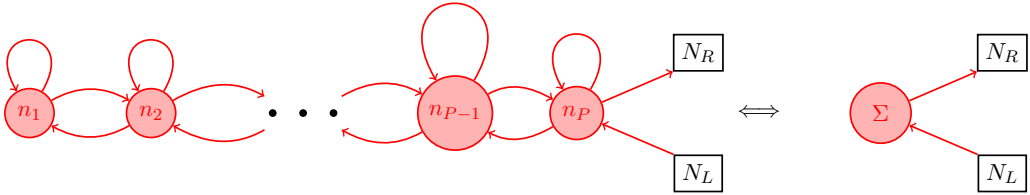

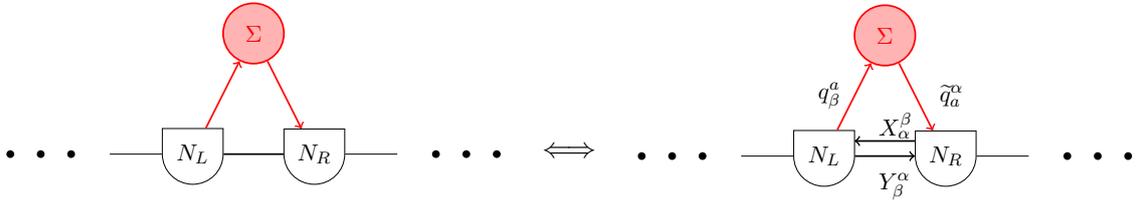
\begin{figure}[htbp]
\begin{center}
\begin{tabular}{ccc}
\scalebox{0.8}{\begin{tikzpicture}[node distance=2cm, nnode/.style={circle,draw,thick, fill, inner sep=1 pt},cnode/.style={circle,draw,thick,minimum size=1.0 cm},snode/.style={rectangle,draw,thick,minimum size=1.0 cm},rnode/.style={red, circle,draw,thick,fill=red!30 ,minimum size=1.0cm}]
\node[nnode] (1) at (0,0){} ;
\node[nnode] (2) at (0.5,0){} ;
\node[nnode] (3) at (1,0){} ;
\node[](4) at (1.5,0){};
\node[cf-group] (5) at (3,0) {\rotatebox{-90}{$N_L$}};
\node[cf-group] (6) at (5,0) {\rotatebox{-90}{$N_R$}};
\node[rnode] (7) at (4,2) {$\Sigma$};
\draw[-] (4) -- (5);
\draw[thick, -] (5) to (6);
\draw[red, thick,->] (5) -- (7);
\draw[red, thick,->] (7) -- (6);
\node[] (9) at (6.5,0){} ;
\node[nnode] (10) at (7,0){} ;
\node[nnode] (11) at (7.5,0){} ;
\node[nnode](12) at (8,0){};
\draw[-] (6) -- (9);
\node[text width=0.2cm] (16) at (4,-0.75){};
\end{tikzpicture}}
&{\begin{tikzpicture}\node[] (17) at (0,0) {$\Longleftrightarrow$}; 
\node[] (18) at (0,-0.6) {};
\end{tikzpicture}}
&\scalebox{0.8}{\begin{tikzpicture}[node distance=2cm, nnode/.style={circle,draw,thick, fill, inner sep=1 pt},cnode/.style={circle,draw,thick,minimum size=1.0 cm},snode/.style={rectangle,draw,thick,minimum size=1.0 cm},rnode/.style={red, circle,draw,thick,fill=red!30 ,minimum size=1.0cm}]
\node[nnode] (1) at (0,0){} ;
\node[nnode] (2) at (0.5,0){} ;
\node[nnode] (3) at (1,0){} ;
\node[](4) at (1.5,0){};
\node[cf-group] (5) at (3,0) {\rotatebox{-90}{$N_L$}};
\node[cf-group] (6) at (5,0) {\rotatebox{-90}{$N_R$}};
\node[rnode] (7) at (4,2) {$\Sigma$};
\draw[-] (4) -- (5);
\draw[thick, ->] (5) to (6);
\draw[thick, ->] (4.5,0.25) to (3.5, 0.25);
\draw[red, thick,->] (5) -- (7);
\draw[red, thick,->] (7) -- (6);
\node[] (9) at (6.5,0){} ;
\node[nnode] (10) at (7,0){} ;
\node[nnode] (11) at (7.5,0){} ;
\node[nnode](12) at (8,0){};
\draw[-] (6) -- (9);
\node[text width=0.2cm] (13) at (3,1){$q^a_\beta$};
\node[text width=0.2cm] (14) at (5,1){$\tq^\alpha_a$};
\node[text width=0.2cm] (15) at (4,0.5){$X^\beta_\alpha$};
\node[text width=0.2cm] (16) at (4,-0.5){$Y^\alpha_\beta$};
\end{tikzpicture}}
\end{tabular}
\caption{\footnotesize{3d and 1d fields in the cubic superpotential.}}
\label{3d1dquiv-gen}
\end{center}
\end{figure}

Next, let us summarize certain basic facts about a vortex defect in the 3d theory. 
For linear quivers with unitary gauge groups, there exists a Type IIB construction that realizes a large class 
of vortex defects \cite{Assel:2015oxa} in terms of coupled 3d-1d system. We will present a brief 
description of these coupled systems.

Given a 3d linear quiver $\CT$, there exists a pair of 3d-1d systems $\CT[V_{Q,R}^{r,l}]$, given 
in \figref{3d-1d-gen}, which realize a vortex defect associated with the gauge node $U(N)$. 
The 1d theory, which we will denote as $\Sigma_{r,l}^{Q,R}$, is a (2,2) SQM and its form is explicitly shown in \figref{1d-gen}. 
The coupled system $\CT[V_{Q,R}^{r,l}]$  is labelled by an integer $Q$ and a representation $R$ of $U(N)$.
The data of the SQM, including the ranks of the 1d gauge nodes, can be read off from the representation $R$.
In addition, one has to specify the signs of the FI parameters associated with the 1d gauge nodes. For the SQM $\Sigma^{Q,R}_{r}$,
all the FI parameters should be taken to be negative, while for the SQM $\Sigma^{Q,R}_{l}$ all the FI parameters should be taken 
to be positive. \\
 
A given coupled system is obtained in two steps. First, the $U(N_L) \times U(N_R)$ flavor symmetry of the SQM due to chiral multiplets 
is weakly gauged by background 3d vector multiplets. 
This is achieved by the following cubic superpotential (localized on the defect world volume) involving the 1d chiral multiplets 
and the 3d hypermultiplets :
\be \label{3d-1dSup}
\wt{W}_0 = q^a_{\,\,\beta}\,X^\beta_{\,\,\alpha}\, \tq^\alpha_{\,\,a}|_{\rm def} + \ldots,
\ee
where the $\ldots$ in the superpotential denote additional terms which contain higher derivatives of the complex scalar $X$ \cite{Dimofte:2019zzj}, 
and the indices run over $a=1,\ldots, n_P, \, \beta= 1,\ldots,N_L, \, \alpha=1,\ldots,N_R$ \footnote{We have decomposed the 
3d $\CN=4$ bifundamental hypermultiplet into a pair of $\CN=2$ chiral multiplets -- $X$ and $Y$, which transform in the 
bifundamental representation of $U(N_L) \times U(N_R)$ and $U(N_R) \times U(N_L)$ respectively}. 
The cubic superpotential, which preserves the (2,2) supersymmetry of the system, reduces the flavor symmetry 
$(U(N_L) \times U(N_R))_{1d} \times (U(N_L) \times U(N_R))_{3d}$ to the diagonal subgroup $G_{3d-1d}=(U(N_L) \times U(N_R))_{3d-1d}$. 

In the next step, one promotes the background vector multiplet 
for a subgroup of $G_{3d-1d}$ to a dynamical vector multiplet. For example, in constructing the 3d-1d 
system $\CT[V_{Q,R}^{r}]$, where $N_L=N$, $N_R=M+Q$, one promotes the background vector multiplet for 
$U(N) \times U(M) \subset G_{3d-1d}$ to a dynamical vector multiplet. In contrast, for the system $\CT[V_{Q,R}^{l}]$, 
we have $N_L=P+K-Q$, $N_R=N$, and we promote $U(P) \times U(N) \subset G_{3d-1d}$ to a dynamical vector multiplet.\\

A given vortex defect may be realized by multiple 3d-1d coupled quivers -- this is referred to as the
hopping duality \cite{Assel:2015oxa}. Using the Type IIB description of the vortex defects in linear quivers, 
it was demonstrated in \cite{Assel:2015oxa} that the two coupled quivers $\CT[V_{Q,R}^r]$ and 
$\CT[V_{Q,R}^l]$ in \figref{3d-1d-gen} describe the same vortex defect. $\CT[V_{Q,R}^r]$ and 
$\CT[V_{Q,R}^l]$ are therefore hopping duals. 

The hopping duality can also be inferred by studying the sphere partition functions of these defects. 
The expectation values of the vortex defect $V^{l,r}_{Q,R}$ in the theory $\CT$ is given as:
\begin{align}\label{vev-V-LQ}
\langle{V^{l,r}_{Q,R}}\rangle_{\CT}= W^{l,r}_{\rm b.g.}(\vec t) \times \frac{1}{Z^{(\CT)}(\vec m; \vec t)} \times \lim_{z\to 1} \int  \Big[d\vec s\Big] \, Z^{(\CT)}_{\rm int}(\vec s, \vec m, \vec t) \cdot \CI^{\Sigma^{Q,R}_{l,r}}(\vec s, \vec m, z| \vec \xi \gtrless 0),
\end{align}
where $W^{l,r}_{\rm b.g.}$ are certain background Wilson defects, $Z^{(\CT)}_{\rm int}$ is the integrand of the sphere partition function arising from the 3d bulk fields, $ \CI^{\Sigma^{Q,R}_{l,r}}$ is the Witten index of the coupled SQM, and 
the superscripts $l,r$ denote the choice of the specific 3d-1d system (left or right) \footnote{The 
correct prescription for taking the limit is to first analytically continue $z \in i \BR$ in the integrand, perform the integration, 
and then finally set $z=1$.}.  
A prescription for the signs of the FI parameters determines the specific chamber 
in which the Witten index is evaluated. The Witten indices for the right and the left SQM associated with the vortex defect 
in \figref{3d-1d-gen} are explicitly given as
\begin{align}
& \CI^{\Sigma^{Q,R}_{r}}=\sum_{w \in R}\, \CF^r(\vec s, z)\, \prod^N_{j=1} \prod^M_{i=1} \frac{\ch{(s^{(N)}_j - s^{(M)}_i)}}{\ch{(s^{(N)}_j  + iw_jz -s^{(M)}_i)}}\,\prod^Q_{a=1}\frac{\ch{(s^{(N)}_j - m_a)}}{\ch{(s^{(N)}_j  + iw_jz -m_a)}} , \label{Z1d-r}\\
& \CI^{\Sigma^{Q,R}_{l}}= \sum_{w \in R}\, \CF^l(\vec s, z)\, \prod^N_{j=1} \prod^P_{l=1} \frac{\ch{(s^{(N)}_j - s^{(P)}_l)}}{\ch{(s^{(N)}_j  - iw_jz -s^{(P)}_l)}}\,\prod^{K-Q}_{b=1}\frac{\ch{(s^{(N)}_j - m_b)}}{\ch{(s^{(N)}_j  - iw_jz -m_b)}} ,\label{Z1d-l}
\end{align}
where $w= (w_1, \ldots, w_N)$ are the weights of the representation $R$ of $U(N)$. The functions $\CF^{r,l}(\vec s, z)$ 
have poles which give zero residues in the limit $z \to 1$, and therefore can be simply replaced by their respective $z \to 1$ limits 
(which turn out to be overall signs) in the formula for the expectation value. The background Wilson defects for the left and the right 
SQM are given by
\be
W^{r}_{\rm b.g.}(\vec t)= e^{2\pi |R| t_r}, \qquad W^{l}_{\rm b.g.}(\vec t)= e^{2\pi |R| t_l},
\ee
where the FI parameter of the $U(N)$ gauge group is given by $\eta^{(N)}= t_{l} -t_r$.  
The integer $|R|$ is the number of boxes in the Young diagram associated with the representation $R$.

The hopping duality can then be understood as a simple change of integration
variables in the defect partition function. 
Consider the partition function of the right 3d-1d quiver, given by \eref{vev-V-LQ}-\eref{Z1d-r}, 
and implement the following transformation:
\be
s^{(N)}_j \to s^{(N)}_j - iw_j\,z ,\qquad \forall j =1,\ldots, N,
\ee
keeping all the other integration variables fixed. Since we are analytically continuing $z \in i \BR$, this transformation 
does not introduce or remove any new poles. One can readily check that the resultant expression gives the partition function of the left 
3d-1d quiver in \eref{vev-V-LQ}-\eref{Z1d-l}.

\subsection{Quivers with multiple edges}\label{VD-edges}

Next, we describe coupled 3d-1d quivers where the 1d theory is coupled to a pair of 3d gauge nodes connected by multiple 
bifundamental hypermultiplets. Let us denote a generic 3d theory in this class as $\CT$, and 
zoom in on a part of the quiver consisting of two consecutive gauge nodes -- $U(N)$ 
and $U(M)$, connected by $p$ bifundamental hypers. These bifundamentals are associated with a $U(p)$ flavor symmetry 
(or, $SU(p)$ flavor symmetry, but this distinction will not be important for our presentation).\\

\begin{figure}[htbp]
\begin{center}
\scalebox{0.8}{\begin{tikzpicture}[
nnode/.style={circle,draw,thick, fill=blue,minimum size= 6mm},cnode/.style={circle,draw,thick,minimum size=4mm},snode/.style={rectangle,draw,thick,minimum size=6mm},rnode/.style={red, circle,draw,thick,fill=red!30 ,minimum size=4mm},rrnode/.style={red, circle,draw,thick,fill=red!30 ,minimum size=1.0cm}]
\node[rnode] (1) at (-4,0) {$n_1$};
\node[rnode] (2) at (-2,0) {$n_2$};
\node[] (3) at (0,0.2){};
\node[] (4) at (0,-0.2){};
\node[circle,draw,thick, fill, inner sep=1 pt] (5) at (0,0){} ;
\node[circle,draw,thick, fill, inner sep=1 pt] (6) at (0.5,0){} ;
\node[circle,draw,thick, fill, inner sep=1 pt] (7) at (1,0){} ;
\node[] (8) at (1,0.2){};
\node[] (9) at (1,-0.2){};
\node[rnode] (10) at (3,0) {$n_{P-1}$};
\node[rnode] (11) at (5,0) {$n_{P}$};
\node[snode] (12) at (7,1) {$N_R$};
\node[snode] (13) at (7,-1) {$N_L$};
\draw[red, thick, ->] (1) to [out=30,in=150] (2);
\draw[red, thick, ->] (2) to [out=210,in=330] (1);
\draw[red, thick, ->] (2) to [out=30,in=150] (3);
\draw[red, thick, ->] (4) to [out=210,in=330] (2);
\draw[red, thick, ->] (8) to [out=30,in=150] (10);
\draw[red, thick, ->] (10) to [out=210,in=330] (9);
\draw[red, thick, ->] (10) to [out=30,in=150] (11);
\draw[red, thick, ->] (11) to [out=210,in=330] (10);
\draw[line width=0.75mm, red, ->] (11) to (12.south west);
\draw[red, thick, ->] (13.north west) to (11);
\draw[red, thick, ->] (1) to [out=60,in=120,looseness=8] (1);
\draw[red, thick, ->] (2) to [out=60,in=120,looseness=8] (2);
\draw[red, thick, ->] (10) to [out=60,in=120,looseness=8] (10);
\draw[red, thick, ->] (11) to [out=60,in=120,looseness=8] (11);
\node[text width=0.2cm] (20) at (6,0.75){$p'$};
\node[] (17) at (8,0) {$\Longleftrightarrow$};
\node[rrnode] (14) at (10,0) {$\Sigma$};
\node[snode] (15) at (12,1) {$N_R$};
\node[snode] (16) at (12,-1) {$N_L$};
\draw[line width=0.75mm, red, ->] (14) to (15.south west);
\draw[red, thick, ->] (16.north west) to (14);
\node[text width=0.2cm] (17) at (11,0.75){$p'$};
\end{tikzpicture}}
\caption{\footnotesize{Note that $N_L=N$, $N_R=M+Q$ for the right SQM. The SQM has a global symmetry 
$U(p') \times U(N_R) \times U(N_L)$.}}
\label{1d-gen-2}
\end{center}
\end{figure}
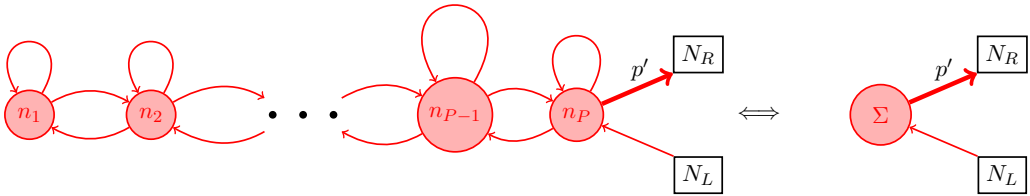
 
Now, consider an SQM of the form in \figref{1d-gen-2} where $N_R=N$, $N_L=Q+M$ and a generic $p'<p$, such that the 1d theory has a 
global symmetry $U(p') \times U(N_R) \times U(N_L)$. One way of obtaining this SQM is to start from an SQM with the 
same gauge group and bifundamental matter, but with $p'N_R$ anti-fundamental chirals and $N_L$ fundamental hypers. 
The global symmetry arising from the anti-fundamental chirals has a subgroup $U(N_R)^{p'}$. Identifying these $p'$ $U(N_R)$
factors to a single $U(N_R)$ factor, we obtain the SQM in \figref{1d-gen-2}. In \appref{sec:WI-app}, we demonstrate how this 
operation can be implemented in terms of the Witten index. 

\begin{figure}[htbp]
\begin{center}
 \scalebox{.7}{\begin{tikzpicture}[node distance=2cm, nnode/.style={circle,draw,thick, fill, inner sep=1 pt},cnode/.style={circle,draw,thick,minimum size=1.0 cm},snode/.style={rectangle,draw,thick,minimum size=1.0 cm},rnode/.style={red, circle,draw,thick,fill=red!30 ,minimum size=1.0cm}]
 \node[cnode] (20) at (-4,0){$P$} ;
 \node[nnode] (1) at (-6.5,0){} ;
\node[nnode] (2) at (-6,0){} ;
\node[](3) at (-5.5,0){};
\node[cf-group] (4) at (-2,0) {\rotatebox{-90}{$N$}};
\node[cf-group] (5) at (2,0) {\rotatebox{-90}{$M$}};
\node[snode] (6) at (-2,-2) {$K-Q$};
\node[snode] (11) at (0,-2) {$Q$};
\draw[thick] (3) -- (20);
\draw[thick] (4) -- (20);
\draw[thick] (4) -- (6);
\node[rnode] (7) at (0,3) {$\Sigma$};
\node[](8) at (3.5,0){};
\node[nnode] (9) at (4,0){} ;
\node[nnode] (10) at (4.5,0){} ;
\draw[thick] (5) -- (8);
\draw[thick] (4) -- (11);
\draw[line width=0.75mm, red] (4.south east) to (5.north east);
\draw[line width=0.75mm, black] (4.south west) to (5.north west);
\draw[red, thick, ->] (4) -- (7);
\draw[line width=0.75mm, red, ->] (7) -- (5);
\draw[red, thick, ->] (7) -- (11);
\node[text width=0.1cm](15) at (0,0.75){$p'$};
\node[text width=0.1cm](16) at (1.5, 2){$p'$};
\node[text width=1cm](17) at (0, -0.5){$p-p'$};
\node[text width=0.1cm](18) at (-1,-3){$(\CT[V_{Q,R,p'}^{(I)\,-}])$};
\end{tikzpicture}}
\caption{\footnotesize{A generic coupled 3d-1d system that realizes a vortex defect for the gauge node $U(N)$. 
The $-$ superscript implies that the signs of the 1d FI parameters are all negative. The thick red line connecting the 
$U(N)$ and $U(M)$ gauge nodes represents the bifundamental hypers that enter the cubic superpotential.}}
\label{3d1d-edges}
\end{center}
\end{figure}
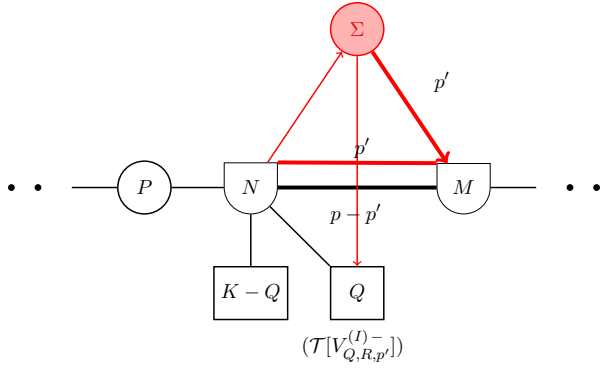

The $U(p') \times U(N_R) \times U(N_L)$ global symmetry is then weakly gauged by 3d background vector 
multiplets, as before. The $U(p')$ factor is identified with a $U(p') \subset U(p)$ subgroup of the 3d global symmetry arising 
from the $p'$ bifundamental hypers. This is achieved by turning on the cubic superpotential
\be \label{3d-1dSup-2}
\wt{W}_0 = q^a_{\,\,\beta}\,X^\beta_{\,\,i\,\alpha}\, \tq^{\alpha\,\,i}_{\,\,a}|_{\rm def} + \ldots,
\ee
where $i=1,\ldots,p'$, and the other indices run over the same range as above. In the final step, one promotes 
the background vector multiplet for a $U(N) \times U(M)$ subgroup of the $U(p') \times U(N_R) \times U(N_L)$ 
global symmetry to a dynamical vector multiplet. This leads to the 3d-1d coupled quiver of the form given in 
\figref{3d1d-edges}.\\

We propose that the coupled 3d-1d quivers realize a class of vortex defects associated with the $U(N)$ gauge 
node in the theory $\CT$ is given in \figref{3d1d-edges}. 
The resultant defects are denoted as $V^{(I)\,-}_{Q,R,p'}$, where $Q$ is an integer and $R$ is a representation of $U(N)$, 
while the integer $p' <p$ depends on the number of bifundamental hypers involved in 
the cubic superpotential \eqref{3d-1dSup-2}. The negative sign in the subscript implies that all the 1d FI 
parameters should be taken to be negative.

The expectation value of the vortex defect $V^{(I)\,-}_{Q,R,p'}$ in the theory $\CT$ on a round three sphere 
is given as:
\begin{align}\label{vev-V-nADE-1}
\langle{V^{(I)\,-}_{Q,R,p'}}\rangle_{\CT}= W^{-}_{\rm b.g.}(\vec \eta) \times \frac{1}{Z^{(\CT)}(\vec m; \vec \eta)} \times \lim_{z\to 1} \int  \Big[d\vec s\Big] \, Z^{(\CT)}_{\rm int}(\vec s, \vec m, \vec \eta) \, \CI^{\Sigma^{Q,R,p'}_{(I)\,-}}(\vec s, \vec m, z| \vec \xi < 0),
\end{align}
where $W^{-}_{\rm b.g.}$ is a background Wilson defect, $Z^{(\CT)}_{\rm int}$ is the integrand of the sphere partition function arising from the 3d bulk fields, 
and $\CI^{\Sigma^{Q,R,p'}_{(I)\,-}}$ is the Witten index of the coupled SQM. The latter is given as :
\begin{align}
\CI^{\Sigma^{Q,R,p'}_{(I)\,-}} = \sum_{w \in R}\,\prod^N_{j=1} \prod^{p'}_{l=1} \prod^M_{i=1} \frac{\ch{(s^{(N)}_j - s^{(M)}_i -\mu^l)}}{\ch{(s^{(N)}_j  + iw_jz -s^{(M)}_i -\mu^l)}}\times \prod^N_{j=1} \prod^Q_{a=1}\frac{\ch{(s^{(N)}_j - m_a)}}{\ch{(s^{(N)}_j  + iw_jz -m_a)}},
\end{align}
where $\{\mu^l\}_{l=1,\ldots,p}$ denote real masses of the $p$ bifundamental hypermultiplets, $\{m_a\}$ denote the real masses of the 
$Q$ hypers in the fundamental representation of $U(N)$, and $R$ is a representation of $U(N)$. Plugging in the expression for the 
Witten index, the expectation value for the vortex defect $V^{(I)\,-}_{Q,R,p'}$ can be written as:
\begin{align}\label{vev-V-nADE-2}
\langle{V^{(I)\,-}_{Q,R,p'}}\rangle_{\CT}= \frac{W^{-}_{\rm b.g.}(\vec \eta)}{Z^{(\CT)}}\, \lim_{z\to 1}\, \sum_{w \in R}\, & \int  \Big[d\vec s\Big] \,
\Big[\ldots \Big]\, \frac{\prod_{k<k'} \sinh^2{\pi s^{(P)}_{kk'}}\,\prod_{j<j'} \sinh^2{\pi s^{(N)}_{jj'}}\, \prod_{i<i'} \sinh^2{\pi s^{(M)}_{ii'}}}
{\prod_{j,k}\ch{(s^{(N)}_{j} - s^{(P)}_{k})}\, \prod^{K-Q}_{b=1} \ch{(s^{(N)}_{j} -\tm_b)}} \nn \\
& \times \frac{1}{\prod_{j,i}\prod^p_{l=p'+1} \ch{(s^{(N)}_j - s^{(M)}_i -\mu^l)}} \nn \\
& \times \frac{1}{\prod^{p'}_{l=1} \ch{(s^{(N)}_j - s^{(M)}_i -\mu^l+ iw_jz)}\,\prod^Q_{a=1}\ch{(s^{(N)}_j -m_a + iw_jz)}},
\end{align}
where $\Big[\ldots \Big]$ denotes the contribution of the rest of the quiver, and won't be relevant for the discussion here.\\

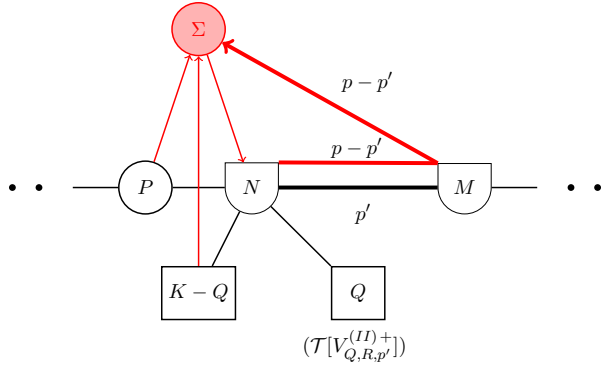
\begin{figure}[htbp]
\begin{center}
 \scalebox{.7}{\begin{tikzpicture}[node distance=2cm, nnode/.style={circle,draw,thick, fill, inner sep=1 pt},cnode/.style={circle,draw,thick,minimum size=1.0 cm},snode/.style={rectangle,draw,thick,minimum size=1.0 cm},rnode/.style={red, circle,draw,thick,fill=red!30 ,minimum size=1.0cm}]
 \node[cnode] (20) at (-4,0){$P$} ;
 \node[nnode] (1) at (-6.5,0){} ;
\node[nnode] (2) at (-6,0){} ;
\node[](3) at (-5.5,0){};
\node[cf-group] (4) at (-2,0) {\rotatebox{-90}{$N$}};
\node[cf-group] (5) at (2,0) {\rotatebox{-90}{$M$}};
\node[snode] (6) at (-3,-2) {$K-Q$};
\node[snode] (11) at (0,-2) {$Q$};
\draw[thick] (3) -- (20);
\draw[thick] (4) -- (20);
\draw[thick] (4) -- (6);
\node[rnode] (7) at (-3,3) {$\Sigma$};
\node[](8) at (3.5,0){};
\node[nnode] (9) at (4,0){} ;
\node[nnode] (10) at (4.5,0){} ;
\draw[thick] (5) -- (8);
\draw[thick] (4) -- (11);
\draw[line width=0.75mm, red] (4.south east) to (5.north east);
\draw[line width=0.75mm, black] (4.south west) to (5.north west);
\draw[red, thick, ->] (20) -- (7);
\draw[red, thick, ->] (6) -- (7);
\draw[red, thick, ->] (7) -- (4);
\draw[line width=0.75mm, red, ->] (5.north east) -- (7);
\node[text width= 1 cm](15) at (0,0.75){$p-p'$};
\node[text width=1 cm](16) at (0.2, 2){$p-p'$};
\node[text width=0.1cm](17) at (0, -0.5){$p'$};
\node[text width=0.1cm](18) at (-1,-3){$(\CT[V_{Q,R,p'}^{(II)\,+}])$};
\end{tikzpicture}}
\caption{\footnotesize{A coupled 3d-1d quiver which is a hopping dual for the quiver in \figref{3d1d-edges}, i.e. it realizes 
the same vortex defect.  
The $+$ subscript implies that the signs of the 1d FI parameters are all positive.}}
\label{3d1d-edges-hd}
\end{center}
\end{figure}

Similar to the case of linear quiver, there may be multiple 3d-1d systems that realize the same vortex defect in the theory $\CT$. 
Generically, there is at least one such hopping dual for the 3d-1d system in \figref{3d1d-edges}. This system can be read off by implementing 
the change of variables 
\be \label{cov-V}
s^{(N)}_j \to s^{(N)}_j - iw_j\,z ,\qquad \forall j =1,\ldots, N,
\ee
in \eqref{vev-V-nADE-2} keeping all the other integration variables fixed, which gives:
\begin{align}\label{vev-V-nADE-3}
\langle{V^{(I)\,-}_{Q,R,p'}}\rangle_{\CT}= \frac{W^{+}_{\rm b.g.}(\vec \eta)}{Z^{(\CT)}}\, \lim_{z\to 1}\, \sum_{w \in R}\, & \int  \Big[d\vec s\Big] \,
\Big[\ldots \Big]\, \frac{\prod_{k<k'} \sinh^2{\pi s^{(P)}_{kk'}}\,\prod_{j<j'} \sinh^2{\pi s^{(N)}_{jj'}}\, \prod_{i<i'} \sinh^2{\pi s^{(M)}_{ii'}}}
{\prod_{j,k}\ch{(s^{(N)}_{j} -iw_j\,z - s^{(P)}_{k})}\, \prod_{b} \ch{(s^{(N)}_{j}-iw_j\,z -\tm_b)}} \nn \\
& \times \frac{1}{\prod_{j,i}\prod^p_{l=p'+1} \ch{(s^{(N)}_j -iw_j\,z - s^{(M)}_i -\mu^l)}} \nn \\
& \times \frac{1}{\prod^{p'}_{l=1} \ch{(s^{(N)}_j - s^{(M)}_i -\mu^l)}\,\prod^Q_{a=1}\ch{(s^{(N)}_j -m_a)}},
\end{align}
where $W^{+}_{\rm b.g.}(\vec \eta)$ is a background Wilson defect. The RHS of the above equation can be readily identified as the partition 
function of another 3d-1d quiver $\Sigma^{Q,R,p'}_{(II)\,+}$ (normalized by the 3d partition function) shown in \figref{3d1d-edges-hd}. 
The $+$ sign in the subscript implies that the Witten index should be computed with all the 1d FI parameters being positive. We will 
denote the associated vortex defect as $V^{(II)\,+}_{Q,R,p'}$. The expression on the RHS of \eref{vev-V-nADE-3} therefore implies:
\be
\boxed{\langle{V^{(I)\,-}_{Q,R,p'}}\rangle_{\CT}= \langle{V^{(II)\,+}_{Q,R,p'}}\rangle_{\CT},}
\ee
which shows that the coupled quivers $\CT[\Sigma^{Q,R,p'}_{(I)\,-}]$ and $\CT[\Sigma^{Q,R,p'}_{(II)\,+}]$ are related by a hopping duality.

In \Secref{sec:Ab-Ex} and \Secref{sec:NAb-Ex}, we will explicitly construct 3d-1d quivers of the form 
\figref{3d1d-edges}-\figref{3d1d-edges-hd} using $S$-type operations, and check that they indeed map 
to Wilson defects under mirror symmetry.


\section{Defects in an Abelian quiver}\label{sec:Ab-Ex}

\begin{figure}[htbp]
\begin{center}
\scalebox{0.6}{\begin{tikzpicture}[node distance=2cm,cnode/.style={circle,draw,thick,minimum size=8mm},snode/.style={rectangle,draw,thick,minimum size=8mm},pnode/.style={rectangle,red,draw,thick,minimum size=8mm}]
\node[cnode] (1) at (-2,0) {$1$};
\node[cnode] (2) at (2,0) {$1$};
\node[snode] (3) at (-2,-2) {$n-p$};
\node[snode] (4) at (2,-2) {$l$};
\draw[thick] (1) -- (3);
\draw[thick] (2) -- (4);
\draw[line width=0.75mm, black] (1) to (2);
\node[text width=0.1cm](15) at (0,0.25){$p$};
\node[text width=0.1cm](16) at (0,-3){$(X')$};
\end{tikzpicture}}
\qquad \qquad
\scalebox{0.5}{\begin{tikzpicture}[node distance=2cm,cnode/.style={circle,draw,thick,minimum size=8mm},snode/.style={rectangle,draw,thick,minimum size=8mm},pnode/.style={rectangle,red,draw,thick,minimum size=8mm}]
\node[snode] (1) at (-2,0) {$1$};
\node[cnode] (2) at (0,0) {$1$};
\node[cnode] (3) at (2,0) {$1$};
\node[] (4) at (3,0) {};
\node[] (5) at (4,0) {};
\node[cnode] (6) at (5,0) {$1$};
\node[cnode] (7) at (7,0) {$1$};
\node[] (8) at (8,0) {};
\node[] (9) at (12,0) {};
\node[cnode] (10) at (13,0) {$1$};
\node[cnode] (11) at (15,0) {$1$};
\node[snode] (12) at (17,0) {$1$};
\node[cnode] (13) at (7,2) {$1$};
\node[cnode] (14) at (9,2) {$1$};
\node[] (15) at (10,2) {};
\node[] (16) at (12,2) {};
\node[cnode] (17) at (13,2) {$1$};
\node[snode] (30) at (15,2) {$1$};
\draw[thick] (1) -- (2);
\draw[thick] (2) -- (3);
\draw[thick] (3) -- (4);
\draw[thick,dashed] (4) -- (5);
\draw[thick] (5) -- (6);
\draw[thick] (6) -- (7);
\draw[thick] (7) -- (8);
\draw[thick,dashed] (8) -- (9);
\draw[thick] (9) -- (10);
\draw[thick] (10) -- (11);
\draw[thick] (11) -- (12);
\draw[thick] (7) -- (13);
\draw[thick] (13) -- (14);
\draw[thick] (14) -- (15);
\draw[thick,dashed] (15) -- (16);
\draw[thick] (16) -- (17);
\draw[thick] (17) -- (30);
\node[text width=0.1cm](20) at (0,-1) {$1$};
\node[text width=0.1cm](21) at (2,-1) {$2$};
\node[text width= 1.5 cm](22) at (5,-1) {$n-p-1$};
\node[text width=1 cm](23) at (7,-1) {$n-p$};
\node[text width=1 cm](24) at (13,-1) {$n-2$};
\node[text width=1 cm](25) at (15,-1) {$n-1$};
\node[text width=0.1 cm](26) at (7,3) {$1$};
\node[text width=0.1 cm](27) at (9,3) {$2$};
\node[text width=1 cm](28) at (13,3) {$l-1$};
\node[text width=0.1cm](30) at (7,-3){$(Y')$};
\end{tikzpicture}}
\caption{\footnotesize{A three-parameter family of dual quiver gauge theory pairs with $U(1)$ gauge groups labelled by the integers $(n,p,l)$ with $n> p>0$, $l \geq 1$,
and $p \geq 2$.}}
\label{AbEx1gen}
\end{center}
\end{figure}
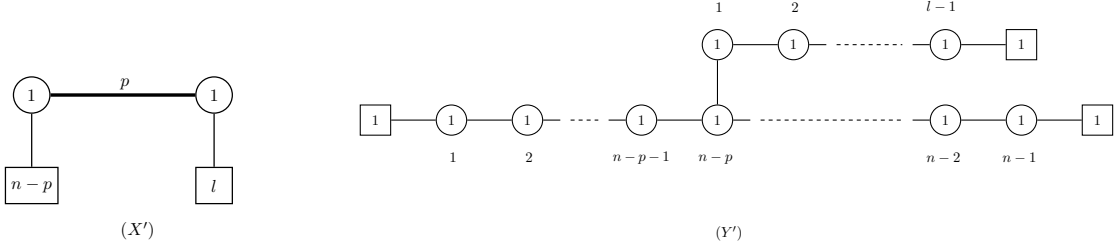

The first example is an infinite family of Abelian mirrors shown in \figref{AbEx1gen}, labelled by three positive integers $(n,l,p)$ 
with the constraints $n> p>0$, $l \geq 1$, and $p \geq 2$. The theory $X'$ consists of two gauge nodes connected 
by multiple edges. The dual theory $Y'$ has the shape of a star-shaped quiver consisting of three linear quivers glued at a 
common $U(1)$ gauge node \footnote{In \cite{Dey:2020hfe}, the quiver $Y'$ was written in a slightly different form. One can readily check that 
the two quivers are related by a simple redefinition of the Abelian vector multiplets.}. The dimensions of the respective Higgs and 
Coulomb branches, and the associated global symmetries are shown in Table \ref{Tab:AbEx1gen}.\\

For the choice $n=2p$ and $l=p$ , the quiver $X'$ reduces to the complete graph quiver -- the 
3d mirror of the $(A_2, A_{3p-1})$ AD theory reduced on a circle. The quiver $Y'$ -- the mirror of the 3d mirror -- 
gives another Lagrangian description of the 3d SCFT.

\begin{center}
\begin{table}[htbp]
\resizebox{\textwidth}{!}{%
\begin{tabular}{|c|c|c|}
\hline
Moduli space data & Theory $X'$ & Theory $Y'$ \\
\hline \hline 
dim\,$\CM_H$ & $n+l-2$ & $2$\\
\hline
dim\,$\CM_C$ & $2$ & $n+l-2$\\
\hline
$G_H$ & $SU(p) \times SU(n-p) \times SU(l) \times U(1) $  & $U(1) \times U(1)$\\
\hline
$G_C$ & $U(1) \times U(1)$ & $SU(p) \times SU(n-p) \times SU(l) \times U(1) $ \\
\hline
\end{tabular}}
\caption{Summary table for the moduli space dimensions and global symmetries for the mirror pair in \figref{AbEx1gen}.}
\label{Tab:AbEx1gen}
\end{table}
\end{center}

\subsection{Duality of the theories without defects}
\begin{figure}[htbp]
\begin{center}
\scalebox{0.6}{\begin{tikzpicture}[node distance=2cm,cnode/.style={circle,draw,thick,minimum size=8mm},snode/.style={rectangle,draw,thick,minimum size=8mm},pnode/.style={rectangle,red,draw,thick,minimum size=8mm}]
\node[cnode] (1) at (0,0) {$1$};
\node[snode] (2) at (0,-2) {$n$};
\draw[thick] (1) -- (2);
\node[text width=0.1cm](3) at (0,-3){$(X)$};
\end{tikzpicture}}
\qquad \qquad
\scalebox{0.6}{\begin{tikzpicture}[node distance=2cm,cnode/.style={circle,draw,thick,minimum size=8mm},snode/.style={rectangle,draw,thick,minimum size=8mm},pnode/.style={rectangle,red,draw,thick,minimum size=8mm}]
\node[cnode] (1) at (0,0) {$1$};
\node[cnode] (2) at (2,0) {$1$};
\node[] (3) at (3,0) {};
\node[] (4) at (4,0) {};
\node[cnode] (5) at (5,0) {$1$};
\node[cnode] (6) at (7,0) {$1$};
\node[snode] (7) at (0,-2) {$1$};
\node[snode] (8) at (7, -2) {$1$};
\draw[thick] (1) -- (2);
\draw[thick] (2) -- (3);
\draw[thick, dashed] (3) -- (4);
\draw[thick] (4) -- (5);
\draw[thick] (5) -- (6);
\draw[thick] (6) -- (8);
\draw[thick] (1) -- (7);
\node[text width=0.1cm](20) at (0,1) {$1$};
\node[text width=0.1cm](21) at (2,1) {$2$};
\node[text width= 1 cm](22) at (5,1) {$n-2$};
\node[text width=1 cm](23) at (7,1) {$n-1$};
\node[text width=0.1cm](20) at (3.5,-3){$(Y)$};
\end{tikzpicture}}
\caption{\footnotesize{A pair of mirror dual Abelian linear quivers.}}
\label{AbEx1LQ}
\end{center}
\end{figure}
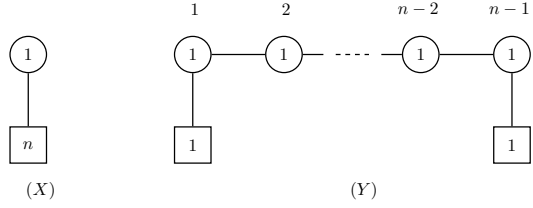

Let us first consider the duality without any half-BPS defects.
The starting point for obtaining the mirror pair $(X',Y')$ by $S$-type operations is a linear mirror pair $(X,Y)$, as 
shown in \figref{AbEx1LQ}. Mirror symmetry of $X$ and $Y$ implies that the partition functions are related in the 
following fashion:
\begin{align}
& Z^{(X)}(\vec m, \vec t)  =C_{XY}(\vec m, \vec t)\, Z^{(Y)}(\vec t, -\vec m), \\
& C_{XY}(\vec m, \vec t)=  e^{2\pi i (m_1\,t_1- m_n\,t_2)},
\end{align}
where $C_{XY}$ is a contact term. 

\begin{figure}[htbp]
\begin{center}
\begin{tabular}{ccccc}
\scalebox{.6}{\begin{tikzpicture}[node distance=2cm,cnode/.style={circle,draw,thick,minimum size=8mm},snode/.style={rectangle,draw,thick,minimum size=8mm},pnode/.style={rectangle,red,draw,thick,minimum size=8mm}]
\node[cnode] (1) at (0,0) {$1$};
\node[snode] (2) at (0,-2) {$n-p$};
\node[snode] (3) at (2,0) {$p$};
\draw[thick] (1) -- (2);
\draw[thick] (1) -- (3);
\node[text width=0.1cm](3) at (0,-3){$(X)$};
\end{tikzpicture}}
& \scalebox{.7}{\begin{tikzpicture} \draw[thick, ->] (0,0) -- (2,0); 
\node[] at (1,-1.8) {};
\end{tikzpicture}}
&\scalebox{.6}{\begin{tikzpicture}[node distance=2cm,cnode/.style={circle,draw,thick,minimum size=8mm},snode/.style={rectangle,draw,thick,minimum size=8mm},pnode/.style={rectangle,green,draw,thick,minimum size=8mm}]
\node[cnode] (1) at (0,0) {$1$};
\node[snode] (2) at (0,-2) {$n-p$};
\node[pnode] (3) at (2,2) {$1$};
\node[pnode] (4) at (2,0.5) {$1$};
\node[pnode] (5) at (2,-2) {$1$};
\draw[thick] (1) -- (2);
\draw[thick] (1) -- (3);
\draw[thick] (1) -- (4);
\draw[thick] (1) -- (5);
\draw[thick, dotted, black] (2, -0.5) -- (2,-1.2);
\node[text width=0.1cm](3) at (0,-3){$(X)$};
\node[text width=0.1cm](15) at (3, 2){$1$};
\node[text width=0.1cm](16) at (3, 0.5){$2$};
\node[text width=0.1cm](17) at (3, -2){$p$};
\end{tikzpicture}}
& \scalebox{.7}{\begin{tikzpicture} \draw[thick, ->] (0,0) -- (2,0); 
\node[] at (1,-1.8) {};
\node[text width=1cm](1) at (1,0.5) {$\CO^\alpha_{\CP}$};
\end{tikzpicture}}
&\scalebox{.6}{\begin{tikzpicture}[node distance=2cm,cnode/.style={circle,draw,thick,minimum size=8mm},snode/.style={rectangle,draw,thick,minimum size=8mm}]
\node[cnode] (1) at (-2,0) {$1$};
\node[cnode] (2) at (2,0) {$1$};
\node[snode] (3) at (-2,-2) {$n-p$};
\node[snode] (4) at (2,-2) {$l$};
\draw[thick] (1) -- (3);
\draw[thick] (2) -- (4);
\draw[line width=0.75mm, black] (1) to (2);
\node[text width=0.1cm](15) at (0,0.25){$p$};
\node[text width=0.1cm](16) at (0,-3){$(X')$};
\end{tikzpicture}}
\end{tabular}
\caption{\footnotesize{The $S$-type operation $\CO^\alpha_{\CP}$ implemented on $X$. The green nodes are identified, followed by a flavoring operation at the 
identified node with $l$ fundamental hypermultiplets, and the identified node is then gauged.}}
\label{SonX-Ab}
\end{center}
\end{figure}
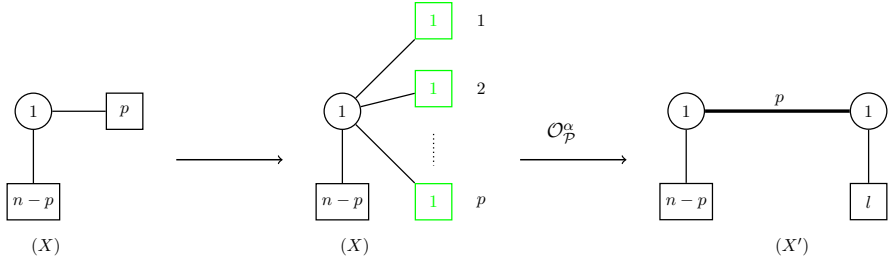

The quiver $X'$ can be constructed by implementing an elementary Abelian identification-flavoring-gauging 
operation $\CO^\alpha_{\CP}$ on the linear quiver $X$:
\begin{align}
\CO^\alpha_{\CP}(X)= G^\alpha_\CP \circ F^\alpha_\CP \circ I^\alpha_\CP(X) , \label{SOp-1a}
\end{align}
where $\CO^\alpha_{\CP}$ is shown explicitly in \figref{SonX-Ab}. The $p$ $U(1)$ nodes being identified are shown in green. 
The mass parameters associated with the $U(1)^p$ global symmetry being identified are chosen as :
\begin{align}
u_j=  m_{n-p+j}, \qquad j=1,\ldots, p. \label{SOp-1b}
\end{align}
Note that \eref{SOp-1b} corresponds to a specific choice of the permutation matrix $\CP$. The resultant quiver $X'$ is shown in \figref{SonX-Ab}
-- it is a $U(1)_1 \times U(1)_2$ gauge theory with $n-p$ hypers in the fundamental of $U(1)_1$, $l$ hypers in the fundamental 
of $U(1)_2$, and $p$ hypers in the bifundamental of $U(1)_1 \times U(1)_2$. The associated real masses are denoted as
$\vec{m}^{(1)}, \vec{m}^{(2)}$ and $\vec m^{\rm bif}$ respectively. In addition, we have the parameters $\vec t$, 
such that the real FI deformations are given as $\eta_1=t_1 -t_2$, and $\eta_2= t_2 - t_3$. 

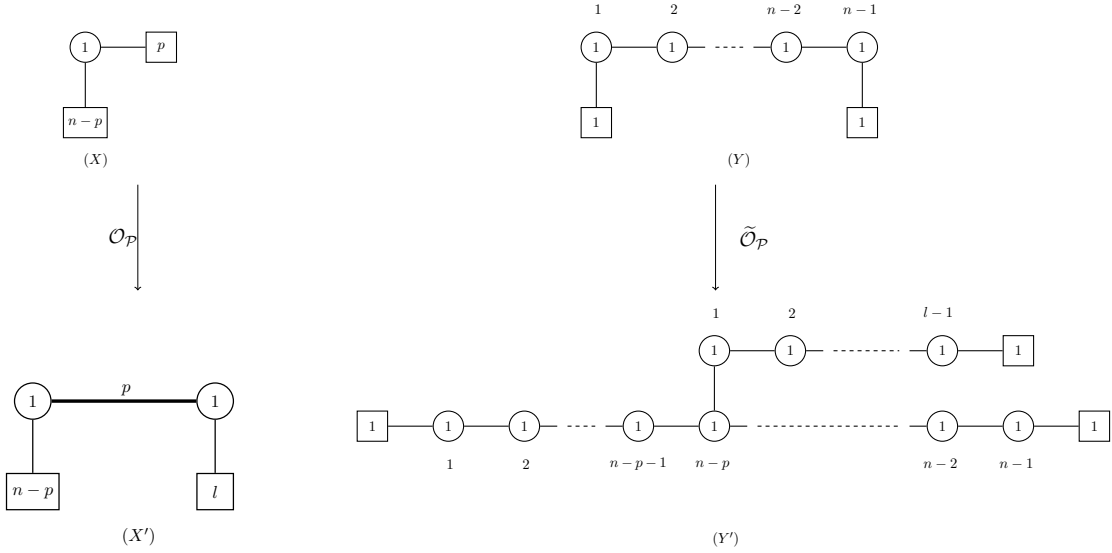
\begin{figure}[htbp]
\begin{center}
\begin{tabular}{ccc}
\scalebox{.5}{\begin{tikzpicture}[node distance=2cm,cnode/.style={circle,draw,thick,minimum size=8mm},snode/.style={rectangle,draw,thick,minimum size=8mm},pnode/.style={rectangle,red,draw,thick,minimum size=8mm}]
\node[cnode] (1) at (0,0) {$1$};
\node[snode] (2) at (0,-2) {$n-p$};
\node[snode] (3) at (2,0) {$p$};
\draw[thick] (1) -- (2);
\draw[thick] (1) -- (3);
\node[text width=0.1cm](3) at (0,-3){$(X)$};
\end{tikzpicture}}
& \qquad \qquad 
& \scalebox{.5}{\begin{tikzpicture}[node distance=2cm,cnode/.style={circle,draw,thick,minimum size=8mm},snode/.style={rectangle,draw,thick,minimum size=8mm},pnode/.style={rectangle,red,draw,thick,minimum size=8mm}]
\node[cnode] (1) at (0,0) {$1$};
\node[cnode] (2) at (2,0) {$1$};
\node[] (3) at (3,0) {};
\node[] (4) at (4,0) {};
\node[cnode] (5) at (5,0) {$1$};
\node[cnode] (6) at (7,0) {$1$};
\node[snode] (7) at (0,-2) {$1$};
\node[snode] (8) at (7, -2) {$1$};
\draw[thick] (1) -- (2);
\draw[thick] (2) -- (3);
\draw[thick, dashed] (3) -- (4);
\draw[thick] (4) -- (5);
\draw[thick] (5) -- (6);
\draw[thick] (6) -- (8);
\draw[thick] (1) -- (7);
\node[text width=0.1cm](20) at (0,1) {$1$};
\node[text width=0.1cm](21) at (2,1) {$2$};
\node[text width= 1 cm](22) at (5,1) {$n-2$};
\node[text width=1 cm](23) at (7,1) {$n-1$};
\node[text width=0.1cm](20) at (3.5,-3){$(Y)$};
\end{tikzpicture}}\\
 \scalebox{.7}{\begin{tikzpicture}
\draw[->] (15,-3) -- (15,-5);
\node[text width=0.1cm](20) at (14.5, -4) {$\CO_{\CP}$};
\end{tikzpicture}}
&\qquad \qquad 
& \scalebox{.7}{\begin{tikzpicture}
\draw[->] (15,-3) -- (15,-5);
\node[text width=0.1cm](29) at (15.5, -4) {$\wt{\CO}_{\CP}$};
\end{tikzpicture}}\\
\scalebox{.6}{\begin{tikzpicture}[node distance=2cm,cnode/.style={circle,draw,thick,minimum size=8mm},snode/.style={rectangle,draw,thick,minimum size=8mm},pnode/.style={rectangle,red,draw,thick,minimum size=8mm}]
\node[cnode] (1) at (-2,0) {$1$};
\node[cnode] (2) at (2,0) {$1$};
\node[snode] (3) at (-2,-2) {$n-p$};
\node[snode] (4) at (2,-2) {$l$};
\draw[thick] (1) -- (3);
\draw[thick] (2) -- (4);
\draw[line width=0.75mm, black] (1) to (2);
\node[text width=0.1cm](15) at (0,0.25){$p$};
\node[text width=0.1cm](16) at (0,-3){$(X')$};
\end{tikzpicture}}
&\qquad \qquad 
& \scalebox{.5}{\begin{tikzpicture}[node distance=2cm,cnode/.style={circle,draw,thick,minimum size=8mm},snode/.style={rectangle,draw,thick,minimum size=8mm},pnode/.style={rectangle,red,draw,thick,minimum size=8mm}]
\node[snode] (1) at (-2,0) {$1$};
\node[cnode] (2) at (0,0) {$1$};
\node[cnode] (3) at (2,0) {$1$};
\node[] (4) at (3,0) {};
\node[] (5) at (4,0) {};
\node[cnode] (6) at (5,0) {$1$};
\node[cnode] (7) at (7,0) {$1$};
\node[] (8) at (8,0) {};
\node[] (9) at (12,0) {};
\node[cnode] (10) at (13,0) {$1$};
\node[cnode] (11) at (15,0) {$1$};
\node[snode] (12) at (17,0) {$1$};
\node[cnode] (13) at (7,2) {$1$};
\node[cnode] (14) at (9,2) {$1$};
\node[] (15) at (10,2) {};
\node[] (16) at (12,2) {};
\node[cnode] (17) at (13,2) {$1$};
\node[snode] (30) at (15,2) {$1$};
\draw[thick] (1) -- (2);
\draw[thick] (2) -- (3);
\draw[thick] (3) -- (4);
\draw[thick,dashed] (4) -- (5);
\draw[thick] (5) -- (6);
\draw[thick] (6) -- (7);
\draw[thick] (7) -- (8);
\draw[thick,dashed] (8) -- (9);
\draw[thick] (9) -- (10);
\draw[thick] (10) -- (11);
\draw[thick] (11) -- (12);
\draw[thick] (7) -- (13);
\draw[thick] (13) -- (14);
\draw[thick] (14) -- (15);
\draw[thick,dashed] (15) -- (16);
\draw[thick] (16) -- (17);
\draw[thick] (17) -- (30);
\node[text width=0.1cm](20) at (0,-1) {$1$};
\node[text width=0.1cm](21) at (2,-1) {$2$};
\node[text width= 1.5 cm](22) at (5,-1) {$n-p-1$};
\node[text width=1 cm](23) at (7,-1) {$n-p$};
\node[text width=1 cm](24) at (13,-1) {$n-2$};
\node[text width=1 cm](25) at (15,-1) {$n-1$};
\node[text width=0.1 cm](26) at (7,3) {$1$};
\node[text width=0.1 cm](27) at (9,3) {$2$};
\node[text width=1 cm](28) at (13,3) {$l-1$};
\node[text width=0.1cm](30) at (7,-3){$(Y')$};
\end{tikzpicture}}
\end{tabular}
\caption{\footnotesize{Construction of a pair of mirror dual theories, involving non-ADE quivers.}}
\label{SimpAbEx1GFI}
\end{center}
\end{figure}

The dual theory can be read off from the dual partition partition which can be computed using the prescription in  \appref{sec:SOp-app} 
and is summarized in \appref{sec:PF-Ab-app}. The dual operation $\wt{\CO}_{\CP}$ and the resultant $Y'$ is shown in \figref{SimpAbEx1GFI}.
The quiver $Y'$ can be decomposed into two linear subquivers:
\begin{itemize}
\item a linear chain (1) consisting of $n-1$ $U(1)$ gauge nodes,
\item a linear chain (2) consisting of $l-1$ $U(1)$ gauge nodes, 
\end{itemize}
where the $(n-p)$-th  gauge node of chain (1) is connected to the first gauge node of chain (2) by a single bifundamental 
hypermultiplet.\\

Let us label the FI parameters of the linear chain (1) of $n-1$ nodes as $\eta'^{(1)}_{k} = t'^{(1)}_k - t'^{(1)}_{k+1}$, 
with $k=1,\ldots, n-1$, and the FI parameters of the linear chain (2) of $l-1$ nodes as  
$\eta'^{(2)}_{b} = t'^{(2)}_b - t'^{(2)}_{b+1}$, with $b=1,\ldots, (l-1)$. The fundamental masses associated with 
chain (1) are denoted as $m'^{(1)}_{1}$ and $m'^{(1)}_{n-1}$, while the lone fundamental mass associated 
with chain (2) is denoted as $m'^{(2)}_{l-1}$.

With this parametrization of masses and FI parameters, 
the partition functions of $X'$ and $Y'$ can be shown to be related as
\begin{align}
& Z^{(X')}(\{ \vec m^{(1)},  \vec m^{(2)}, \vec m^{\rm bif} \}, \vec t ) = C_{X'Y'} \cdot 
\,Z^{(Y')}(\{m'^{(1)}_{1}, m'^{(1)}_{n-1}, m'^{(2)}_{l-1} \}, \{-\vec{t'}^{(1)}, -\vec{t'}^{(2)} \}), \label{MS-Pf-Ab-main} \\
& C_{X'Y'} = e^{-2\pi i \,t_3\, m^{(2)}_l}\, e^{2\pi i (m^{(1)}_1 t_1 - m^{\rm bif}_p t_2)}, \label{contact-Ab-main}
\end{align}
where on the RHS of \eref{MS-Pf-Ab-main}, we have chosen to write the independent mass parameters in terms 
of the fundamental masses of $Y'$, setting the bifundamental masses to zero. The mirror map relating the FI 
parameters of $Y'$ to the masses of $X'$ is given as:
\begin{subequations}
\begin{empheq}[box=\widefbox]{align}
\vec{t'}^{(1)} = & \{ m^{(1)}_1, \ldots, m^{(1)}_{n-p}, m^{\rm bif}_{1} + m^{(2)}_1, \ldots, m^{\rm bif}_{p} + m^{(2)}_1\}, \label{mm-Ab-1a}\\
 \vec{t}'^{(2)} = & \{ m^{(2)}_1,\ldots, m^{(2)}_l \}, \label{mm-Ab-1b}
\end{empheq}
\end{subequations}
while the mirror relating the fundamental masses of $Y'$ to FI parameters of $X'$ is then given as:
\begin{subequations}
\begin{empheq}[box=\widefbox]{align}
& m'^{(1)}_{1} = t_1, \\
&  m'^{(1)}_{n-1} = t_2,\\
& m'^{(2)}_{l-1} = t_3. 
\end{empheq}
\end{subequations}



\subsection{Vortex defects : Hopping duals and mirror maps}

\begin{figure}[htbp]
\begin{center}
\begin{tabular}{ccccc}
\scalebox{.6}{\begin{tikzpicture}[node distance=2cm,cnode/.style={circle,draw,thick,minimum size=8mm},snode/.style={rectangle,draw,thick,minimum size=8mm},pnode/.style={rectangle,red,draw,thick,minimum size=8mm}, nnode/.style={circle, red, draw,thick,minimum size=1.0cm}, lnode/.style={shape = rounded rectangle, minimum size= 1cm, rotate=90, rounded rectangle right arc = none, draw}]
\node[lnode] (1) at (0,0) {\rotatebox{-90}{1}};
\node[snode] (2) at (0,-2) {$n-p'$};
\node[snode] (3) at (2,0) {$p'$};
\draw[thick] (1) -- (2);
\draw[thick] (1) -- (3);
\node[nnode] (5) at (1,2) {$k$};
\draw[red, thick, ->] (5) to [out=150,in=210,looseness=8] (5);
\draw[red, ->] (1) -- (5);
\draw[red, ->] (5) -- (3);
\node[text width=1cm](3) at (0,-3){$(X[V_{p',k}^r])$};
\end{tikzpicture}}
& \scalebox{.7}{\begin{tikzpicture} \draw[thick, ->] (0,0) -- (2,0); 
\node[] at (1,-1.8) {};
\end{tikzpicture}}
&\scalebox{.6}{\begin{tikzpicture}[node distance=2cm,cnode/.style={circle,draw,thick,minimum size=8mm},snode/.style={rectangle,draw,thick,minimum size=8mm},pnode/.style={rectangle,green,draw,thick,minimum size=8mm}, nnode/.style={circle, red, draw,thick,minimum size=1.0cm}, lnode/.style={shape = rounded rectangle, minimum size= 1cm, rotate=90, rounded rectangle right arc = none, draw}]
\node[lnode] (1) at (0,0) {\rotatebox{-90}{1}};
\node[snode] (2) at (-2,0) {$n-p$};
\node[pnode] (3) at (2,2) {$1$};
\node[pnode] (4) at (2,0.5) {$1$};
\node[pnode] (5) at (2,-2) {$1$};
\node[nnode] (6) at (0,2) {$k$};
\draw[red, thick, ->] (6) to [out=150,in=210,looseness=8] (6);
\node[pnode] (7) at (-2,-2) {$1$};
\node[pnode] (8) at (0,-2) {$1$};
\draw[thick] (1) -- (2);
\draw[thick] (1) -- (3);
\draw[thick] (1) -- (4);
\draw[thick] (1) -- (5);
\draw[thick] (1) -- (7);
\draw[thick] (1) -- (8);
\draw[thick, red, ->] (1) -- (6);
\draw[thick, red, ->] (6) -- (3);
\draw[thick, red, ->] (6) -- (4);
\draw[thick, red, ->] (6) -- (5);
\draw[thick, dotted, black] (2, -0.5) -- (2,-1.2);
\draw[thick, dotted, black] (-1.3, -2) -- (-0.7, -2);
\node[text width=1cm](3) at (0,-3){$(X[V_{p',k}^r])$};
\node[text width=0.1cm](15) at (3, 2){$1$};
\node[text width=0.1cm](16) at (3, 0.5){$2$};
\node[text width=0.1cm](17) at (3, -2){$p'$};
\node[text width=0.1cm](18) at (-2, -1.3){$1$};
\node[text width=1cm](19) at (0, -1.3){$p'-p$};
\end{tikzpicture}}
& \scalebox{.7}{\begin{tikzpicture} \draw[thick, ->] (0,0) -- (2,0); 
\node[] at (1,-1.8) {};
\node[text width=1cm](1) at (1,0.5) {$\CO^\alpha_{\CP}$};
\end{tikzpicture}}
& \scalebox{.6}{\begin{tikzpicture}[node distance=2cm,cnode/.style={circle,draw,thick,minimum size=8mm},snode/.style={rectangle,draw,thick,minimum size=8mm},pnode/.style={rectangle,red,draw,thick,minimum size=8mm}, nnode/.style={circle, red, draw,thick,minimum size=1.0cm}, lnode/.style={shape = rounded rectangle, minimum size= 1cm, rotate=90, rounded rectangle right arc = none, draw}]
\node[lnode] (1) at (-2,0) {\rotatebox{-90}{1}};
\node[lnode] (2) at (2,0) {\rotatebox{-90}{1}};
\node[snode] (3) at (-2,-2) {$n-p$};
\node[snode] (4) at (2,-2) {$l$};
\draw[thick] (1) -- (3);
\draw[thick] (2) -- (4);
\node[nnode] (5) at (0,3) {$k$};
\draw[red, thick, ->] (5) to [out=150,in=210,looseness=8] (5);
\draw[line width=0.75mm, red] (1.south east) to (2.north east);
\draw[line width=0.75mm, black] (1.south west) to (2.north west);
\draw[red, ->] (1) -- (5);
\draw[line width=0.75mm, red, ->] (5) -- (2);
\node[text width=0.1cm](15) at (0,0.75){$p'$};
\node[text width=0.1cm](16) at (1.5, 2){$p'$};
\node[text width=1cm](17) at (0, -0.5){$p-p'$};
\node[text width=0.1cm](18) at (-1,-3){$(X'[V^{(I)\,-}_{0,k,p'}])$};
\end{tikzpicture}}
\end{tabular}
\caption{\footnotesize{The $S$-type operation $\CO^\alpha_{\CP}$ implemented on the 3d-1d quiver $X[V_{p',k}^r]$ leading to the new 3d-1d quiver 
$X'[V^{(I)\,-}_{0,k,p'}]$. The green nodes are identified, followed by a flavoring operation at the identified node with $l$ fundamental 
hypermultiplets, and the identified node is then gauged.}}
\label{SonX-Ab-def}
\end{center}
\end{figure}
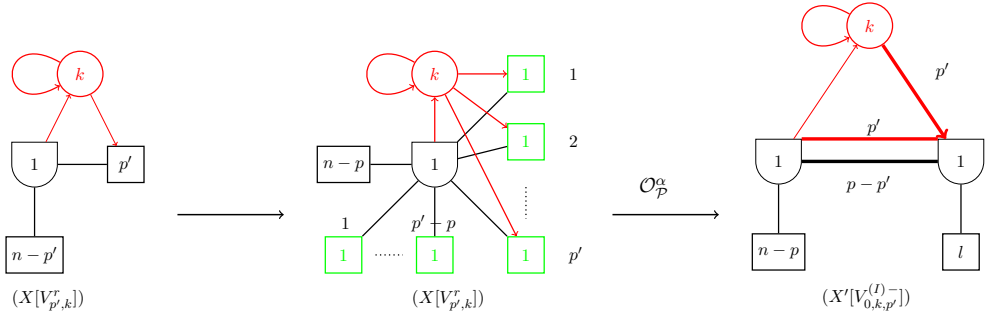

In this section, we construct the 3d-1d coupled quivers of the form given in \figref{3d1d-edges} that realize half-BPS vortex 
defects in the quiver $X'$. We then study the hopping dualities of these coupled systems and their mirror maps. 
The starting point of the construction is the coupled quiver in the theory $X$, as shown on the extreme 
left of \figref{SonX-Ab-def}. In the notation of \Secref{VD-LQ}, this quiver realizes a vortex defect $V^r_{p',k}$, 
where $p'$ and $k$ are positive integers. The superscript $r$ implies that the single 1d FI parameter associated 
with the $U(k)$ gauge node should be chosen to be negative.

We now implement the $S$-type operation $\CO^\alpha_\CP$, as defined in \eref{SOp-1a}-\eref{SOp-1b}, 
on the 3d-1d quiver $(X[V_{p',k}^r])$, assuming $p > p'$. The procedure involves identifying $p$ $U(1)$ 
flavor nodes, which are shown in green in \figref{SonX-Ab-def}, followed by a flavoring and gauging operation of the 
identified node. Of these $p$ $U(1)$ flavor nodes, $p'$ are coupled to the 1d quiver as shown, while the remaining 
$p-p'$ nodes are not. The resultant coupled quiver is shown in the extreme right of \figref{SonX-Ab-def}. In the 
notation of \Secref{VD-edges}, this realizes the vortex defect $V^{(I)\,-}_{0,k,p'}$. Following the general 
prescription of \eref{S-Op-Agen}, the defect partition function can be written as:
\begin{align}
Z^{(X'[V^{(I)\,-}_{0,k,p'}])} = W_{\rm bg}(t_2,k)\, \lim_{z\to 1}\, \int\, \prod^2_{k=1} ds_{k}\,Z^{(X')}_{\rm int}(\vec s, \{\vec m^{(a)} \}, \vec m^{\rm bif}, \vec t, \eta_\alpha)
\cdot \CI^{\Sigma_{(I)\,-}^{0,k,p'}}(\vec s,\vec m^{\rm bif}, z|\xi <0), \label{VortexAb1-Pf}
\end{align}
where the Witten index is being computed in the negative FI chamber. 
The 3d matrix model integrand $Z^{(X')}_{\rm int}$ and the Witten index $\CI^{\Sigma_{(I)\,-}^{0,k,p'}}$ of the SQM are given as:
\begin{align}
& Z^{(X')}_{\rm int}=\frac{e^{2\pi i \,(t_1- t_2)\,s_1} \, e^{2\pi i \,(t_2- t_3)\,s_2}}{\prod^{n-p}_{i=1} \ch{(s_1-m^{(1)}_i)}\, \prod^{l}_{a=1} \ch{(s_2-m^{(2)}_a)}\, \prod^{p}_{j=1} \ch{(s_1- s_2 -m^{\rm bif}_j)}}, \\
& \CI^{\Sigma_{(I)\,-}^{0,k,p'}}(\vec s,\vec m^{\rm bif}, z|\xi <0) = \frac{\prod^{p}_{j=p-p'+1} \ch{(s_1- s_2 -m^{\rm bif}_j)}}{ \prod^{p}_{j=p-p'+1} \ch{(s_1- s_2 -m^{\rm bif}_j + ikz)}}, \qquad W_{\rm bg}(t_2,k)= e^{2\pi t_2 k} .
\end{align}
The defect partition function can then be simplified to the following form:
\be
Z^{(X'[V^{(I)\,-}_{0,k,p'}])}= W_{\rm bg}(t_2,k)\, \lim_{z\to 1}\, Z^{(X')}(\vec m^{(1)}, \vec m^{(2)}, \{m^{\rm bif}_j \}^{p-p'}_{j=1}, \{m^{\rm bif}_j -ikz\}^{p}_{j=p-p'+1}; \vec t).
\ee

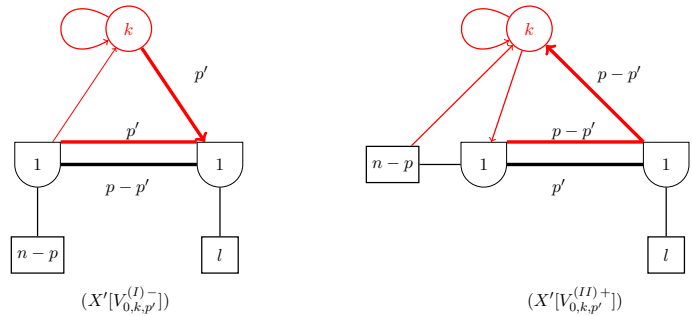
\begin{figure}[htbp]
\begin{center}
\begin{tabular}{ccc}
 \scalebox{.6}{\begin{tikzpicture}[node distance=2cm,cnode/.style={circle,draw,thick,minimum size=8mm},snode/.style={rectangle,draw,thick,minimum size=8mm},pnode/.style={rectangle,red,draw,thick,minimum size=8mm}, nnode/.style={circle, red, draw,thick,minimum size=1.0cm}, lnode/.style={shape = rounded rectangle, minimum size= 1cm, rotate=90, rounded rectangle right arc = none, draw}]
\node[lnode] (1) at (-2,0) {\rotatebox{-90}{1}};
\node[lnode] (2) at (2,0) {\rotatebox{-90}{1}};
\node[snode] (3) at (-2,-2) {$n-p$};
\node[snode] (4) at (2,-2) {$l$};
\draw[thick] (1) -- (3);
\draw[thick] (2) -- (4);
\node[nnode] (5) at (0,3) {$k$};
\draw[red, thick, ->] (5) to [out=150,in=210,looseness=8] (5);
\draw[line width=0.75mm, red] (1.south east) to (2.north east);
\draw[line width=0.75mm, black] (1.south west) to (2.north west);
\draw[red, ->] (1) -- (5);
\draw[line width=0.75mm, red, ->] (5) -- (2);
\node[text width=0.1cm](15) at (0,0.75){$p'$};
\node[text width=0.1cm](16) at (1.5, 2){$p'$};
\node[text width=1cm](17) at (0, -0.5){$p-p'$};
\node[text width=0.1cm](18) at (-1,-3){$(X'[V^{(I)\,-}_{0,k,p'}])$};
\end{tikzpicture}}
& \qquad \qquad
& \scalebox{.6}{\begin{tikzpicture}[node distance=2cm,cnode/.style={circle,draw,thick,minimum size=8mm},snode/.style={rectangle,draw,thick,minimum size=8mm},pnode/.style={rectangle,red,draw,thick,minimum size=8mm}, nnode/.style={circle, red, draw,thick,minimum size=1.0cm}, lnode/.style={shape = rounded rectangle, minimum size= 1cm, rotate=90, rounded rectangle right arc = none, draw}]
\node[lnode] (1) at (-2,0) {\rotatebox{-90}{1}};
\node[lnode] (2) at (2,0) {\rotatebox{-90}{1}};
\node[snode] (3) at (-4,0) {$n-p$};
\node[snode] (4) at (2,-2) {$l$};
\draw[thick] (1) -- (3);
\draw[thick] (2) -- (4);
\node[nnode] (5) at (-1,3) {$k$};
\draw[red, thick, ->] (5) to [out=150,in=210,looseness=8] (5);
\draw[line width=0.75mm, red] (1.south east) to (2.north east);
\draw[line width=0.75mm, black] (1.south west) to (2.north west);
\draw[red, thick, ->] (5) -- (1);
\draw[line width=0.75mm, red, ->] (2) -- (5);
\draw[red, thick, ->] (3) -- (5);
\node[text width=1cm](15) at (0,0.75){$p-p'$};
\node[text width=1cm](16) at (1, 2){$p-p'$};
\node[text width=1cm](17) at (0, -0.5){$p'$};
\node[text width=0.1cm](18) at (-1,-3){$(X'[V^{(II)\,+}_{0,k,p'}])$};
\end{tikzpicture}}
\end{tabular}
\caption{\footnotesize{The hopping duals for the vortex defect.}}
\label{VD-Ab-HD}
\end{center}
\end{figure}

A hopping dual of this coupled quiver can be read off from the defect partition function after implementing the change of variable:
\be
s_1 \to s_1 - i\,k\,z.
\ee
Following the analysis in \Secref{VD-edges}, the coupled quiver is shown on the right of \figref{VD-Ab-HD}, 
which is manifestly of the type in \figref{3d1d-edges-hd}. The associated vortex defect is denoted as $V^{(II)\,+}_{0,k,p'}$.

Let us now determine the mirror map of this vortex defect. The dual defect partition 
function can be written down following the general prescription in \eref{PF-wtOPgenD-A2B}-\eref{CZ-wtOPDD-A2B}. 
In the present case, we can simply use the relation \eref{MS-Pf-Ab-main} between the 
partition functions of $(X',Y')$, and the mirror map of FI parameters/masses in 
\eref{mm-Ab-1a}-\eref{mm-Ab-1b} to write down the dual defect partition function. 
This leads to the following relation:
\begin{align}
Z^{(X'[V^{(I)\,-}_{0,k,p'}])} = & C_{X'Y'} \cdot \int \,  \prod^{n-1}_{k=1}\,  d\s_k\,\prod^{l-1}_{b=1}\, d\tau_b\cdot e^{2\pi k \s_{n-p'}}\, Z^{(Y')}_{\rm int}(\vec \s, \vec \tau, \vec m', \vec \eta'), \\
= & C_{X'Y'} \cdot Z^{(Y'[W^{(1)}_{k,\, n-p'}])}(\vec m', \vec \eta'),
\end{align}
where $W^{(1)}_{k,\, n-p'}$ is a Wilson defect of charge $k$ in the $n-p'$-th $U(1)$ gauge node of  the linear subquiver (1) of $Y'$, 
and $C_{X'Y'}$ is the contact term given in \eref{contact-Ab-main}. We therefore have the mirror map:
\be
\boxed{\langle{V^{(I)\,-}_{Q,R,p'}}\rangle_{X'}= \langle{V^{(II)\,+}_{Q,R,p'}}\rangle_{X'} = \langle{W^{(1)}_{k,\, n-p'}}\rangle_{Y'},}
\ee
and the dual Wilson defect in the theory $Y'$ is shown in \figref{Abdual1}.


\begin{figure}[htbp]
\begin{center}
\scalebox{0.6}{\begin{tikzpicture}[node distance=2cm,cnode/.style={circle,draw,thick,minimum size=8mm},snode/.style={rectangle,draw,thick,minimum size=8mm},pnode/.style={rectangle,red,draw,thick,minimum size=8mm}]
\node[snode] (1) at (-2,0) {$1$};
\node[cnode] (2) at (0,0) {$1$};
\node[cnode] (3) at (2,0) {$1$};
\node[] (4) at (3,0) {};
\node[] (5) at (4,0) {};
\node[cnode] (6) at (5,0) {$1$};
\node[cnode] (7) at (7,0) {$1$};
\node[] (8) at (8,0) {};
\node[] (9) at (10,0) {};
\node[cnode] (10) at (12,0) {$1$};
\node[] (40) at (14,0) {};
\node[] (41) at (15,0) {};
\node[cnode] (11) at (17,0) {$1$};
\node[snode] (12) at (19,0) {$1$};
\node[cnode] (13) at (7,2) {$1$};
\node[cnode] (14) at (9,2) {$1$};
\node[] (15) at (10,2) {};
\node[] (16) at (12,2) {};
\node[cnode] (17) at (13,2) {$1$};
\node[snode] (30) at (15,2) {$1$};
\draw[thick] (1) -- (2);
\draw[thick] (2) -- (3);
\draw[thick] (3) -- (4);
\draw[thick,dashed] (4) -- (5);
\draw[thick] (5) -- (6);
\draw[thick] (6) -- (7);
\draw[thick] (7) -- (8);
\draw[thick,dashed] (8) -- (9);
\draw[thick] (9) -- (10);
\draw[thick] (10) -- (40);
\draw[thick, dashed] (40) -- (41);
\draw[thick] (41) -- (11);
\draw[thick] (11) -- (12);
\draw[thick] (7) -- (13);
\draw[thick] (13) -- (14);
\draw[thick] (14) -- (15);
\draw[thick,dashed] (15) -- (16);
\draw[thick] (16) -- (17);
\draw[thick] (17) -- (30);
\node[text width=0.1cm](20) at (0,-1) {$1$};
\node[text width=0.1cm](21) at (2,-1) {$2$};
\node[text width= 1.5 cm](22) at (5,-1) {$n-p-1$};
\node[text width=1 cm](23) at (7,-1) {$n-p$};
\node[text width=1 cm](24) at (12,-1) {$n-p'$};
\node[text width=1 cm](31) at (12,1) {$W^{(1)}_{k,\, n-p'}$};
\node[text width=1 cm](25) at (17,-1) {$n-1$};
\node[text width=0.1 cm](26) at (7,3) {$1$};
\node[text width=0.1 cm](27) at (9,3) {$2$};
\node[text width=1 cm](28) at (13,3) {$l-1$};
\node[text width=0.1cm](30) at (7,-3){$(Y'[W^{(1)}_{k,\, n-p'}])$};
\end{tikzpicture}}
\caption{\footnotesize{Wilson defect in the theory $Y'$ mirror dual to the vortex defects given in \figref{VD-Ab-HD} in the theory $X'$.}}
\label{Abdual1}
\end{center}
\end{figure}

The computation in this section can be easily extended to construct the vortex defects $V^{(I)\, -}_{Q,k,p'}$ with $Q<n$, 
starting from the 3d-1d coupled quiver $X[\Sigma^{p',k}_r]$ via a similar $S$-type operation.

\section{Defects in a non-Abelian quiver}\label{sec:NAb-Ex}

\begin{figure}[htbp]
\begin{center}
\begin{tabular}{ccc}
\scalebox{.7}{\begin{tikzpicture}[
cnode/.style={circle,draw,thick, minimum size=1.0cm},snode/.style={rectangle,draw,thick,minimum size=1cm}]
\node[cnode] (9) at (0,1){1};
\node[snode] (10) at (0,-1){1};
\node[cnode] (11) at (2, 0){2};
\node[cnode] (12) at (4, 1){1};
\node[cnode] (13) at (4, -1){1};
\node[snode] (14) at (6, 1){$p_2$};
\node[snode] (15) at (6, -1){$p_3$};
\draw[-] (9) -- (11);
\draw[-] (10) -- (11);
\draw[-] (12) -- (11);
\draw[-] (13) -- (11);
\draw[-] (12) -- (14);
\draw[-] (13) -- (15);
\draw[line width=0.75mm, black] (12) to (13);
\node[text width=0.1cm](20) at (4.5,0){$p_1$};
\node[text width=0.1cm](31)[below=0.5 cm of 13]{$(X')$};
\end{tikzpicture}}
&\qquad
&\scalebox{.6}{\begin{tikzpicture}[node distance=2cm,cnode/.style={circle,draw,thick,minimum size=8mm},snode/.style={rectangle,draw,thick,minimum size=8mm},pnode/.style={rectangle,red,draw,thick,minimum size=8mm}]
\node[cnode] (1) at (-3,0) {$2$};
\node[cnode] (2) at (-2,2) {$1$};
\node[cnode] (3) at (0,2.5) {$1$};
\node[cnode] (4) at (2,2) {$1$};
\node[cnode] (5) at (3,0) {$1$};
\node[cnode] (6) at (2,-2) {$1$};
\node[cnode] (7) at (0,-2.5) {$1$};
\node[cnode] (8) at (-2,-2) {$1$};
\node[snode] (10) at (-5,0) {$3$};
\node[cnode] (11) at (4,0) {$1$};
\node[cnode] (12) at (5,0) {$1$};
\node[cnode] (13) at (7.5,0) {$1$};
\node[snode] (14) at (9,0) {$1$};
\draw[thick, blue] (1) to [bend left=40] (2);
\draw[thick] (2) to [bend left=40] (3);
\draw[thick,dashed] (3) to [bend left=40] (4);
\draw[thick] (4) to [bend left=40] (5);
\draw[thick] (5) to [bend left=40] (6);
\draw[dashed] (6) to [bend left=40] (7);
\draw[thick] (7) to [bend left=40] (8);
\draw[thick] (8) to [bend left=40] (1);
\draw[thick] (1) -- (10);
\draw[thick] (5) -- (11);
\draw[thick] (11) -- (12);
\draw[dashed] (12) -- (13);
\draw[thick] (13) -- (14);
\node[text width=1cm](11) at (-2,2.5) {$1$};
\node[text width=1cm](12) at (0, 3) {$2$};
\node[text width=1cm](13) at (2, 2.5) {$p_2-1$};
\node[text width=1cm](15) at (2,-2.5) {$p_1$};
\node[text width=1.5 cm](16) at (0,-3.2){$2$};
\node[text width=1cm](17) at (-2.5,-2.5){$1$};
\node[text width=0.1cm](18) at (4, 0.6){$1$};
\node[text width=0.1cm](19) at (5, 0.6){$2$};
\node[text width=1cm](20) at (7.5, 0.6){$p_3-1$};
\node[text width=0.1cm](30) at (0,-4){$(Y')$};
\end{tikzpicture}}
\end{tabular}
\caption{\footnotesize{An infinite family of mirror duals labelled by the positive integers $(p_1,p_2,p_3)$ subject to the constraints
$p_1\geq 1$, $p_2 \geq 1$, and $p_3>1$. The blue line denotes a hypermultiplet transforming in the determinant representation 
of $U(2)$ and having charge 1 under $U(1)_1$.}}
\label{fig: AbEx4gen}
\end{center}
\end{figure}
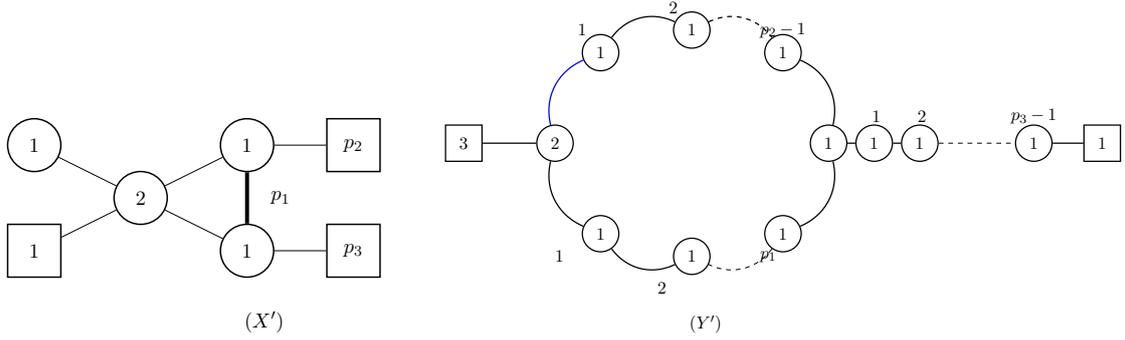

The second example is an infinite family of Abelian mirrors shown in \figref{fig: AbEx4gen}, labelled by three positive integers 
$(p_1,p_2,p_3)$ subject to the constraints $p_1\geq 1$, $p_2 \geq 1$, and $p_3>1$. 
Two of the gauge nodes in the theory $X'$ are connected by $p_1\geq 1$ bifundamental hypers. 
The dual theory $Y'$ is a non-ADE-type quiver gauge theory built out of unitary gauge nodes, 
and fundamental/bifundamental matter, along with a single hypermultiplet which transforms 
in the determinant representation of $U(2)$ gauge group and has charge 1 under one of the 
adjacent $U(1)$ gauge nodes, as denoted by the blue line in the figure. 
The dimensions of the respective Higgs and Coulomb branches, and the associated global symmetries 
are shown in Table \ref{Tab:AbEx4}.\\

\begin{center}
\begin{table}[htbp]
\resizebox{\textwidth}{!}{%
\begin{tabular}{|c|c|c|}
\hline
Moduli space data & Theory $X'$ & Theory $Y'$ \\
\hline \hline 
dim\,$\CM_H$ & $1+p_1+p_2 +p_3$ & 5\\
\hline
dim\,$\CM_C$ & 5 & $1+p_1+p_2 +p_3$\\
\hline
$G_H$ & $SU(p_1) \times SU(p_2) \times SU(p_3) \times U(1)^3$ &  $SU(3) \times U(1)^2$\\
\hline
$G_C$ & $SU(3) \times U(1)^2$ &  $SU(p_1) \times SU(p_2) \times SU(p_3) \times U(1)^3$ \\
\hline
\end{tabular}}
\caption{\footnotesize{Summary table for the moduli space dimensions and global symmetries for the mirror pair in \figref{fig: AbEx4gen}.}}
\label{Tab:AbEx4}
\end{table}
\end{center}

For keeping the presentation simple, we will consider the case $p_1=p_2 = p_3=2$. The dual pair $(X',Y')$ are mirror Lagrangians 
associated with the circle reduction of the AD theory $D_9(SU(3))$. 

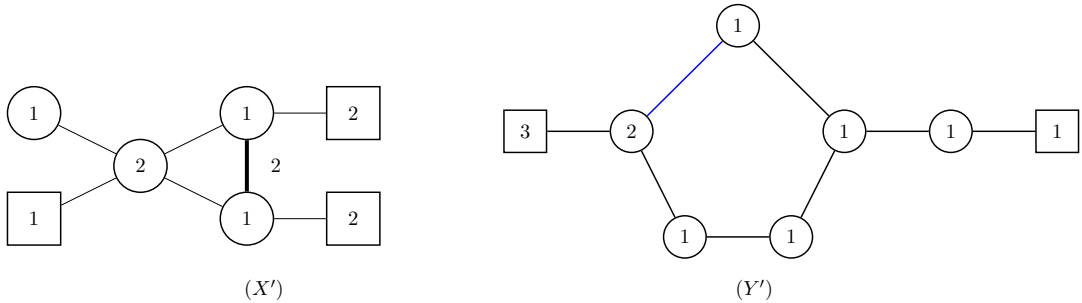
\begin{figure}[htbp]
\begin{center}
\begin{tabular}{ccc}
\scalebox{.7}{\begin{tikzpicture}[
cnode/.style={circle,draw,thick, minimum size=1.0cm},snode/.style={rectangle,draw,thick,minimum size=1cm}]
\node[cnode] (9) at (0,1){1};
\node[snode] (10) at (0,-1){1};
\node[cnode] (11) at (2, 0){2};
\node[cnode] (12) at (4, 1){1};
\node[cnode] (13) at (4, -1){1};
\node[snode] (14) at (6, 1){$2$};
\node[snode] (15) at (6, -1){$2$};
\draw[-] (9) -- (11);
\draw[-] (10) -- (11);
\draw[-] (12) -- (11);
\draw[-] (13) -- (11);
\draw[-] (12) -- (14);
\draw[-] (13) -- (15);
\draw[line width=0.75mm, black] (12) to (13);
\node[text width=0.1cm](20) at (4.5,0){$2$};
\node[text width=0.1cm](31)[below=0.5 cm of 13]{$(X')$};
\end{tikzpicture}}
&\qquad  \qquad
&\scalebox{.7}{\begin{tikzpicture}[node distance=2cm,cnode/.style={circle,draw,thick,minimum size=8mm},snode/.style={rectangle,draw,thick,minimum size=8mm},pnode/.style={rectangle,red,draw,thick,minimum size=8mm}]
\node[cnode] (1) at (-3,0) {$2$};
\node[snode] (2) at (-5,0) {$3$};
\node[cnode] (3) at (-1,2) {$1$};
\node[cnode] (4) at (-2,-2) {$1$};
\node[cnode] (5) at (0,-2) {$1$};
\node[cnode] (6) at (1,0) {$1$};
\node[cnode] (7) at (3,0) {$1$};
\node[snode] (8) at (5,0) {$1$};
\draw[thick] (1) -- (2);
\draw[thick, blue] (1) -- (3);
\draw[thick] (1) -- (4);
\draw[thick] (4) -- (5);
\draw[thick] (5) -- (6);
\draw[thick] (6) -- (7);
\draw[thick] (7) -- (8);
\draw[thick] (3) -- (6);
\node[text width=0.1cm](30) at (-1,-3){$(Y')$};
\end{tikzpicture}}
\end{tabular}
\caption{\footnotesize{Dual theories for $p_1=p_2=p_3=2$.}}
\label{fig: AbEx4}
\end{center}
\end{figure}

\subsection{Duality of the theories without defects}\label{NAb-nodef}

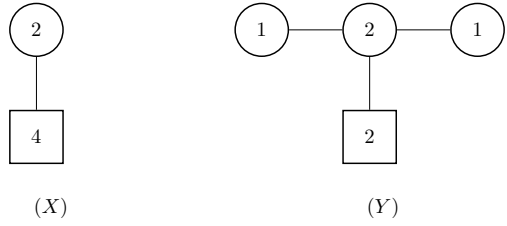
\begin{figure}[htbp]
\begin{center}
\scalebox{.7}{\begin{tikzpicture}[cnode/.style={circle,draw,thick, minimum size=1.0cm},snode/.style={rectangle,draw,thick,minimum size=1cm}]
\node[cnode] (1) {2};
\node[snode] (2) [below=1cm of 1]{4};
\draw[-] (1) -- (2);
\node[text width=0.1cm](20)[below=0.5 cm of 2]{$(X)$};
\end{tikzpicture}
\qquad \qquad \qquad \qquad 
\begin{tikzpicture}[
cnode/.style={circle,draw,thick, minimum size=1.0cm},snode/.style={rectangle,draw,thick,minimum size=1cm}]
\node[cnode] (1) {1};
\node[cnode] (2) [right=1cm  of 1]{2};
\node[cnode] (3) [right=1cm  of 2]{1};
\node[snode] (4) [below=1cm of 2]{2};
\node[text width=0.1cm](20)[below=0.5 cm of 4]{$(Y)$};
\draw[-] (1) -- (2);
\draw[-] (2)-- (3);
\draw[-] (2)-- (4);
\end{tikzpicture}}
\end{center}
\caption{\footnotesize{A pair of dual linear quivers with unitary gauge groups.}}
\label{LQ-basic}
\end{figure}

We first consider the duality without any half-BPS defects.
The starting point for obtaining the mirror pair $(X',Y')$ by an $S$-type operation is the linear mirror pair $(X,Y)$, as 
shown in \figref{LQ-basic}. Mirror symmetry of $X$ and $Y$ implies that
\begin{align}
& Z^{(X)}(\vec m ; \vec t) =C_{XY}(\vec m, \vec t) \, Z^{(Y)}(\vec t; - \vec m), \\
& C_{XY}(\vec m, \vec t)= e^{2\pi i t_1(m_1+ m_2)} e^{-2\pi i t_2(m_3+m_4)}.
\end{align}
where $C_{XY}$ is a contact term.

\begin{figure}[htbp]
\begin{center}
\begin{tabular}{ccc}
\scalebox{.7}{\begin{tikzpicture}[
cnode/.style={circle,draw,thick, minimum size=1.0cm},snode/.style={rectangle,draw,thick,minimum size=1cm}]
\node[snode] (9) at (0,1){1};
\node[snode] (10) at (0,-1){1};
\node[cnode] (11) at (2, 0){2};
\node[snode, green] (12) at (4, 1){1};
\node[snode] (13) at (4, -1){1};
\draw[-] (9) -- (11);
\draw[-] (10) -- (11);
\draw[-] (12) -- (11);
\draw[-] (13) -- (11);
\node[text width=0.1cm](21)[above=0.2 cm of 9]{3};
\node[text width=0.1cm](23)[above=0.2 cm of 12]{1};
\node[text width=0.1cm](24)[below=0.05 cm of 13]{2};
\node[text width=0.1cm](31)[below=0.5 cm of 11]{$(X)$};
\end{tikzpicture}}
&  \scalebox{.7}{\begin{tikzpicture} \draw[thick, ->] (0,0) -- (2,0); 
\node[] at (1,-1.8) {};
\node[text width=1cm](1) at (1,0.5) {$\CO^1_{\CP_1}$};
\end{tikzpicture}}
& \scalebox{.7}{\begin{tikzpicture}[
cnode/.style={circle,draw,thick, minimum size=1.0cm},snode/.style={rectangle,draw,thick,minimum size=1cm}]
\node[snode] (9) at (0,1){1};
\node[snode] (10) at (0,-1){1};
\node[cnode] (11) at (2, 0){2};
\node[cnode] (12) at (4, 1){1};
\node[snode, green] (13) at (4, -1){1};
\node[snode] (14) at (6, 2.5){2};
\node[snode, green] (15) at (6, 1){1};
\node[snode, green] (16) at (6, -0.5){1};
\draw[-] (9) -- (11);
\draw[-] (10) -- (11);
\draw[-] (12) -- (11);
\draw[-] (13) -- (11);
\draw[-] (12) -- (14);
\draw[-] (12) -- (15);
\draw[-] (12) -- (16);
\node[text width=0.1cm](21)[above=0.2 cm of 9]{3};
\node[text width=0.1cm](23)[above=0.2 cm of 12]{1};
\node[text width=0.1cm](24)[below=0.05 cm of 13]{2};
\node[text width=1cm](31)[below=0.5 cm of 13]{$\CO^1_{\CP_1}(X)$};
\end{tikzpicture}}\\
 \scalebox{.7}{\begin{tikzpicture}
\draw[thick, ->] (15,-3) -- (15,-5);
\node[text width=0.1cm](20) at (14.0, -4) {$\CO_{\vec \CP}$};
\end{tikzpicture}}
&\qquad \qquad \qquad
& \scalebox{.7}{\begin{tikzpicture}
\draw[thick,->] (15,-3) -- (15,-5);
\node[text width=0.1cm](29) at (15.5, -4) {${\CO}^2_{\CP_2}$};
\end{tikzpicture}}\\
\scalebox{.7}{\begin{tikzpicture}[
cnode/.style={circle,draw,thick, minimum size=1.0cm},snode/.style={rectangle,draw,thick,minimum size=1cm}]
\node[cnode] (9) at (0,1){1};
\node[snode] (10) at (0,-1){1};
\node[cnode] (11) at (2, 0){2};
\node[cnode] (12) at (4, 1){1};
\node[cnode] (13) at (4, -1){1};
\node[snode] (14) at (6, 1){$2$};
\node[snode] (15) at (6, -1){$2$};
\draw[-] (9) -- (11);
\draw[-] (10) -- (11);
\draw[-] (12) -- (11);
\draw[-] (13) -- (11);
\draw[-] (12) -- (14);
\draw[-] (13) -- (15);
\draw[line width=0.75mm, black] (12) to (13);
\node[text width=0.1cm](20) at (4.5,0){$2$};
\node[text width=0.1cm](21)[above=0.2 cm of 9]{3};
\node[text width=0.1cm](23)[above=0.2 cm of 12]{1};
\node[text width=0.1cm](24)[below=0.05 cm of 13]{2};
\node[text width=0.1cm](31)[below=0.5 cm of 13]{$(X')$};
\end{tikzpicture}}
& \scalebox{.7}{\begin{tikzpicture} \draw[thick, ->] (2,0) -- (0,0); 
\node[] at (1,-1.8) {};
\node[text width=1cm](1) at (1,-0.5) {$\CO^3_{\CP_3}$};
\end{tikzpicture}}
& \scalebox{.7}{\begin{tikzpicture}[
cnode/.style={circle,draw,thick, minimum size=1.0cm},snode/.style={rectangle,draw,thick,minimum size=1cm}]
\node[snode, green] (9) at (0,1){1};
\node[snode] (10) at (0,-1){1};
\node[cnode] (11) at (2, 0){2};
\node[cnode] (12) at (4, 1){1};
\node[cnode] (13) at (4, -1){1};
\node[snode] (14) at (6, 1){$2$};
\node[snode] (15) at (6, -1){$2$};
\draw[-] (9) -- (11);
\draw[-] (10) -- (11);
\draw[-] (12) -- (11);
\draw[-] (13) -- (11);
\draw[-] (12) -- (14);
\draw[-] (13) -- (15);
\draw[line width=0.75mm, black] (12) to (13);
\node[text width=0.1cm](20) at (4.5,0){$2$};
\node[text width=0.1cm](21)[above=0.2 cm of 9]{3};
\node[text width=0.1cm](23)[above=0.2 cm of 12]{1};
\node[text width=0.1cm](24)[below=0.05 cm of 13]{2};
\node[text width=3 cm](31)[below=0.5 cm of 13]{$\CO^2_{\CP_2} \circ \CO^1_{\CP_1}(X)$};
\end{tikzpicture}}
\end{tabular}
\caption{\footnotesize{The construction of the quiver gauge theory $X'$ of \figref{fig: AbEx4gen} by a sequence of 
three elementary Abelian $S$-type operation. The labels $i=1,2,3$ on the flavor nodes correspond to the mass parameters 
$u_i$}. At each step, the flavor node on which the elementary operation acts is shown in green.}
\label{S-Op-II}
\end{center}
\end{figure}

The quiver $X'$ can be obtained from the linear quiver $X$ by implementing an $S$-type operation $\CO_{\vec \CP}$, 
which includes three elementary Abelian $S$-type operations, i.e.
\begin{align}
\CO_{\vec \CP} (X)= \CO^3_{\CP_3} \circ \CO^2_{\CP_2} \circ \CO^1_{\CP_1}(X), \label{SOp-2a}
\end{align}
where $ \CO^i_{\CP_i}$ ($i=1,2,3$) are elementary Abelian $S$-type operations shown in \figref{S-Op-II}. 
The mass parameters $\{u_i\}$ associated the $S$-type operation $\CO^i_{\CP_i}$, are given as
\be
u_1=m_3, \, u_2=m_4, \, u_3=m_1, \, v=m_2. \label{SOp-2b}
\ee
Note that $\CO^1_{\CP_1}$ is a flavoring-gauging operation, $\CO^2_{\CP_2}$ is an identification-flavoring-gauging 
operation, while $\CO^3_{\CP_3}$ is a gauging operation. At each step, the flavor node(s) on which the $S$-type 
operation acts is marked in green. The resultant quiver $X'$ is shown in \figref{S-Op-II} -- it is a 
$U(2)_0 \times \prod^3_{i=1}U(1)_i$ gauge theory (where the subscripts are node labels) with fundamental and 
bifundamental matter as shown. In particular, the $U(1)_1$ and $U(1)_2$ nodes are connected by two bifundamental 
hypers. 

The real mass deformations are labelled as follows. Let $m^{(0)}$, $\vec m^{(1)}$, and $\vec m^{(2)}$ be the masses 
of fundamental hypers associated with the gauge nodes $U(2)_0$, $U(1)_1$ and $U(1)_2$ respectively, and 
$\vec m^{\rm bif}$ be the masses for the $U(1)_1 \times U(1)_2$ bifundamental hypers. The masses of other 
bifundamental hypers can be set to zero by appropriately shifting the integration variables in the matrix model.
The FI parameters are labelled as $\eta_{0}=t_1 -t_2$, and $\{\eta_i\}^3_{i=1}$.

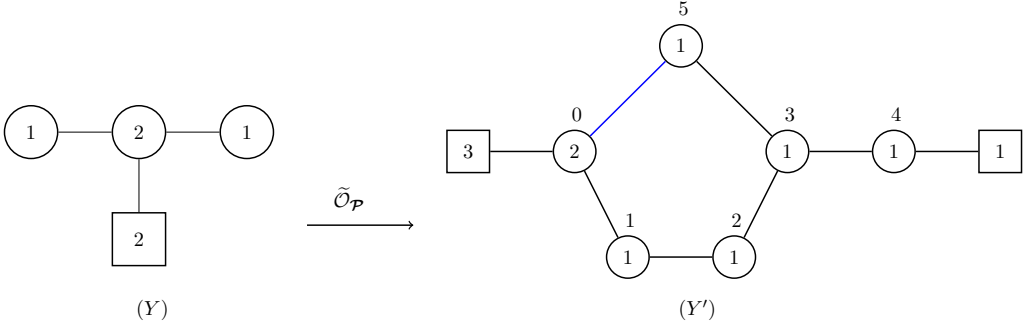
\begin{figure}[htbp]
\begin{center}
\begin{tabular}{ccc}
\scalebox{.7}{\begin{tikzpicture}[
cnode/.style={circle,draw,thick, minimum size=1.0cm},snode/.style={rectangle,draw,thick,minimum size=1cm}]
\node[cnode] (1) {1};
\node[cnode] (2) [right=1cm  of 1]{2};
\node[cnode] (3) [right=1cm  of 2]{1};
\node[snode] (4) [below=1cm of 2]{2};
\node[text width=0.1cm](20)[below=0.5 cm of 4]{$(Y)$};
\draw[-] (1) -- (2);
\draw[-] (2)-- (3);
\draw[-] (2)-- (4);
\end{tikzpicture}}
&  \scalebox{.7}{\begin{tikzpicture} \draw[thick, ->] (0,0) -- (2,0); 
\node[] at (1,-1.8) {};
\node[text width=1cm](1) at (1,0.5) {$\wt{\CO}_{\vec \CP}$};
\end{tikzpicture}}
& \scalebox{.7}{\begin{tikzpicture}[node distance=2cm,cnode/.style={circle,draw,thick,minimum size=8mm},snode/.style={rectangle,draw,thick,minimum size=8mm},pnode/.style={rectangle,red,draw,thick,minimum size=8mm}]
\node[cnode] (1) at (-3,0) {$2$};
\node[snode] (2) at (-5,0) {$3$};
\node[cnode] (3) at (-1,2) {$1$};
\node[cnode] (4) at (-2,-2) {$1$};
\node[cnode] (5) at (0,-2) {$1$};
\node[cnode] (6) at (1,0) {$1$};
\node[cnode] (7) at (3,0) {$1$};
\node[snode] (8) at (5,0) {$1$};
\draw[thick] (1) -- (2);
\draw[thick, blue] (1) -- (3);
\draw[thick] (1) -- (4);
\draw[thick] (4) -- (5);
\draw[thick] (5) -- (6);
\draw[thick] (6) -- (7);
\draw[thick] (7) -- (8);
\draw[thick] (3) -- (6);
\node[text width=0.1cm](31) at (-3, 0.7){0};
\node[text width=0.1cm](32) at (-2, -1.3){1};
\node[text width=0.1cm](33) at (0, -1.3){2};
\node[text width=0.1cm](34) at (1, 0.7){3};
\node[text width=0.1cm](35) at (3, 0.7){4};
\node[text width=0.1cm](36) at (-1, 2.7){5};
\node[text width=0.1cm](30) at (-1,-3){$(Y')$};
\end{tikzpicture}}
\end{tabular}
\caption{\footnotesize{The construction of the quiver gauge theory $(Y')$ from the theory $(Y)$ by the action of the dual operation $\wt{\CO}_{\vec \CP}$.}}
\label{SimpAbEx2GFI}
\end{center}
\end{figure}

The dual theory can be read off from the dual partition partition which can be computed using the prescription in appendix \Secref{sec:SOp-app} 
and is summarized in appendix \Secref{sec:PF-NAb-app}. The dual operation $\wt{\CO}_{\CP}$ and the resultant $Y'$ is shown in 
\figref{SimpAbEx2GFI}. The FI parameters are given as $\{\eta'_{j}\}^5_{j=0}$, where the superscript $j$ coincides with the node 
label in the quiver $Y'$. Similarly, the mass deformations for the fundamental hypers are denoted as $\vec m'^{(0)}$ and $m'^{(4)}$, 
while that for the Abelian hypermultiplet is $m_{\rm Ab}$. 

With this parametrization of FI parameters and masses, the partition functions of $X'$ and $Y'$ are related as
\begin{align}
& Z^{(X')}(m^{(0)}, \vec{m}^{(1)}, \vec{m}^{(2)}, m^{\rm bif}; \vec t, \{\eta_{i}\}) = 
C_{X'Y'}\cdot Z^{(Y')} (\vec m'^{(0)},m'^{(4)}, m_{\rm Ab} ; - \{\eta'_{j}\}^5_{j=0}), \label{MS-Pf-NAb-main}\\
& C_{X'Y'}= e^{2\pi i (2t_1+\eta_3)m^{(0)}}\, e^{2\pi i(\eta_2-t_2)m^{(2)}_1}\, e^{2\pi i(\eta_1-t_2)m^{(1)}_2}, \label{contact-NAb-main}
\end{align}
where $C_{X'Y'}$ should be identified as the new contact term. The mirror map relating the FI parameters of the theory $Y'$ with the masses 
of the theory $X'$ is given as follows:
\begin{subequations}
\begin{empheq}[box=\widefbox]{align}
& \eta'_0= (m^{(0)} - m^{(2)}_1), \label{mm-NAb-1a}\\
&  \eta'_1= m^{\rm bif}_1, \label{mm-NAb-1b}\\
& \eta'_2= (m^{\rm bif}_1- m^{\rm bif}_2), \label{mm-NAb-1c}\\
& \eta'_3= (m^{\rm bif}_2 - m^{(1)}_1 + m^{(2)}_2 ), \label{mm-NAb-1d}\\
& \eta'_4= (m^{(1)}_1 - m^{(1)}_2), \label{mm-NAb-1e}\\
& \eta'_5= (m^{(2)}_1 - m^{(2)}_2). \label{mm-NAb-1f}
\end{empheq}
\end{subequations}
The mirror map relating the mass parameters of the theory $Y'$ with the FI parameters of the theory $X'$ can also 
be read off from \eref{NAb1-DualPf}, but it won't be relevant for the rest of this paper.


\subsection{Vortex defects in $X'$ : Mirror Maps and Hopping dualities}\label{NAb-VD}

In this section, we construct two distinct types of vortex defects in the quiver gauge theory $X'$ as coupled 3d-1d quivers, 
using $S$-type operations. In \Secref{VD-U(2)}, we consider a vortex defect associated with the central $U(2)$ gauge node of $X'$,
while in \Secref{VD-medges}, we consider a defect associated with edge of the quiver with multiplicity 2. In each case, we discuss the 
hopping duals, and obtain the dual Wilson defects in the quiver $Y'$.

\subsubsection{Vortex defect in the central $U(2)$ node of $X'$}\label{VD-U(2)}
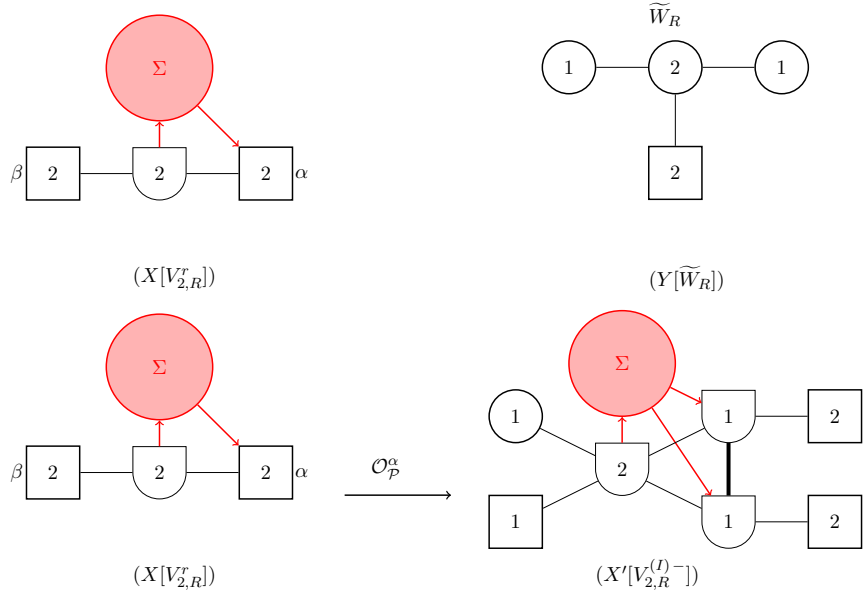
\begin{figure}[htbp]
\begin{center}
\begin{tabular}{ccc}
\scalebox{.7}{\begin{tikzpicture}[node distance=2cm,
cnode/.style={circle,draw,thick, minimum size=1.0cm},snode/.style={rectangle,draw,thick,minimum size=1.0cm}, nnode/.style={red, circle,draw,thick,fill=red!30 ,minimum size=2.0cm}]
\node[snode] (1) at (0,0) {2} ;
\node[cf-group] (2) at (2,0) {\rotatebox{-90}{2}};
\node[snode] (3) at (4,0) {2};
\node[nnode] (4) at (2,2) {$\Sigma$};
\node[text width=1cm](7) at (2, -2) {$(X[V^r_{2,R}])$};
\node[text width=0.1cm](8) at (4.6, 0) {$\alpha$};
\node[text width=0.1cm](9) at (-.75, 0) {$\beta$};
\draw[red, thick, ->] (2)--(4);
\draw[red, thick, ->] (4)--(3);
\draw[-] (1) -- (2);
\draw[-] (2) -- (3);
\end{tikzpicture}}
& \qquad \qquad \qquad
& \scalebox{.7}{\begin{tikzpicture}[node distance=2cm,
cnode/.style={circle,draw,thick, minimum size=1.0cm},snode/.style={rectangle,draw,thick,minimum size=1cm}, pnode/.style={red,rectangle,draw,thick, minimum size=1.0cm}]
\node[cnode] (1) at (0,0) {1} ;
\node[cnode] (2) at (2,0) {2} ;
\node[cnode] (3) at (4,0) {1} ;
\node[snode] (4) at (2,-2) {2};
\draw[-] (1) -- (2);
\draw[-] (2) -- (3);
\draw[-] (2) -- (4);
\node[text width=1cm](5) at (2, 1) {$\wt{W}_{R}$};
\node[text width=1cm](6) at (2, -4) {$(Y[\wt{W}_{R}])$};
\end{tikzpicture}}\\
\scalebox{.7}{\begin{tikzpicture}[node distance=2cm,
cnode/.style={circle,draw,thick, minimum size=1.0cm},snode/.style={rectangle,draw,thick,minimum size=1.0cm}, nnode/.style={red, circle,draw,thick,fill=red!30 ,minimum size=2.0cm}]
\node[snode] (1) at (0,0) {2} ;
\node[cf-group] (2) at (2,0) {\rotatebox{-90}{2}};
\node[snode] (3) at (4,0) {2};
\node[nnode] (4) at (2,2) {$\Sigma$};
\node[text width=1cm](7) at (2, -2) {$(X[V^r_{2,R}])$};
\node[text width=0.1cm](8) at (4.6, 0) {$\alpha$};
\node[text width=0.1cm](9) at (-.75, 0) {$\beta$};
\draw[red, thick, ->] (2)--(4);
\draw[red, thick, ->] (4)--(3);
\draw[-] (1) -- (2);
\draw[-] (2) -- (3);
\end{tikzpicture}}
&\scalebox{.7}{\begin{tikzpicture} \draw[thick, ->] (0,0) -- (2,0); 
\node[] at (1,-1.8) {};
\node[text width=1cm](1) at (1,0.5) {$\CO^\alpha_{\CP}$};
\end{tikzpicture}}
&\scalebox{.7}{\begin{tikzpicture}[node distance=2cm, nnode/.style={circle,draw,thick, red, fill=red!30, minimum size=2.0 cm},cnode/.style={circle,draw,thick,minimum size=1.0 cm},snode/.style={rectangle,draw,thick,minimum size=1.0 cm}]
\node[cnode] (1) at (0,1) {1} ;
\node[cf-group] (2) at (2,0) {\rotatebox{-90}{2}};
\node[snode] (3) at (0,-1) {1};
\node[cf-group] (5) at (4, 1) {\rotatebox{-90}{1}};
\node[cf-group] (6) at (4, -1) {\rotatebox{-90}{1}};
\node[nnode] (7) at (2,2) {$\Sigma$};
\node[snode] (8) at (6,1) {2};
\node[snode] (9) at (6,-1) {2};
\draw[red, thick, ->] (2)--(7);
\draw[red, thick, ->] (7)--(5);
\draw[red, thick, ->] (7)--(6);
\draw[-] (1) -- (2);
\draw[-] (2) -- (3);
\draw[-] (2) -- (5);
\draw[-] (2) -- (6);
\draw[-] (5) -- (8);
\draw[-] (6) -- (9);
\draw[line width=0.75mm, black] (5) to (6);
\node[text width=1cm](9) at (2, -2) {$(X'[V^{(I)\,-}_{2,R}])$};
\end{tikzpicture}}
\end{tabular}
\caption{\footnotesize{Top: a vortex defect in a linear quiver and its dual Wilson defect. Bottom: the $S$-type operation on the linear 3d-1d coupled 
quiver leading to a coupled quiver in a non-ADE 3d theory.}}
\label{AbU2Nf4}
\end{center}
\end{figure}

The starting point of the construction is the mirror pair of quiver gauge theories with defects $(X[V^r_{2,R}], Y[\wt{W}_{R}])$, 
as shown on the top line of \figref{AbU2Nf4}, where $R$ is a representation of $U(2)$. The superscript $r$ implies that the 1d FI 
parameters should be chosen to be negative. 
We then implement the $S$-type operation \eref{SOp-2a}-\eref{SOp-2b} on the coupled quiver $X[V_{2,R}^r]$, 
which leads to the quiver $X'[V^{(I)\,-}_{2,R}]$.

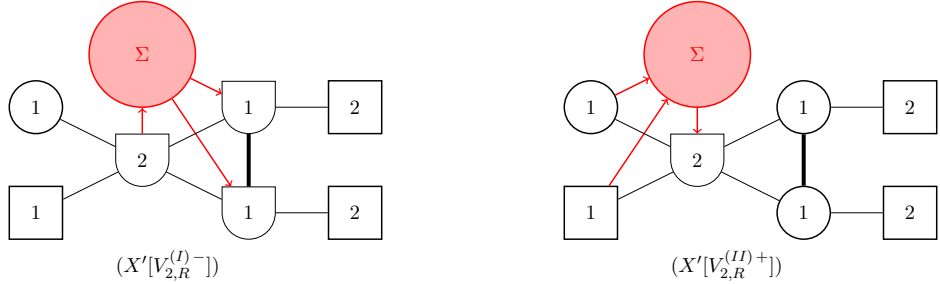
\begin{figure}[htbp]
\begin{center}
\begin{tabular}{ccc}
\scalebox{.7}{\begin{tikzpicture}[node distance=2cm, nnode/.style={circle,draw,thick, red, fill=red!30, minimum size=2.0 cm},cnode/.style={circle,draw,thick,minimum size=1.0 cm},snode/.style={rectangle,draw,thick,minimum size=1.0 cm}]
\node[cnode] (1) at (0,1) {1} ;
\node[cf-group] (2) at (2,0) {\rotatebox{-90}{2}};
\node[snode] (3) at (0,-1) {1};
\node[cf-group] (5) at (4, 1) {\rotatebox{-90}{1}};
\node[cf-group] (6) at (4, -1) {\rotatebox{-90}{1}};
\node[nnode] (7) at (2,2) {$\Sigma$};
\node[snode] (8) at (6,1) {2};
\node[snode] (9) at (6,-1) {2};
\draw[red, thick, ->] (2)--(7);
\draw[red, thick, ->] (7)--(5);
\draw[red, thick, ->] (7)--(6);
\draw[-] (1) -- (2);
\draw[-] (2) -- (3);
\draw[-] (2) -- (5);
\draw[-] (2) -- (6);
\draw[-] (5) -- (8);
\draw[-] (6) -- (9);
\draw[line width=0.75mm, black] (5) to (6);
\node[text width=1cm](9) at (2, -2) {$(X'[V^{(I)\,-}_{2,R}])$};
\end{tikzpicture}}
&\qquad  \qquad \qquad
&\scalebox{.7}{\begin{tikzpicture}[node distance=2cm, nnode/.style={circle,draw,thick, red, fill=red!30, minimum size=2.0 cm},cnode/.style={circle,draw,thick,minimum size=1.0 cm},snode/.style={rectangle,draw,thick,minimum size=1.0 cm}]
\node[cnode] (1) at (0,1) {1} ;
\node[cf-group] (2) at (2,0) {\rotatebox{-90}{2}};
\node[snode] (3) at (0,-1) {1};
\node[cnode] (5) at (4, 1) {1};
\node[cnode] (6) at (4, -1) {1};
\node[nnode] (7) at (2,2) {$\Sigma$};
\node[snode] (8) at (6,1) {2};
\node[snode] (9) at (6,-1) {2};
\draw[red, thick, ->] (7)--(2);
\draw[red, thick, ->] (3)--(7);
\draw[red, thick, ->] (1)--(7);
\draw[-] (1) -- (2);
\draw[-] (2) -- (3);
\draw[-] (2) -- (5);
\draw[-] (2) -- (6);
\draw[-] (5) -- (8);
\draw[-] (6) -- (9);
\draw[line width=0.75mm, black] (5) to (6);
\node[text width=1cm](9) at (2, -2) {$(X'[V^{(II)\,+}_{2,R}])$};
\end{tikzpicture}}
\end{tabular}
\caption{\footnotesize{The hopping duals for the vortex defect.}}
\label{VD-NAb1-HD}
\end{center}
\end{figure}

Following the general prescription of \eref{S-Op-Agen}, the defect partition function in the theory $X'$ can be written as:
\begin{align}
Z^{(X'[V^{(I)\,-}_{2,R}])} = W_{\rm bg} (\vec t, R) \,\lim_{z\to 1} \, \int \,\frac{d^2s_0}{2!}\,\prod^3_{i=1}\, d{s}_i \,
Z^{(X')}_{\rm int}(\vec s, m^{(0)}, \vec{m}^{(1)}, \vec{m}^{(2)}, \vec m^{\rm bif}; \vec t, \{\eta_{i}\})\,\CI^{\Sigma^{2,R}_{(I)\,-}}(\vec s, z|\vec \xi <0),
\end{align}
where the 3d matrix model integrand and the Witten index of the SQM are given as
\begin{align}
& Z^{(X')}_{\rm int}=\frac{e^{2\pi i \sum_i \,\eta_i s_i}}{\prod^2_{a=1}\ch{(s_1-m^{(1)}_a)}\, 
\prod^2_{b=1}\ch{(s_1-s_2 -m^{\rm bif}_b)}\, \prod^2_{c=1}\ch{(s_2-m^{(2)}_c)}} \nn \\
& \qquad \qquad \times \frac{e^{2\pi i \tr{\vec{s}_0} ({t}_1- {t}_2)}  \sinh^2{\pi(s_{0\,1}-s_{0\,2})}}{\prod^2_{j=1}\prod^3_{i=1} \cosh{\pi(s_{0\,j} - s_i)}\cdot \ch{(s_{0\,j}- m^{(0)})}}, \label{NAb-Zint-1}\\
& \CI^{\Sigma^{2,R}_{(I)\,-}}(\vec s, z|\vec \xi <0)=\sum_{w \in R} \prod^2_{j=1} \prod^2_{i=1} \frac{\ch{(s_{0\,j} - s_i)}}{\ch{(s_{0\,j} + i w_j z - s_i)}}, \qquad  
W_{\rm bg} (\vec t, R)= e^{2\pi t_2 |R|}.
\end{align}
The hopping dual of the system can be obtained by the standard transformation $s_{0\,j} \to s_{0\,j} - i w_j z$, and is shown on the 
right in \figref{VD-NAb1-HD}. If one had chosen to implement the $S$-type operation on the ``left" quiver $X[V^l_{2,R}]$ (instead of the 
``right" quiver  $X[V^r_{2,R}]$), one would obtain the quiver $X'[V^{(II)\,+}_{2,R}]$ directly. Note that the superscript $+$ indicates that 
the Witten index should be computed in the positive chamber.

Let us now determine the mirror map of this vortex defects in \figref{VD-NAb1-HD}. The dual defect partition 
function can be written down following the general prescription in \eref{PF-wtOPgenD-A2B}-\eref{CZ-wtOPDD-A2B}:
\begin{align}
Z^{(X'[V^{(I)\,-}_{2,R}])}= & C_{X'Y'}\cdot \int \frac{d^2\s_0}{2!}\,\prod^5_{k=1} d \s_k\, 
Z^{(Y')}_{\rm int}(\vec\s, \vec m', \vec \eta')\,\sum_{w \in R} \, e^{2\pi \sum_j w_j \s_{0\,j} } \nn \\
= & C_{X'Y'}\cdot Z^{(Y'[\wt{W}'^{(0)}_R])}_{\rm int}( \vec m', \vec \eta'),
\end{align}
where in the second step one can identify the RHS as the partition function of theory $Y'$ with a 
Wilson defect for the $U(2)$ gauge node in the representation $R$, up to a contact term $C_{X'Y'}$ 
defined in \eref{contact-NAb-main}. Therefore, the final mirror map may be 
summarized as
\be
\boxed{\langle V^{(I)\,-}_{2,R} \rangle_{X'} = \langle V^{(II)\,+}_{2,R} \rangle_{X'}= \langle \wt{W}'^{(0)}_{R} \rangle_{Y'} .}
\ee
and the dual Wilson defect in the theory $Y'$ is shown in \figref{D9SU3-d2}.

\begin{figure}[htbp]
\begin{center}
\scalebox{.7}{\begin{tikzpicture}[node distance=2cm,cnode/.style={circle,draw,thick,minimum size=8mm},snode/.style={rectangle,draw,thick,minimum size=8mm},pnode/.style={rectangle,red,draw,thick,minimum size=8mm}]
\node[cnode] (1) at (-3,0) {$2$};
\node[snode] (2) at (-5,0) {$3$};
\node[cnode] (3) at (-1,2) {$1$};
\node[cnode] (4) at (-2,-2) {$1$};
\node[cnode] (5) at (0,-2) {$1$};
\node[cnode] (6) at (1,0) {$1$};
\node[cnode] (7) at (3,0) {$1$};
\node[snode] (8) at (5,0) {$1$};
\draw[thick] (1) -- (2);
\draw[thick, blue] (1) -- (3);
\draw[thick] (1) -- (4);
\draw[thick] (4) -- (5);
\draw[thick] (5) -- (6);
\draw[thick] (6) -- (7);
\draw[thick] (7) -- (8);
\draw[thick] (3) -- (6);
\node[text width=1cm](36) at (-2, 0) {$\wt{W}'^{(0)}_{R}$};
\node[text width=0.1cm](40) at (-3, -0.6){0};
\node[text width=0.1cm](41) at (-2, -2.6){1};
\node[text width=0.1cm](42) at (0, -2.6){2};
\node[text width=0.1cm](43) at (1, -0.6){3};
\node[text width=0.1cm](44) at (3, -0.6){4};
\node[text width=0.1cm](45) at (-1, 2.6){5};
\node[text width=0.1cm](30) at (-1,-3){$(Y'[\wt{W}'^{(0)}_{R}])$};
\end{tikzpicture}}
\caption{\footnotesize{Dual Wilson defect for the vortex defects realized by the 3d-1d quivers in \figref{VD-NAb1-HD}.}}
\label{D9SU3-d}
\end{center}
\end{figure}
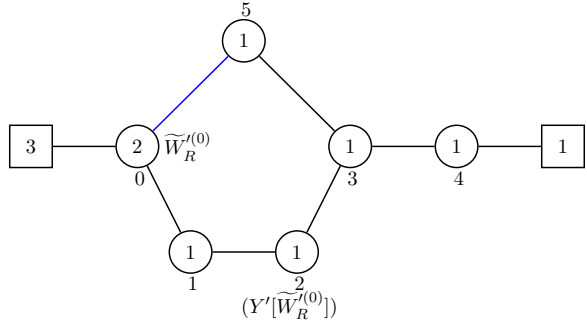


\subsubsection{Vortex defect associated with the edge of multiplicity 2}\label{VD-medges}

\begin{figure}[htbp]
\begin{center}
\begin{tabular}{ccc}
\scalebox{.7}{\begin{tikzpicture}[
cnode/.style={circle,draw,thick, minimum size=1.0cm},snode/.style={rectangle,draw,thick,minimum size=1cm}]
\node[snode] (9) at (0,1){1};
\node[snode] (10) at (0,-1){1};
\node[cnode] (11) at (2, 0){2};
\node[snode, green] (12) at (4, 1){1};
\node[snode] (13) at (4, -1){1};
\draw[-] (9) -- (11);
\draw[-] (10) -- (11);
\draw[-] (12) -- (11);
\draw[-] (13) -- (11);
\node[text width=0.1cm](21)[above=0.2 cm of 9]{3};
\node[text width=0.1cm](23)[above=0.2 cm of 12]{1};
\node[text width=0.1cm](24)[below=0.05 cm of 13]{2};
\node[text width=0.1cm](31)[below=0.5 cm of 11]{$(X)$};
\end{tikzpicture}}
&  \scalebox{.7}{\begin{tikzpicture} \draw[thick, ->] (0,0) -- (2,0); 
\node[] at (1,-1.8) {};
\node[text width=1cm](1) at (1,0.5) {$\CO^1_{\CP_1}$};
\end{tikzpicture}}
& \scalebox{.7}{\begin{tikzpicture}[nnode/.style={circle,draw,thick, red, fill=red!30, minimum size=1.0 cm},
cnode/.style={circle,draw,thick, minimum size=1.0cm},snode/.style={rectangle,draw,thick,minimum size=1cm}]
\node[snode] (9) at (0,1){1};
\node[snode] (10) at (0,-1){1};
\node[cnode] (11) at (2, 0){2};
\node[cf-group] (12) at (4, 1){\rotatebox{-90}{1}};
\node[snode, green] (13) at (4, -1){1};
\node[snode] (14) at (3, 2.5){2};
\node[snode, green] (15) at (7, 1){1};
\node[snode, green] (16) at (6, -0.5){1};
\node[nnode] (17) at (6,2.5){$k$};
\draw[red, thick, ->] (17) to [out=60,in=120,looseness=8] (17);
\draw[-] (9) -- (11);
\draw[-] (10) -- (11);
\draw[-] (12) -- (11);
\draw[-] (13) -- (11);
\draw[-] (12) -- (14);
\draw[-, red] (12) -- (15);
\draw[-, red] (12) -- (16);
\draw[red, ->] (12) -- (17);
\draw[red, ->] (17) -- (15);
\draw[red, ->] (17) -- (16);
\node[text width=0.1cm](21)[above=0.2 cm of 9]{3};
\node[text width=0.1cm](23) at (4,2){1};
\node[text width=0.1cm](24) [below=0.05 cm of 13]{2};
\node[text width=1cm](31)[below=0.5 cm of 13]{$\CO^1_{\CP_1}(X)$};
\end{tikzpicture}}\\
 \scalebox{.7}{\begin{tikzpicture}
\draw[thick, ->] (15,-3) -- (15,-5);
\node[text width=0.1cm](20) at (14.0, -4) {$\CO_{\vec \CP}$};
\end{tikzpicture}}
&\qquad \qquad \qquad
& \scalebox{.7}{\begin{tikzpicture}
\draw[thick,->] (15,-3) -- (15,-5);
\node[text width=0.1cm](29) at (15.5, -4) {${\CO}^2_{\CP_2}$};
\end{tikzpicture}}\\
\scalebox{.7}{\begin{tikzpicture}[nnode/.style={circle,draw,thick, red, fill=red!30, minimum size=1.0 cm},
cnode/.style={circle,draw,thick, minimum size=1.0cm},snode/.style={rectangle,draw,thick,minimum size=1cm}]
\node[cnode] (9) at (0,1){1};
\node[snode] (10) at (0,-1){1};
\node[cnode] (11) at (2, 0){2};
\node[cf-group] (12) at (4, 1){\rotatebox{-90}{1}};
\node[cf-group] (13) at (4, -1){\rotatebox{-90}{1}};
\node[snode] (14) at (6, 2.5){$2$};
\node[snode] (15) at (6, -2.5){$2$};
\node[nnode] (16) at (6,0){$k$};
\draw[red, thick, ->] (16) to [out=60,in=120,looseness=8] (16);
\draw[-] (9) -- (11);
\draw[-] (10) -- (11);
\draw[-] (12) -- (11);
\draw[-] (13) -- (11);
\draw[-] (12) -- (14);
\draw[-] (13) -- (15);
\draw[red,->] (12) -- (16);
\draw[red, line width=0.75mm, ->] (16) -- (13);
\draw[line width=0.75mm, red] (12) to (13);
\node[text width=0.1cm](20) at (4.5,0){$2$};
\node[text width=0.1cm](21)[above=0.2 cm of 9]{3};
\node[text width=0.1cm](23) at (4,2){1};
\node[text width=0.1cm](24) at (4,-2){2};
\node[text width=0.1cm](25) at (5,-1){2};
\node[text width=1.3 cm](31) at (3,-3){$(X'[V^{(I)\,-}_{0,k,2}])$};
\end{tikzpicture}}
& \scalebox{.7}{\begin{tikzpicture} \draw[thick, ->] (2,0) -- (0,0); 
\node[] at (1,-1.8) {};
\node[text width=1cm](1) at (1,-0.5) {$\CO^3_{\CP_3}$};
\end{tikzpicture}}
& \scalebox{.7}{\begin{tikzpicture}[nnode/.style={circle,draw,thick, red, fill=red!30, minimum size=1.0 cm},
cnode/.style={circle,draw,thick, minimum size=1.0cm},snode/.style={rectangle,draw,thick,minimum size=1cm}]
\node[snode, green] (9) at (0,1){1};
\node[snode] (10) at (0,-1){1};
\node[cnode] (11) at (2, 0){2};
\node[cf-group] (12) at (4, 1){\rotatebox{-90}{1}};
\node[cf-group] (13) at (4, -1){\rotatebox{-90}{1}};
\node[snode] (14) at (6, 2.5){$2$};
\node[snode] (15) at (6, -2.5){$2$};
\node[nnode] (16) at (6,0){$k$};
\draw[red, thick, ->] (16) to [out=60,in=120,looseness=8] (16);
\draw[-] (9) -- (11);
\draw[-] (10) -- (11);
\draw[-] (12) -- (11);
\draw[-] (13) -- (11);
\draw[-] (12) -- (14);
\draw[-] (13) -- (15);
\draw[red,->] (12) -- (16);
\draw[red, line width=0.75mm, ->] (16) -- (13);
\draw[line width=0.75mm, red] (12) to (13);
\node[text width=0.1cm](20) at (4.5,0){$2$};
\node[text width=0.1cm](21)[above=0.2 cm of 9]{3};
\node[text width=0.1cm](23) at (4,2){1};
\node[text width=0.1cm](24) at (4,-2){2};
\node[text width=0.1cm](25) at (5,-1){2};
\node[text width=3 cm](31) at (3,-3){$\CO^2_{\CP_2} \circ \CO^1_{\CP_1}(X)$};
\end{tikzpicture}}
\end{tabular}
\caption{\footnotesize{The construction of the quiver gauge theory $X'$ with a vortex defect by a sequence of 
three elementary Abelian $S$-type operation. The labels $i=1,2,3$ on the flavor nodes correspond to the mass parameters 
$u_i$}. At each step, the flavor node on which the elementary operation acts is shown in blue.}
\label{S-Op-IIA}
\end{center}
\end{figure}
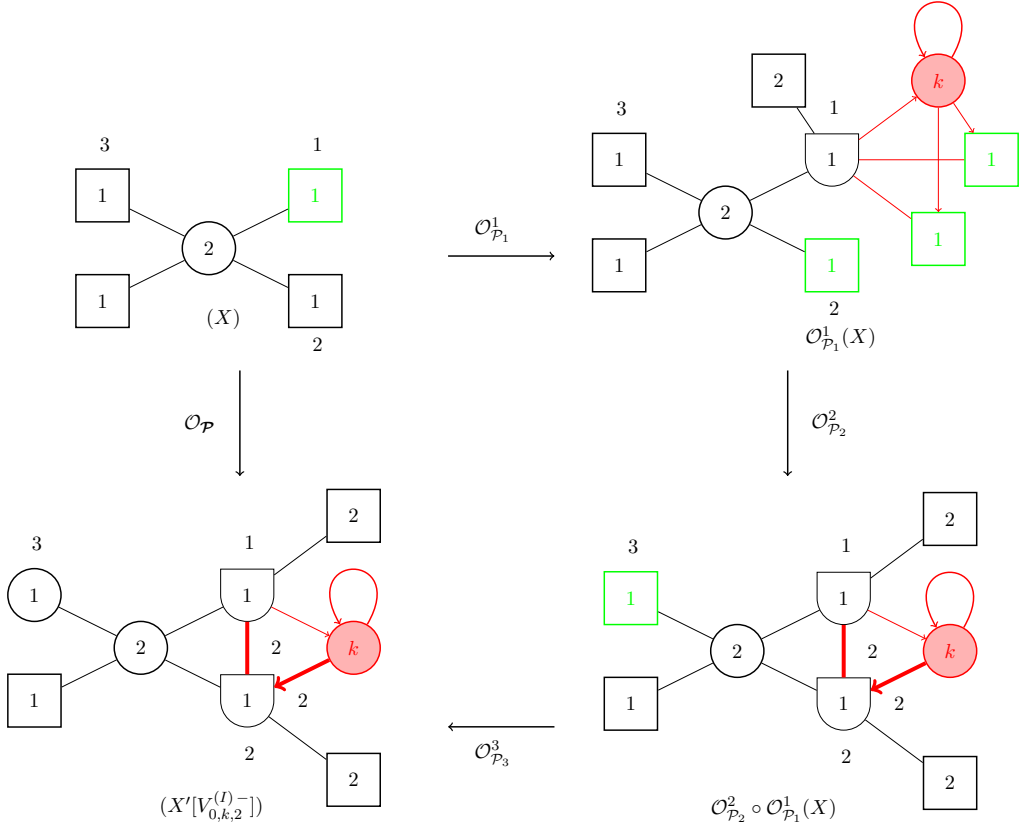

The starting point of the construction is the mirror pair of quiver gauge theories without defects $(X, Y)$, 
as shown in \figref{LQ-basic}. We implement an $S$-type operation on the quiver $X$ of the following form
\be
\CO_{\vec \CP} (X)= \CO^3_{\CP_3} \circ \CO^2_{\CP_2} \circ \CO^1_{\CP_1}(X),
\ee
where the $ \CO^i_{\CP_i}$ ($i=1,2,3$) are elementary Abelian $S$-type operations shown in \figref{S-Op-IIA}.
The mass parameters $\{u_i\}$ associated the $S$-type operation $\CO^i_{\CP_i}$, are given as
\be
u_1=m_3, \, u_2=m_4, \, u_3=m_1, \, v=m_2. \label{SOp-2c}
\ee
The first operation $\CO^1_{\CP_1}$ is a flavoring-defect-gauging operation, where the defect operation involves 
coupling a (2,2) SQM to the 3d theory, as shown in the top right quiver of \figref{S-Op-IIA}. The FI parameter of the 
coupled SQM is chosen to be in the negative chamber, so that the coupled quiver realizes a vortex defect of charge 
$k$ for the $U(1)_1$ gauge node. The second operation $\CO^2_{\CP_2}$ is an identification-flavoring-gauging 
operation and the third operation $\CO^3_{\CP_3}$ is a gauging operation as shown. The resultant vortex defect 
is denoted as $V^{(I)\,-}_{0,k,2}$, following the notation of \Secref{VD-edges}.

\begin{figure}[htbp]
\begin{center}
\begin{tabular}{ccc}
\scalebox{.7}{\begin{tikzpicture}[nnode/.style={circle,draw,thick, red, fill=red!30, minimum size=1.0 cm},
cnode/.style={circle,draw,thick, minimum size=1.0cm},snode/.style={rectangle,draw,thick,minimum size=1cm}]
\node[cnode] (9) at (0,1){1};
\node[snode] (10) at (0,-1){1};
\node[cnode] (11) at (2, 0){2};
\node[cf-group] (12) at (4, 1){\rotatebox{-90}{1}};
\node[cf-group] (13) at (4, -1){\rotatebox{-90}{1}};
\node[snode] (14) at (6, 2.5){$2$};
\node[snode] (15) at (6, -2.5){$2$};
\node[nnode] (16) at (6,0){$k$};
\draw[red, thick, ->] (16) to [out=60,in=120,looseness=8] (16);
\draw[-] (9) -- (11);
\draw[-] (10) -- (11);
\draw[-] (12) -- (11);
\draw[-] (13) -- (11);
\draw[-] (12) -- (14);
\draw[-] (13) -- (15);
\draw[red,->] (12) -- (16);
\draw[red, line width=0.75mm, ->] (16) -- (13);
\draw[line width=0.75mm, red] (12) to (13);
\node[text width=0.1cm](20) at (4.5,0){$2$};
\node[text width=0.1cm](21)[above=0.2 cm of 9]{3};
\node[text width=0.1cm](23) at (4,2){1};
\node[text width=0.1cm](24) at (4,-2){2};
\node[text width=0.1cm](25) at (5,-1){2};
\node[text width=1.3 cm](31) at (3,-3){$(X'[V_{0,k,2}^{(I)\,-}])$};
\end{tikzpicture}}
& \qquad \qquad \qquad
& \scalebox{.7}{\begin{tikzpicture}[nnode/.style={circle,draw,thick, red, fill=red!30, minimum size=1.0 cm},
cnode/.style={circle,draw,thick, minimum size=1.0cm},snode/.style={rectangle,draw,thick,minimum size=1cm}]
\node[cnode] (9) at (0,1){1};
\node[snode] (10) at (0,-1){1};
\node[cf-group] (11) at (2, 0){\rotatebox{-90}{2}};
\node[cf-group] (12) at (4, 1){\rotatebox{-90}{1}};
\node[cnode] (13) at (4, -1){1};
\node[snode] (14) at (6, 2.5){$2$};
\node[snode] (15) at (6, -2.5){$2$};
\node[nnode] (16) at (2, 2.5){$k$};
\draw[red, thick, ->] (16) to [out=60,in=120,looseness=8] (16);
\draw[-] (9) -- (11);
\draw[-] (10) -- (11);
\draw[-] (12) -- (11);
\draw[-] (13) -- (11);
\draw[-] (12) -- (14);
\draw[-] (13) -- (15);
\draw[red,thick, ->] (16) -- (12);
\draw[line width=0.75mm] (12) to (13);
\draw[red, thick, ->] (11) -- (16);
\draw[red, thick, ->] (14) -- (16);
\node[text width=0.1cm](20) at (4.5,0){$2$};
\node[text width=0.1cm](21)[above=0.2 cm of 9]{3};
\node[text width=0.1cm](23)at (4,2){1};
\node[text width=0.1cm](24)[below=0.05 cm of 13]{2};
\node[text width=1.3 cm](31) at (3,-3){$(X'[V_{0,k,2}^{(II)\,+}])$};
\end{tikzpicture}}
\end{tabular}
\caption{\footnotesize{The hopping duals for the vortex defect.}}
\label{VD-NAb2-HD}
\end{center}
\end{figure}
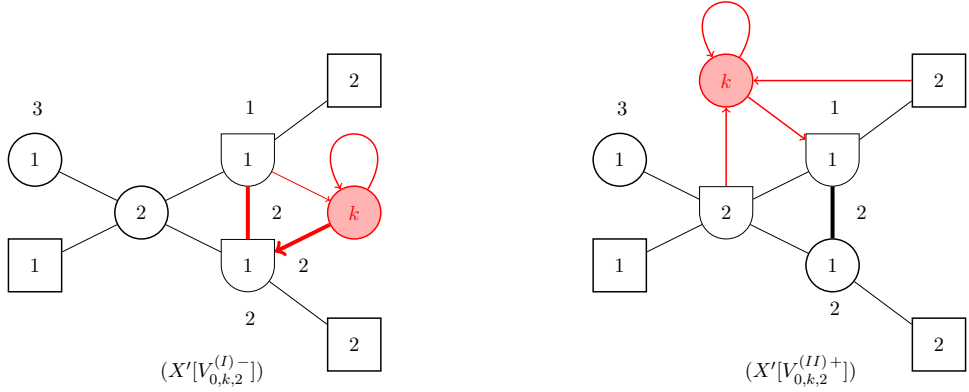

Following the general prescription of \eref{S-Op-Agen}, the defect partition function can be written as:
\begin{align}
Z^{(X'[V^{(I)\,-}_{0,k,2}])} =  \lim_{z\to 1} \, \int \,\frac{d^2s_0}{2!}\,\prod^3_{i=1}\, d{s}_i \,
Z^{(X')}_{\rm int}(\vec s, \ldots)\,\CI^{\Sigma^{0,k,2}_{(I)\,-}}(\vec s, \vec m^{\rm bif}, z|\vec \xi <0),
\end{align}
where $Z^{(X')}_{\rm int}$ is given in \eref{NAb-Zint-1}, and the Witten index $\CI^{\Sigma^{0,k,2}_{(I)\,-}}$ is:
\be
\CI^{\Sigma^{0,k,2}_{(I)\,-}}(\vec s, \vec m^{\rm bif}, z|\vec \xi <0) = \frac{\prod^2_{b=1} \ch{(s_1-s_2 - m^{\rm bif}_b )}}{\prod^2_{b=1} \ch{(s_1-s_2 - m^{\rm bif}_b +ikz)}}.
\ee
The hopping dual of the system can be obtained by the standard transformation $s_{1} \to s_{1} - i k\,z$, and is shown on the 
right in \figref{VD-NAb2-HD}. We denote the associated vortex defect as $V^{(II)\,+}_{0,k,2}$, where the superscript $+$ indicates 
that the Witten index should be computed in the positive chamber. \\


The mirror dual of this vortex defect can be read off from the dual defect partition 
function as follows.
Starting from the vortex defect partition function, we have
\begin{align}
&Z^{(X'[V^{0,k,2}_{(I)\,-}])}=   \lim_{z \to 1} \, Z^{(X')}(m^{(0)},\vec m^{(1)}, \vec m^{(2)}, \vec m^{\rm bif}-ikz; \vec t, \vec \eta) \nn \\
& =  \lim_{z \to 1}\, C_{X'Y'}(\vec m^{\rm bif}-ikz, \ldots)\, Z^{(Y')}(\vec t, \vec \eta ; m^{(0)},\vec m^{(1)}, \vec m^{(2)}, \vec m^{\rm bif}-ikz),
\end{align}
where in the second step we have used the relation between the partition functions of $X'$ and $Y'$ due to mirror symmetry. 
Since $C_{X'Y'}$ does not depend on the mass parameters $\vec m^{\rm bif}$ (as can be seen from the explicit expression 
in \eref{contact-NAb-main}), the above equation reduces to
\be
Z^{(X'[V^{0,k,2}_{(I)\,-}])}= C_{X'Y'} \cdot Z^{(Y')}(\vec t, \vec \eta ; m^{(0)},\vec m^{(1)}, \vec m^{(2)}, \vec m^{\rm bif}-ik).
\ee
Using the mirror map \eref{mm-NAb-1a}-\eref{mm-NAb-1f} relating masses of $X'$ to FI parameters of $Y'$, we obtain the 
following relation:
\begin{align}
Z^{(X'[V^{0,k,2}_{(I)\,-}])} =&  C_{X'Y'} \, \int \frac{d^2\s_0}{2!}\,\prod^5_{k=1} d\s_k\, Z^{(Y')}_{\rm int} (\vec \s, \vec \tau, \vec m', -\vec \eta') \,
Z_{\rm Wilson}(\s_1, k)\, Z_{\rm Wilson}(\s_3, -k), \nn \\
=& C_{X'Y'} \, Z^{(Y'[W'^{(1)}_k \cdot W'^{(3)}_{-k}])}(\vec m', -\vec \eta').
\end{align}
where in the second step one can identify the RHS as the partition function of theory $Y'$ with a Wilson defect of 
charge $k$ for the gauge node $U(1)_1$ and charge $-k$ for the gauge node $U(1)_3$. This is shown in \figref{D9SU3-d-1}.
One can therefore summarize the mirror map as:
\be
\boxed{\langle V^{0,k,2}_{(I)\,-} \rangle_{X'} = \langle V^{0,k,2}_{(II)\,+} \rangle_{X'} =\langle \wt{W}'^{(1)}_k \cdot \wt{W}'^{(3)}_{-k} \rangle_{Y'}  .}
\ee

\begin{figure}[htbp]
\begin{center}
\scalebox{.7}{\begin{tikzpicture}[node distance=2cm,cnode/.style={circle,draw,thick,minimum size=8mm},snode/.style={rectangle,draw,thick,minimum size=8mm},pnode/.style={rectangle,red,draw,thick,minimum size=8mm}]
\node[cnode] (1) at (-3,0) {$2$};
\node[snode] (2) at (-5,0) {$3$};
\node[cnode] (3) at (-1,2) {$1$};
\node[cnode] (4) at (-2,-2) {$1$};
\node[cnode] (5) at (0,-2) {$1$};
\node[cnode] (6) at (1,0) {$1$};
\node[cnode] (7) at (3,0) {$1$};
\node[snode] (8) at (5,0) {$1$};
\draw[thick] (1) -- (2);
\draw[thick, blue] (1) -- (3);
\draw[thick] (1) -- (4);
\draw[thick] (4) -- (5);
\draw[thick] (5) -- (6);
\draw[thick] (6) -- (7);
\draw[thick] (7) -- (8);
\draw[thick] (3) -- (6);
\node[text width=1cm](36) at (-1.8, -1.2) {$\wt{W}'^{(1)}_k$};
\node[text width=1cm](37) at (0.2, 0) {$\wt{W}'^{(3)}_{-k}$};
\node[text width=0.1cm](40) at (-3, -0.6){0};
\node[text width=0.1cm](41) at (-2, -2.6){1};
\node[text width=0.1cm](42) at (0, -2.6){2};
\node[text width=0.1cm](43) at (1, -0.6){3};
\node[text width=0.1cm](44) at (3, -0.6){4};
\node[text width=0.1cm](45) at (-1, 2.6){5};
\node[text width=3 cm](30) at (-1,-3.5){$(Y'[\wt{W}'^{(1)}_k \cdot \wt{W}'^{(3)}_{-k}])$};
\end{tikzpicture}}
\caption{\footnotesize{The Wilson defect dual to the vortex defects in \figref{VD-NAb2-HD}.}}
\label{D9SU3-d-1}
\end{center}
\end{figure}


\subsection{Mapping the defects to the $U-SU$ mirror}\label{NAb-VD-SU}

In \cite{Dey:2021rxw}, an IR duality was conjectured where the dual theories $(\CT, \CT^\vee)$ belong to the following classes of quiver gauge 
theories respectively:

\begin{itemize}

\item Theory $\CT$: A linear chain of unitary and special unitary gauge groups with fundamentals and bifundamentals.

\item Theory $\CT^\vee$: A generically non-linear quiver involving only unitary gauge groups, decorated with Abelian matter 
in addition to fundamental/bifundamental hypers.

\end{itemize}

In particular, it was shown that an infinite family of such pairs $(\CT, \CT^\vee)$ arise as two different Lagrangians for the 
3d SCFT obtained by circle reduction of a $D_p(SU(N))$ Argyres-Douglas theory. It was also shown that the pair $(\CT, \CT^\vee)$ has a mirror 
dual which is a generically non-ADE quiver gauge theory with unitary gauge nodes and multiple edges. 
For the case of $p=9$, $N=3$, we have the following IR dual pair:

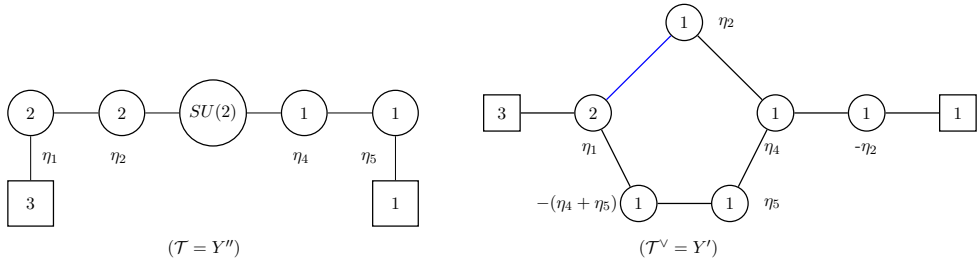
\begin{center}
\begin{tabular}{ccc}
\scalebox{.6}{\begin{tikzpicture}[
cnode/.style={circle,draw,thick, minimum size=1.0cm},snode/.style={rectangle,draw,thick,minimum size=1cm},pnode/.style={circle,double,draw,thick, minimum size=1.0cm}]
\node[cnode] (1) at (0,0){2};
\node[snode] (2) at (0,-2){3};
\node[cnode] (3) at (2,0){2};
\node[cnode] (4) at (4,0){$SU(2)$};
\node[cnode] (5) at (6,0){1};
\node[cnode] (6) at (8,0){1};
\node[snode] (7) at (8,-2){1};
\draw[-] (1) -- (2);
\draw[-] (1)-- (3);
\draw[-] (3)-- (4);
\draw[-] (4)-- (5);
\draw[-] (5)-- (6);
\draw[-] (6)-- (7);
\node[text width=2cm](30) at (4,-3){$(\CT=Y'')$};
\node[text width=0.5cm](11) at (0.5, -1){$\eta_1$};
\node[text width=0.5cm](12) at (2, -1){$\eta_2$};
\node[text width=0.5cm](13) at (6, -1){$\eta_4$};
\node[text width=0.5cm](14) at (7.5, -1){$\eta_5$};
\end{tikzpicture}}
&\qquad
&\scalebox{.6}{\begin{tikzpicture}[node distance=2cm,cnode/.style={circle,draw,thick,minimum size=8mm},snode/.style={rectangle,draw,thick,minimum size=8mm},pnode/.style={rectangle,red,draw,thick,minimum size=8mm}]
\node[cnode] (1) at (-3,0) {$2$};
\node[snode] (2) at (-5,0) {$3$};
\node[cnode] (3) at (-1,2) {$1$};
\node[cnode] (4) at (-2,-2) {$1$};
\node[cnode] (5) at (0,-2) {$1$};
\node[cnode] (6) at (1,0) {$1$};
\node[cnode] (7) at (3,0) {$1$};
\node[snode] (8) at (5,0) {$1$};
\draw[thick] (1) -- (2);
\draw[thick, blue] (1) -- (3);
\draw[thick] (1) -- (4);
\draw[thick] (4) -- (5);
\draw[thick] (5) -- (6);
\draw[thick] (6) -- (7);
\draw[thick] (7) -- (8);
\draw[thick] (3) -- (6);
\node[text width=2cm](30) at (-1,-3){$(\CT^\vee =Y')$};
\node[text width=0.5cm](11) at (-3, -0.8){$\eta_1$};
\node[text width=2.5 cm](12) at (-3, -2){$-(\eta_4 +\eta_5)$};
\node[text width=0.5cm](13) at (1, -2){$\eta_5$};
\node[text width=0.5cm](14) at (0, 2){$\eta_2$};
\node[text width=0.5cm](15) at (1, -0.8){$\eta_4$};
\node[text width=0.5cm](16) at (3, -0.8){-$\eta_2$};
\end{tikzpicture}}
\end{tabular}
\end{center}

In the above quivers, we have indicated how the FI parameters on the left map to FI parameters on the right. 
In \cite{Dey:2021rxw}, this result was explicitly obtained by checking the proposed duality at the level of the three sphere 
partition function. 

The quiver of the class $\CT^\vee$ in this case can be evidently identified as the quiver $Y'$ in \figref{SimpAbEx2GFI}, 
which implies that the theory $\CT$ (which we will refer to as $Y''$ henceforth) is mirror dual to the theory $X'$. 
It is therefore natural to ask which defect operators in $Y''$ do the vortex defects considered in \Secref{VD-U(2)} 
and \Secref{VD-medges} map to. Computing the partition 
functions of the theories $\CT$ and $Y'$ with Wilson defects, one can readily show that:
\be
\langle \wt{W}'^{(0)}_{R} \rangle_{Y'} = \langle \wt{W}''^{(1)}_{R} \rangle_{Y''},
\ee
where $\wt{W}'^{(0)}_{R}$ is a Wilson defect for the $U(2)$ gauge node in the representation $R$ in the theory $Y'$,
while $\wt{W}''^{(1)}_{R}$ is a Wilson defect for the leftmost $U(2)$ gauge node in the theory $Y''$ in the same 
representation. The vortex defect $V^{(I)\,-}_{2,R}$ studied in \Secref{VD-U(2)} therefore maps to the following Wilson defect in the 
theory $Y''$:
\be
\boxed{\langle V^{(I)\,-}_{2,R} \rangle_{X'} = \langle V^{(II)\,+}_{2,R} \rangle_{X'}= 
\langle \wt{W}'^{(0)}_{R} \rangle_{Y'} = \langle \wt{W}''^{(1)}_{R} \rangle_{Y''},}
\ee
as shown in \figref{D9SU3-d2}.\\

For the vortex defect $V^{(I)\,-}_{0,k,2}$ studied in \Secref{VD-medges}, the computation proceeds in an analogous fashion.
Using the map of FI parameters between the theory $\CT$ and $Y'(\CT^\vee)$ as shown above, one can show that at 
the level of the sphere partition function:
\be
\langle \wt{W}'^{(1)}_{k}\cdot \wt{W}'^{(3)}_{-k} \rangle_{Y'} = \langle \wt{W}''^{(4)}_{-k} \rangle_{\CT},
\ee
where $\wt{W}'^{(1)}_{k}\cdot \wt{W}'^{(3)}_{-k}$ is the Wilson defect in $Y'$ shown in \figref{D9SU3-d-1}, while $\wt{W}''^{(4)}_{-k}$ 
is a Wilson defect in theory $\CT$ of charge $k$ for the $U(1)$ gauge node which is fourth from the left (see the figure on the RHS 
of \figref{D9SU3-d2-2}). The vortex defect $V^{(I)\,-}_{0,k,2}$ studied in \Secref{VD-medges} therefore maps to following Wilson defect in the 
theory $\CT$:
\be
\boxed{\langle V^{0,k,2}_{(I)\,-} \rangle_{X'} = \langle V^{0,k,2}_{(II)\,+} \rangle_{X'} =\langle \wt{W}'^{(1)}_k \cdot \wt{W}'^{(3)}_{-k} \rangle_{Y'}
= \langle \wt{W}''^{(4)}_{-k} \rangle_{\CT} .}
\ee
as shown in \figref{D9SU3-d2-2}.

\begin{figure}[htbp]
\begin{center}
\begin{tabular}{ccc}
\scalebox{.7}{\begin{tikzpicture}[node distance=2cm, nnode/.style={circle,draw,thick, red, fill=red!30, minimum size=2.0 cm},cnode/.style={circle,draw,thick,minimum size=1.0 cm},snode/.style={rectangle,draw,thick,minimum size=1.0 cm}]
\node[cnode] (1) at (0,1) {1} ;
\node[cf-group] (2) at (2,0) {\rotatebox{-90}{2}};
\node[snode] (3) at (0,-1) {1};
\node[cf-group] (5) at (4, 1) {\rotatebox{-90}{1}};
\node[cf-group] (6) at (4, -1) {\rotatebox{-90}{1}};
\node[nnode] (7) at (2,2) {$\Sigma^{2,R}$};
\node[snode] (8) at (6,1) {2};
\node[snode] (9) at (6,-1) {2};
\draw[red, thick, ->] (2)--(7);
\draw[red, thick, ->] (7)--(5);
\draw[red, thick, ->] (7)--(6);
\draw[-] (1) -- (2);
\draw[-] (2) -- (3);
\draw[-] (2) -- (5);
\draw[-] (2) -- (6);
\draw[-] (5) -- (8);
\draw[-] (6) -- (9);
\draw[line width=0.75mm, black] (5) to (6);
\node[text width=1cm](9) at (2, -2) {$(X'[V^{(I)\,-}_{2,R}])$};
\end{tikzpicture}}
& \qquad \qquad \qquad
& \scalebox{.7}{\begin{tikzpicture}[node distance=2cm,cnode/.style={circle,draw,thick,minimum size=8mm},snode/.style={rectangle,draw,thick,minimum size=8mm},pnode/.style={rectangle,red,draw,thick,minimum size=8mm}]
\node[cnode] (1) at (-3,0) {$2$};
\node[snode] (2) at (-5,0) {$3$};
\node[cnode] (3) at (-1,0) {$2$};
\node[cnode] (4) at (1,0) {$SU(2)$};
\node[cnode] (5) at (3,0) {$1$};
\node[cnode] (6) at (5,0) {$1$};
\node[snode] (7) at (7,0) {$1$};
\draw[thick] (1) -- (2);
\draw[thick] (1) -- (3);
\draw[thick] (3) -- (4);
\draw[thick] (4) -- (5);
\draw[thick] (5) -- (6);
\draw[thick] (6) -- (7);
\node[text width=1cm](36) at (-3, 0.8) {$\wt{W}''^{(1)}_{R}$};
\node[text width=0.1cm](50) at (-3, -0.8) {1};
\node[text width=0.1cm](50) at (-1, -0.8) {2};
\node[text width=0.1cm](50) at (1, -1) {3};
\node[text width=0.1cm](50) at (3, -0.8) {4};
\node[text width=0.1cm](50) at (5, -0.8) {5};
\node[text width=0.1cm](30) at (-1,-3){$(Y''[\wt{W}''^{(1)}_{R}])$};
\end{tikzpicture}}
\end{tabular}
\caption{\footnotesize{A vortex defect in a non-ADE quiver and its dual Wilson defect in the unitary-special unitary mirror.}}
\label{D9SU3-d2}
\end{center}
\end{figure}

\begin{figure}[htbp]
\begin{center}
\begin{tabular}{ccc}
\scalebox{.7}{\begin{tikzpicture}[nnode/.style={circle,draw,thick, red, fill=red!30, minimum size=1.0 cm},
cnode/.style={circle,draw,thick, minimum size=1.0cm},snode/.style={rectangle,draw,thick,minimum size=1cm}]
\node[cnode] (9) at (0,1){1};
\node[snode] (10) at (0,-1){1};
\node[cnode] (11) at (2, 0){2};
\node[cnode] (12) at (4, 1){1};
\node[cnode] (13) at (4, -1){1};
\node[snode] (14) at (6, 2.5){$2$};
\node[snode] (15) at (6, -2.5){$2$};
\node[nnode] (16) at (6,0){$k$};
\draw[red, thick, ->] (16) to [out=60,in=120,looseness=8] (16);
\draw[-] (9) -- (11);
\draw[-] (10) -- (11);
\draw[-] (12) -- (11);
\draw[-] (13) -- (11);
\draw[-] (12) -- (14);
\draw[-] (13) -- (15);
\draw[red,->] (12) -- (16);
\draw[red, line width=0.75mm, ->] (16) -- (13);
\draw[line width=0.75mm, red] (12) to (13);
\node[text width=0.1cm](20) at (4.5,0){$2$};
\node[text width=0.1cm](21)[above=0.2 cm of 9]{3};
\node[text width=0.1cm](23)[above=0.2 cm of 12]{1};
\node[text width=0.1cm](24)[below=0.05 cm of 13]{2};
\node[text width=0.1cm](25) at (5,-1){2};
\node[text width=1.3 cm](31)[below=0.5 cm of 13]{$(X'[V^{(I)\,-}_{0,k,2}])$};
\end{tikzpicture}}
& \qquad \qquad \qquad
& \scalebox{.7}{\begin{tikzpicture}[node distance=2cm,cnode/.style={circle,draw,thick,minimum size=8mm},snode/.style={rectangle,draw,thick,minimum size=8mm},pnode/.style={rectangle,red,draw,thick,minimum size=8mm}]
\node[cnode] (1) at (-3,0) {$2$};
\node[snode] (2) at (-5,0) {$3$};
\node[cnode] (3) at (-1,0) {$2$};
\node[cnode] (4) at (1,0) {$SU(2)$};
\node[cnode] (5) at (3,0) {$1$};
\node[cnode] (6) at (5,0) {$1$};
\node[snode] (7) at (7,0) {$1$};
\draw[thick] (1) -- (2);
\draw[thick] (1) -- (3);
\draw[thick] (3) -- (4);
\draw[thick] (4) -- (5);
\draw[thick] (5) -- (6);
\draw[thick] (6) -- (7);
\node[text width=1cm](36) at (3, 0.8) {$\wt{W}''^{(4)}_{-k}$};
\node[text width=0.1cm](50) at (-3, -0.8) {1};
\node[text width=0.1cm](50) at (-1, -0.8) {2};
\node[text width=0.1cm](50) at (1, -1) {3};
\node[text width=0.1cm](50) at (3, -0.8) {4};
\node[text width=0.1cm](50) at (5, -0.8) {5};
\node[text width=0.1cm](30) at (-1,-3){$(Y''[\wt{W}''^{(4)}_{-k}])$};
\end{tikzpicture}}
\end{tabular}
\caption{\footnotesize{A vortex defect in a non-ADE quiver and its dual Wilson defect in the unitary-special unitary mirror.}}
\label{D9SU3-d2-2}
\end{center}
\end{figure}

\section* {Acknowledgements}
The author would like to thank the organizers of the Simons Summer Workshop 2021, where part of the work was completed.
This work is partially supported at the Johns Hopkins University by NSF grant PHY-1820784, and the Simons Collaboration 
on Global Categorical Symmetries.

\appendix

\section{Witten Index of SQMs} \label{sec:WI-app}

A detailed treatment of Witten indices for (2,2) SQMs and 1d dualities can be found in \cite{Hori:2014tda}.
The computation of Witten indices, relevant for vortex defects in linear quivers, can be found in appendix B 
of \cite{Assel:2015oxa}. In this appendix, we briefly review a few points relevant to our discussion in \Secref{VD-LQ} 
and \Secref{VD-edges}.\\

Consider a (2,2) SQM of the following form:

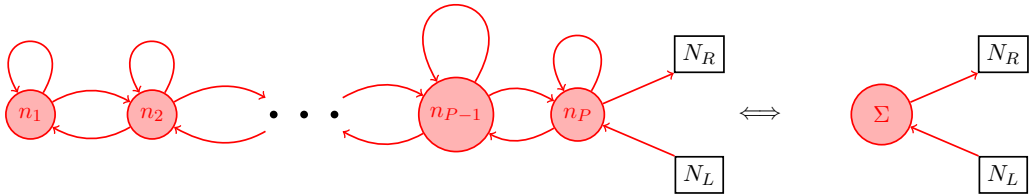
\begin{center}
\scalebox{0.8}{\begin{tikzpicture}[
nnode/.style={circle,draw,thick, fill=blue,minimum size= 6mm},cnode/.style={circle,draw,thick,minimum size=4mm},snode/.style={rectangle,draw,thick,minimum size=6mm},rnode/.style={red, circle,draw,thick,fill=red!30 ,minimum size=4mm},rrnode/.style={red, circle,draw,thick,fill=red!30 ,minimum size=1.0cm}]
\node[rnode] (1) at (-4,0) {$n_1$};
\node[rnode] (2) at (-2,0) {$n_2$};
\node[] (3) at (0,0.2){};
\node[] (4) at (0,-0.2){};
\node[circle,draw,thick, fill, inner sep=1 pt] (5) at (0,0){} ;
\node[circle,draw,thick, fill, inner sep=1 pt] (6) at (0.5,0){} ;
\node[circle,draw,thick, fill, inner sep=1 pt] (7) at (1,0){} ;
\node[] (8) at (1,0.2){};
\node[] (9) at (1,-0.2){};
\node[rnode] (10) at (3,0) {$n_{P-1}$};
\node[rnode] (11) at (5,0) {$n_{P}$};
\node[snode] (12) at (7,1) {$N_R$};
\node[snode] (13) at (7,-1) {$N_L$};
\draw[red, thick, ->] (1) to [out=30,in=150] (2);
\draw[red, thick, ->] (2) to [out=210,in=330] (1);
\draw[red, thick, ->] (2) to [out=30,in=150] (3);
\draw[red, thick, ->] (4) to [out=210,in=330] (2);
\draw[red, thick, ->] (8) to [out=30,in=150] (10);
\draw[red, thick, ->] (10) to [out=210,in=330] (9);
\draw[red, thick, ->] (10) to [out=30,in=150] (11);
\draw[red, thick, ->] (11) to [out=210,in=330] (10);
\draw[red, thick, ->] (11) to (12.south west);
\draw[red, thick, ->] (13.north west) to (11);
\draw[red, thick, ->] (1) to [out=60,in=120,looseness=8] (1);
\draw[red, thick, ->] (2) to [out=60,in=120,looseness=8] (2);
\draw[red, thick, ->] (10) to [out=60,in=120,looseness=8] (10);
\draw[red, thick, ->] (11) to [out=60,in=120,looseness=8] (11);
\node[] (17) at (8,0) {$\Longleftrightarrow$};
\node[rrnode] (14) at (10,0) {$\Sigma$};
\node[snode] (15) at (12,1) {$N_R$};
\node[snode] (16) at (12,-1) {$N_L$};
\draw[red, thick, ->] (14) to (15.south west);
\draw[red, thick, ->] (16.north west) to (14);
\end{tikzpicture}}
\end{center}

The gauge node $U(n_P)$ has $N_L$ fundamental and $N_R$ anti-fundamental chiral multiplets. 
The integers $\{n_1,n_2,\ldots,n_P\}$ encode a representation $R$ of $U(N_L)$ and a representation
$R'$ of $U(N_R)$. For simplicity, consider the case of $P=1$, with $n_1=k$. In this case, the weights
of the representation $R=\CS_k$ are in one-to-one correspondence to the $N_L$-partitions of the integer $k$,
where $\CS_k$ is the $k$-th symmetric representation of $U(N_L)$.
Similarly, the weights of the representation $R'=\CS'_k$ are in one-to-one correspondence to the $N_R$-partitions 
of the integer $k$, where $\CS'_k$ is the $k$-th symmetric representation of $U(N_R)$.

For a generic $P$, the representations have the form:
\be
R = \otimes^P_{i=1} \CS_{k_i}, \qquad R' = \otimes^P_{i=1} \CS'_{k_i},
\ee
where the weights of $\CS_{k_i}$ and $\CS'_{k_i}$ are in one-to-one correspondence to the 
$N_L$-partitions and the $N_R$-partitions respectively of the integer $k_i=n_i-n_{i-1}$. Antisymmetric 
representations can be included in the discussion by appropriately tweaking the SQM \cite{Assel:2015oxa}. \\

In the chamber where all FI parameters are negative, the JK-residue formula for the Witten index of the SQM \cite{Hori:2014tda}  
gets non-zero residues from poles associated with the fundamental chiral multiplets only, and leads to the 
following expression:
\begin{align}
 \CI^{\Sigma}(\vec \zeta <0)=\sum_{w \in R[U(N_L)]}\, \CF(\vec s^{(N_L)}, z)\, \prod^{N_L}_{j=1} \prod^{N_R}_{i=1} \frac{\ch{(s^{(N_L)}_j - s^{(N_R)}_i)}}{\ch{(s^{(N_L)}_j  + iw_jz -s^{(N_R)}_i)}}. \label{Z1d-minus}
\end{align}
In the chamber where all FI parameters are positive, the index is given as:
\begin{align}
\CI^{\Sigma}(\vec \zeta >0)= \sum_{w' \in R'[U(N_R)]}\, \wt{\CF}(\vec s^{(N_R)}, z)\, \prod^{N_L}_{j=1} \prod^{N_R}_{i=1} \frac{\ch{(s^{(N_L)}_j - s^{(N_R)}_i)}}{\ch{(s^{(N_L)}_j  + iw'_iz -s^{(N_R)}_i)}}. \label{Z1d-plus}
\end{align}

Note that the two expressions are very different for a generic SQM, and therefore the choice of the FI chamber is an important 
piece of data in specifying the vortex defect. For a vortex defect associated with a gauge group $U(N)$, one may pick either 
$N_L=N$ or $N_R=N$ and then gauge the corresponding flavor with dynamical 3d vector multiplets. The first choice will 
require that one takes $\vec \zeta <0$ for the coupled SQM, while the second choice will require that $\vec \zeta > 0$. 
For a linear quiver, these choices correspond to the right and the left coupled quivers respectively. \\

Now, let us write down the Witten index of the SQM in \figref{3d1d-edges} in the negative chamber, which allows us to 
compute the partition function of the 3d-1d quiver. We start from the SQM above, with $N_L \to N$, 
and $N_R=p'M +Q$. We then identify the flavor chemical potentials (or masses) associated to the anti-fundamental 
chiral multiplets as follows:
\begin{align}
& s^{(N_R)}_{i + nM} = s^{(M)}_i + \mu^{n+1}, \quad n=0,1,\ldots,p'-1, \quad i=1,\ldots,M,\\
& s^{(N_R)}_{a+ p'M}= m_a, \quad a=1,\ldots, Q.
\end{align}
The above identification leads to the SQM $\Sigma^{Q,R,p'}_{(I)}$ of the following form:
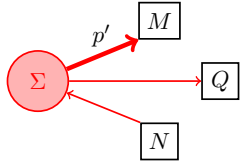
\begin{center}
\scalebox{0.8}{\begin{tikzpicture}[
nnode/.style={circle,draw,thick, fill=blue,minimum size= 6mm},cnode/.style={circle,draw,thick,minimum size=4mm},snode/.style={rectangle,draw,thick,minimum size=6mm},rnode/.style={red, circle,draw,thick,fill=red!30 ,minimum size=4mm},rrnode/.style={red, circle,draw,thick,fill=red!30 ,minimum size=1.0cm}]
\node[rrnode] (14) at (10,0) {$\Sigma$};
\node[snode] (15) at (12,1) {$M$};
\node[snode] (16) at (12,-1) {$N$};
\node[snode] (18) at (13, 0) {$Q$};
\draw[line width=0.75mm, red, ->] (14) to (15.south west);
\draw[red, thick, ->] (16.north west) to (14);
\node[text width=0.2cm] (17) at (11,0.75){$p'$};
\draw[red, thick, ->] (14) to (18);
\end{tikzpicture}}
\end{center}
and its Witten index in the negative chamber can be written as:
\begin{align}
\CI^{\Sigma^{Q,R,p'}_{(I)\,-}} = \sum_{w \in R[U(N)]}\,\prod^N_{j=1} \prod^{p'}_{l=1} \prod^M_{i=1} \frac{\ch{(s^{(N)}_j - s^{(M)}_i -\mu^l)}}{\ch{(s^{(N)}_j  + iw_jz -s^{(M)}_i -\mu^l)}}\times \prod^N_{j=1} \prod^Q_{a=1}\frac{\ch{(s^{(N)}_j - m_a)}}{\ch{(s^{(N)}_j  + iw_jz -m_a)}}.
\end{align}

\section{Construction of 3d-1d coupled quivers from $S$-type operations}\label{sec:SOp-app}

In this section, we briefly review the construction of coupled 3d-1d systems realizing vortex defects using 
$S$-type operations. We refer the reader to Section 3 of the paper \cite{Dey:2021jbf} for a more detailed 
discussion.

\subsection{$S$-type operations on a quiver with defects} \label{Rev-SOps}

\begin{figure}[htbp]
\begin{center}
\begin{tabular}{ccc}
\scalebox{0.7}{\begin{tikzpicture}[
cnode/.style={circle,draw,thick,minimum size=4mm},snode/.style={rectangle,draw,thick,minimum size=8mm},pnode/.style={rectangle,red,draw,thick,minimum size=1.0cm}, bnode/.style={circle,draw, thick, fill=black!30,minimum size=4cm}]
\node[bnode] (1) at (0,0){$X$} ;
\node[snode] (2) [right=1.5cm  of 1]{$M_\alpha$} ;
\draw[-] (1)--(2);
\end{tikzpicture}}
&\scalebox{0.7}{\begin{tikzpicture}
\draw[->] (5,0) -- (7,0); 
\node[text width=1cm](6) at (6,0.5){$G^\alpha_\CP$};
\node[] at (6, -2){};
\end{tikzpicture}}
& \scalebox{0.7}{\begin{tikzpicture}[
cnode/.style={circle,draw,thick,minimum size=4mm},snode/.style={rectangle,draw,thick,minimum size=8mm},pnode/.style={rectangle,red,draw,thick,minimum size=1.0cm}, bnode/.style={circle,draw, thick, fill=black!30,minimum size=4cm}]\node[bnode] (3) at (10,0){$X$} ;
\node[snode] (4) at (14, 1){$M_\alpha- r_\alpha$} ;
\node[cnode] (5) at (14,-1){$r_\alpha$} ;
\draw[-] (3)--(4);
\draw[-] (3)--(5);
\end{tikzpicture}} \\
\qquad & \qquad & \qquad \\
\scalebox{0.7}{\begin{tikzpicture}[
cnode/.style={circle,draw,thick,minimum size=4mm},snode/.style={rectangle,draw,thick,minimum size=8mm},pnode/.style={rectangle,red,draw,thick,minimum size=1.0cm}, bnode/.style={circle,draw, thick, fill=black!30,minimum size=4cm}]
\node[bnode] (1) at (0,0){$X$} ;
\node[snode] (2) [right=1.5cm  of 1]{$M_\alpha$} ;
\draw[-] (1)--(2);
\end{tikzpicture}}
& \scalebox{0.7}{\begin{tikzpicture}
\draw[->] (5,0) -- (7,0); 
\node[text width=1cm](10) at (6,0.5){$F^\alpha_\CP$};
\node[] at (6, -2){};
\end{tikzpicture}}
&\scalebox{0.7}{\begin{tikzpicture}[
cnode/.style={circle,draw,thick,minimum size=4mm},snode/.style={rectangle,draw,thick,minimum size=8mm},pnode/.style={rectangle,red,draw,thick,minimum size=1.0cm}, bnode/.style={circle,draw, thick, fill=black!30,minimum size=4cm}]
\node[bnode] (3) at (10,0){$X$} ;
\node[snode] (4) at (14, 1){$M_\alpha- r_\alpha$} ;
\node[snode] (5) at (14,-1){$r_\alpha$} ;
\node[snode] (6) at (16,-1){$G^\alpha_{\rm F}$} ;
\draw[-] (3)--(4);
\draw[-] (3)--(5);
\draw[-] (5)--(6);
\node[text width=1cm](11) at (15.5, - 0.5){$\CR_\alpha$};
\end{tikzpicture}}\\
\qquad & \qquad & \qquad \\
\scalebox{0.7}{\begin{tikzpicture}[
cnode/.style={circle,draw,thick,minimum size=4mm},snode/.style={rectangle,draw,thick,minimum size=8mm},pnode/.style={rectangle,red,draw,thick,minimum size=1.0cm}, bnode/.style={circle,draw, thick, fill=black!30,minimum size=4cm}]
\node[bnode] (1) at (0,0){$X$} ;
\node[snode] (2) at (4,-2) {$M_\alpha$} ;
\node[snode] (3) at (4, 2) {$M_{\alpha -1}$} ;
\draw[-] (1)--(2);
\draw[-] (1)--(3);
\end{tikzpicture}}
& \scalebox{0.7}{\begin{tikzpicture}
\draw[->] (5,0) -- (7,0);
\node[text width=1cm](6) at (6,0.5){$I^\alpha_\CP$};
\node[] at (6, -2){};
\end{tikzpicture}}
& \scalebox{0.7}{\begin{tikzpicture}[
cnode/.style={circle,draw,thick,minimum size=4mm},snode/.style={rectangle,draw,thick,minimum size=8mm},pnode/.style={rectangle,red,draw,thick,minimum size=1.0cm}, bnode/.style={circle,draw, thick, fill=black!30,minimum size=4cm}]
\node[bnode] (3) at (10,0){$X$} ;
\node[snode] (4) at (12, 3){$M_{\alpha-1}- r_\alpha$} ;
\node[snode] (5) at (14,0){$r_\alpha$} ;
\node[snode] (6) at (12,-3){$M_\alpha- r_\alpha$} ;
\draw[-] (3)--(4);
\draw[-] (3.north east) to  (5);
\draw[-] (3.south east) to (5);
\draw[-] (3)--(6);
\end{tikzpicture}}
\end{tabular}
\end{center}
\caption{\footnotesize{The gauging, flavoring and identification quiver operations on a generic quiver $X$ of class $\CU$. 
The identification operation may involve two or more $U(r_\alpha)$ flavor nodes. The operation $D^\alpha_\CP$ involves 
turning on a vortex or a Wilson-type defect for the flavor node $U(r_\alpha)$ in the theory $(X, \CP)$.}}
\label{fig: QuivOps}
\end{figure}
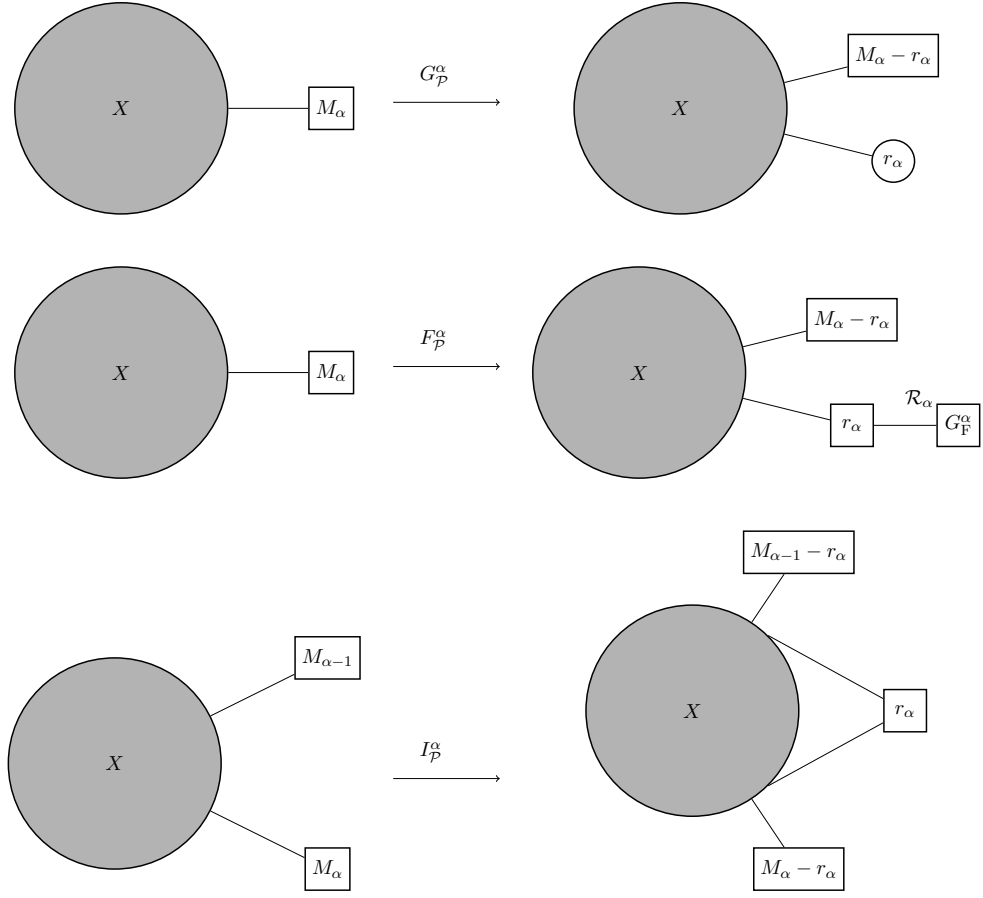

Consider a class of 3d quiver gauge theories for which the Higgs branch 
global symmetry has a subgroup $G^{\rm sub}_{\rm global}=\prod_\gamma U(M_\gamma) \subset G_H$.  
In addition, we will demand that these are good theories in the Gaiotto-Witten sense \cite{Gaiotto:2008ak}.
We will refer to them as class $\CU$. 
Let $X[\wh{\vec A}, \CD]$ be a generic theory in this class decorated by a line defect $\CD$, 
where $\wh{\vec A}$ collectively denotes the background vector multiplets. 
Given such a quiver $X[\wh{\vec A},\CD]$, one can define a 
set of four basic quiver operations, as shown in \figref{fig: QuivOps}:
\begin{enumerate}
\item {\bf{Gauging ($G^\alpha_\CP$)}}
\item {\bf{Flavoring ($F^\alpha_\CP$)}}
\item \textbf{Identification($I^\alpha_{\vec \CP}$)}
\item {\bf{Defect ($D^\alpha_\CP$)}}
\end{enumerate}

The superscript $\alpha$ specifies the flavor node at which the $S$-type operation is being implemented. 
Each operation involves two steps. First, one splits the flavor node 
$U(M_\alpha)$ into two, corresponding to $U(r_\alpha) \times U(M_\alpha - r_\alpha)$ flavor nodes.
The $U(1)^{M_\alpha}$ masses are related to the $U(1)^{r_\alpha} \times U(1)^{M_\alpha-r_\alpha}$ 
masses by the following map:
\be \label{uvdef0}
\overrightarrow{m^\alpha}_{i_\alpha} = \CP_{i_\alpha i} \, \overrightarrow{u}^\alpha_i + \CP_{i_\alpha \, r_\alpha + j} \, \overrightarrow{v}^\alpha_j, \quad( i_\alpha=1,\ldots, M_\alpha,\quad i=1,\ldots, r_\alpha, \quad j=1,\ldots, M_\alpha - r_\alpha),
\ee
where $\CP$ is a permutation matrix of order $M_\alpha$. The theory deformed by the $U(r_\alpha) \times U(M_\alpha - r_\alpha)$ 
mass parameters is denoted as $(X[\wh{\vec A}, \CD], \CP)$. 
In the second step, one gauges, adds flavor to, or identifies the $U(r_\alpha)$ flavor node with other flavor nodes. 
The operation $D^\alpha_\CP$ involves turning on a vortex or a Wilson-type defect for the flavor node 
$U(r_\alpha)$ in the theory $(X[\wh{\vec A}, \CD], \CP)$.

\begin{definition}
\textbf{An elementary $S$-type operation} $\CO^\alpha_{\vec \CP}$ on $X$ at a flavor node $\alpha$, is defined as any possible combination of the 
identification ($I^\alpha_{\vec \CP}$), the  flavoring ($F^\alpha_{\CP}$) and the defect ($D^\alpha_\CP$) operations followed by a single 
gauging operation $G^\alpha_{\CP}$, i.e.
\be \label{Sbasic-def}
\boxed{\CO^\alpha_{\vec \CP}(X) := (G^\alpha_{\vec \CP})\circ  (D^\alpha_{\vec \CP})^{n_3} \circ (F^\alpha_{\vec \CP})^{n_2} \circ (I^\alpha_{\vec \CP})^{n_1}(X), 
\quad (n_i=0,1, \,\, \forall i).}
\ee
\end{definition}

The operation $\CO^\alpha_{\vec \CP}$ generically maps a quiver gauge theory with defect $X[\wh{\vec A}, \CD]$ to a 
new quiver gauge theory with defect $X'[\wh{\vec B},\CD']$:
\be 
\CO^\alpha_{\vec \CP} : X[\wh{\vec A}, \CD] \mapsto X'[\wh{\vec B}, \CD'],
\ee
where the final defect $\CD'$ is built out of the original defect $\CD$ and the defect operation $D^\alpha_{\vec \CP}$.

One can classify the elementary $S$-type operations into four distinct types, 
depending on the constituent quiver operations:
\begin{itemize}
\item Gauging.

\item Flavoring-gauging.

\item Identification-gauging.

\item Identification-flavoring-gauging.

\end{itemize}

Each one can be further combined with a defect operation. An elementary Abelian $S$-type operation is one which involves 
gauging a $U(1)$ global symmetry. Finally, any number of elementary $S$-type operations can be combined to form a 
generic $S$-type operation.

\subsection{Constructing vortex defects and the dual defects}\label{SOps-defects}

The $S$-type operation $\CO^\alpha_{\vec \CP}$ can be realized in terms of supersymmetric observables.
In this work, we will focus on the three sphere partition function. We will explicitly discuss the case of a vortex defect 
which can be realized as a coupled 3d-1d quiver, noting that Wilson defects can be treated in an analogous fashion. 
The partition function for the coupled 3d-1d quiver associated with $X[\wh{\vec A}, \CD]$ is given as 
\begin{align}
Z^{(X[\wh{\vec A}, \CD], \{P_\beta \})}(\{ \vec{u}^\beta \}, \ldots; \vec \eta)=& \lim_{z\to 1} \int \,\Big[d\vec s\Big] \, Z^{(X[\wh{\vec A}, \CD], \{P_\beta \})}_{\rm int}(\vec s, \{ \vec{u}^\beta \}, \ldots, \vec \eta,z) \label{Z-XV-1}\\
=: & \lim_{z\to 1} Z^{(X[\wh{\vec A}, \CD], \{P_\beta \})}(\{ \vec{u}^\beta \}, \ldots; \vec \eta | z), \label{Z-XV-2}
\end{align}
where the function $Z^{(X[\wh{\vec A}, \CD], \{P_\beta \})}_{\rm int}$ has the schematic form:
\be \label{Z-int-XV}
Z^{(X[\wh{\vec A}, \CD], \{P_\beta \})}_{\rm int} = W_{\rm b.g.}(\vec \eta) \cdot Z^{(X[\wh{\vec A}], \{P_\beta \})}_{\rm int}(\vec s, \{ \vec{u}^\beta \}, \ldots, \vec \eta) \cdot \CI^{\Sigma}(\vec s, \{ \vec{u}^\beta \}, \ldots, z| \vec \xi).
\ee
The $S$-type operation $\CO^\alpha_{\vec \CP}$ can then be implemented in terms of the sphere partition function as follows:
\begin{empheq}[box=\widefbox]{align}\label{S-Op-Agen}
&Z^{\CO^\alpha_{\vec \CP}(X[\wh{\vec A},\CD])} (\vec{m}^{\CO^\alpha_{\vec \CP}}, \ldots; \vec \eta, \eta_\alpha)
=: Z^{(X'[\wh{\vec B},\CD'])} (\vec{m}^{\CO^\alpha_{\vec \CP}}, \ldots; \vec \eta, \eta_\alpha) \nn \\
=& \lim_{z\to 1} \int \Big[d\vec{u}^\alpha\Big] \, \CZ_{\CO^\alpha_{\vec \CP}(X)}(\vec u^{\alpha}, \{\vec{u}^\beta\}_{\beta \neq \alpha}, \eta_\alpha, \vec{m}^{\CO^\alpha_{\vec \CP}},z| \Sigma') \, \cdot \, Z^{(X[\wh{\vec A},\CD], \{P_\beta \})}( \{ \vec{u}^\beta \}, \ldots, \vec \eta | z),
\end{empheq}
where $\eta_\alpha$ is an FI parameter associated with gauging, and $\vec{m}^{\CO^\alpha_{\vec \CP}}$ are hypermultiplet masses associated 
with flavoring and/or identification operations. The operator $\CZ_{\CO^\alpha_{\vec \CP}(X)}$ can be constituted from the partition function contributions of gauging ($G^\alpha_\CP$) , flavoring ($F^\alpha_\CP$), identification ($I^\alpha_{\vec \CP}$), and defect ($D^\alpha_\CP$) operations as follows:
\begin{align} 
&\CZ_{\CO^\alpha_{\vec \CP}(X)}(\vec u^{\alpha}, \{\vec{u}^\beta\}_{\beta \neq \alpha}, \eta_\alpha, \vec{m}^{\CO^\alpha_{\vec \CP}}, z | \Sigma') 
=\CZ_{G^\alpha_\CP(X)} \cdot \Big(\CZ_{D^\alpha_\CP(X)}\Big) \cdot \Big(\CZ_{F^\alpha_\CP(X)}\Big)^{n_2} \cdot \Big(\CZ_{I^\alpha_\CP(X)}\Big)^{n_1}, 
 \label{CZ-OP-wDA}
 \end{align}
where $n_1,n_2=0,1$, and the constituent operators are given as
 \begin{align}
& \CZ_{D^\alpha_\CP(X)}(\vec{u}^\alpha, \vec{m}^\alpha_F, \eta_\alpha, z) = W_{\rm b.g.}(\eta_\alpha, \vec \eta) \cdot \CI^{\Sigma'} (\vec{u}^\alpha, \vec{m}^\alpha_F, z| \vec \xi'), \label{CZ-defectA} \\
& \CZ_{G^\alpha_\CP(X)}(\vec{u}^\alpha,\eta_\alpha) = Z_{\rm FI} (\vec{u}^\alpha,\eta_\alpha) \,Z_{\rm 1-loop} ^{\rm vector} (\vec{u}^\alpha)=e^{2\pi i \eta_\alpha \sum_i u^\alpha_i} \, \prod_{i < j} \sinh^2{\pi (u^\alpha_i - u^\alpha_j)}, \label{CZ-gauging}\\
& \CZ_{F^\alpha_\CP(X)}(\vec{u}^\alpha, \vec{m}^\alpha_F)= Z_{\rm 1-loop} ^{\rm hyper} (\vec{u}^\alpha, \vec{m}^\alpha_F)=\prod_{\rho(\CR_\alpha)}\frac{1}{ 
\ch{\rho(\vec{u}^\alpha, \vec{m}^\alpha_F)}}, \label{CZ-flavoring}\\
&\CZ_{I^\alpha_\CP(X)}(\vec u^{\alpha}, \{\vec{u}^\beta\}, \vec \mu) = \int \prod^{p}_{j=1} \prod^{r_\alpha}_{i=1} d {u^{\gamma_j}}_i \, \prod^{p}_{j=1} \delta^{(r_\alpha)}\Big(\vec{u}^{\alpha} - \vec{u}^{\gamma_{j}} + {\mu}^{\gamma_j} \Big). \label{CZ-identification}
\end{align}
In \eref{CZ-defectA}, $\Sigma'$ denotes the SQM being coupled to the 3d theory by the defect operation $D^\alpha_\CP$, with 
$\CI^{\Sigma'}$ being the corresponding Witten index. 
The parameters $\vec{m}^\alpha_F$ in \eref{CZ-flavoring} denote the masses for the hypers added in the flavoring operation.
In \eref{CZ-identification}, the identified nodes (we assume that there are $p$ of them) are labelled as 
$\{ \gamma_j \}_{j=1,\ldots, p}$, while $\{ \mu^{\gamma_j} \}_{j=1,\ldots, p}$ denote the masses introduced by the identification 
operation.\\

The dual of the 3d-1d coupled system $X'[\wh{\vec B},\CD']$ is then given as follows. 
Let the theory dual to $X[\wh{\vec A},\CD]$ be denoted by $Y[\wh{\vec A},{\CD}^\vee]$, which is the quiver gauge theory $Y$ decorated by a 
Wilson line defect ${\CD}^\vee$. The dual Wilson defect ${\CD}^\vee$ is of the generic form:
\be
{\CD}^\vee = \sum_\kappa c_\kappa \, \wt{W}^{\rm flavor}_\kappa \cdot \wt{W}_{\wt{R}_\kappa(\wt{G})}, 
\ee 
where $\wt{W}_{\wt{R}_\kappa(\wt{G})}$ is a Wilson defect in a representation $\wt{R}_\kappa$ of the gauge 
group $\wt{G}$ of the theory $Y$, and $\wt{W}^{\rm flavor}_\kappa$ is a background Wilson defect associated 
with the hypermultiplets.\\

Given an operation $\CO^\alpha_{\vec \CP}$ on $X[\wh{\vec A},\CD]$, one can define a dual operation on 
$Y[\wh{\vec A},{\CD}^\vee]$:
\be
\wt{\CO}^\alpha_{\vec \CP} :  Y[\wh{\vec A},{\CD}^\vee] \mapsto Y'[\wh{\vec B},\CD'^\vee],
\ee
such that the theories $(X'[\wh{\vec B},\CD'], Y'[\wh{\vec B},D'^\vee])$ are IR dual. 
We can write down the partition function of $Y'[\wh{\vec B},D'^\vee]$ in terms of the theory $Y[\wh{\vec A},{\CD}^\vee]$.
Mirror symmetry implies the following identity involving the defect partition functions:
\begin{align}\label{MS-XYD}
Z^{(X[\wh{\vec A}, \CD],\{ P_\beta\})} (\{\vec{u}^\beta\},\ldots; \vec \eta)
 = C_{XY}(\{\vec{u}^\beta\},\ldots, \vec \eta)\, Z^{(Y[\wh{\vec A}, {\CD}^\vee],\{ P_\beta\})} (\vec{m}^Y(\vec{\eta}) ; \vec{\eta}^Y(\{\vec{u}^\beta\}, \ldots)),
\end{align}
where $\CD$ is a vortex defect and $\CD^\vee$ is a Wilson defect. 
However, as discussed in \cite{Dey:2020hfe}, there exists a more refined $z$-dependent identity involving the two partition functions i.e.
\begin{align}\label{MS-XYD-1}
Z^{(X[\wh{\vec A},\CD],\{ P_\beta\})} (\{\vec{u}^\beta\},\ldots; \vec \eta |z) 
& = C_{XY}(\{\vec{u}^\beta\},\ldots, \vec \eta)\,Z^{(Y[\wh{\vec A},{\CD}^\vee],\{ P_\beta\})} (\vec{m}^Y(\vec{\eta}) ; \vec{\eta}^Y(\{\vec{u}^\beta\}, \ldots)|z) \nn \\
&= C_{XY}\cdot \,\int \prod_{\gamma'}  \Big[d\vec{\s}^{\gamma'} \Big]\,Z^{(Y[\wh{\vec A},{\CD}^\vee],\{ P_\beta\})}_{\rm int} (\{\vec{\s}^{\gamma'} \}, \vec{m}^Y(\vec{\eta}), \vec{\eta}^Y(\{\vec{u}^\beta\}, \ldots), z),
\end{align}
where the function $Z^{(X[\wh{\vec A},\CD],\{ P_\beta\})} (\ldots |z) $ is defined in \eref{Z-XV-2} in terms of the vortex defect 
partition function. 
The function $Z^{(Y[\wh{\vec A},{\CD}^\vee],\{ P_\beta\})} (\ldots |z)$ has the property that it reduces to the Wilson defect partition function 
in the limit $z \to 1$, i.e.
\be
\lim_{z\to 1}\,Z^{(Y[\wh{\vec A},{\CD}^\vee],\{ P_\beta\})} (\vec{m}^Y(\vec{\eta}) ; \vec{\eta}^Y(\{\vec{u}^\beta\}, \ldots)|z)
= Z^{(Y[\wh{\vec A},{\CD}^\vee],\{ P_\beta\})} (\vec{m}^Y(\vec{\eta}) ; \vec{\eta}^Y(\{\vec{u}^\beta\}, \ldots)).
\ee

Using the identity \eref{MS-XYD-1}, the dual defect partition function is given by 
the following formula (up to certain contact terms): 
\begin{empheq}[box=\widefbox]{align}\label{PF-wtOPgenD-A2B}
Z^{\wt{\CO}^\alpha_{\vec \CP}(Y[\wh{\vec A}, {\CD}^\vee])}
=& \sum_\kappa c_\kappa \lim_{z\to 1} \int \prod_{\gamma'}  \Big[d\vec{\s}^{\gamma'} \Big]\,  \CZ^\kappa_{\wt{\CO}^\alpha_{\vec \CP}(Y)}(\{\s^{\gamma'}\},\vec{m}^{{\CO}^\alpha_{\vec\CP}}, \eta_{\alpha},\vec \eta,z) \cdot C_{XY}(\{\vec{u}^\beta=0 \},\ldots)  \nn \\
\times & Z_{\wt{W}^{\rm flavor}_\kappa}(\vec{m}^Y(\vec{\eta})|z) \cdot Z^{(Y[\wt{W}_{\wt{R}_\kappa(\wt{G})}],\{\CP_\beta\})}_{\rm int}(\{\vec \s^{\gamma'} \}, \vec{m}^Y(\vec{\eta}), \vec{\eta}^Y(\{\vec{u}^\beta =0 \},\ldots), z),
\end{empheq}
where the function $Z^{(Y[\wt{W}_{\wt{R}_\kappa(\wt{G})}],\{\CP_\beta\})}_{\rm int}$ and the function $Z_{\wt{W}^{\rm flavor}_\kappa}$ 
can be read off from the second line of \eref{MS-XYD-1}. The functions $\CZ^\kappa_{\wt{\CO}^\alpha_{\vec \CP}(Y)}$ are given by a 
formal Fourier transformation of the operator $\CZ_{\CO^\alpha_{\vec \CP}(X)}$ defined in \eref{CZ-OP-wDA} :
\begin{align} \label{CZ-wtOPDD-A2B}
\CZ^\kappa_{\wt{\CO}^\alpha_{\vec \CP}(Y)}
= \int \Big[d\vec{u}^\alpha\Big] \, \CZ_{\CO^\alpha_{\vec \CP}(X)}(\vec u^{\alpha}, \{\vec{u}^\beta\}_{\beta \neq \alpha}, \eta_\alpha, \vec{m}^{\CO^\alpha_{\vec \CP}}, z| \Sigma')\, \cdot e^{2\pi i \sum_{i,\beta}(g^i_\beta (\{\vec \s^{\gamma'} \}, \CP_\beta) +\sum_l b^{il}_\beta \eta_l)\,u^{\beta}_i}.
\end{align}
The functions $g^i_\beta(\{\vec \s^{\gamma'} \}, \CP_\beta)$ and the parameters $\{b^{il}_\beta\}$ that enter on the RHS of \eref{CZ-wtOPDD-A2B} 
are defined as follows:
\begin{align}
& C_{XY}(\{\vec u^\beta\},\ldots, \vec \eta) =e^{2\pi i \sum_{i,l,\beta}\,b^{il}_\beta u^\beta_i \eta_l} \cdot  C_{XY}(\{\vec u^\beta=0\},\ldots, \vec \eta), \\
& Z^{(Y,\{\CP_\beta\})}_{\rm int}
= e^{2\pi i \,\sum_{i,\beta}g^i_\beta (\{\vec \s^{\gamma'} \}, \CP_\beta)\,u^{\beta}_i}\,Z^{(Y,\{\CP_\beta\})}_{\rm int}(\{\vec \s^{\gamma'} \}, \vec{m}^Y(\vec{\eta}), \vec{\eta}^Y(\{\vec{u}^\beta =0 \},\ldots)).
\end{align}

To summarize: the expression of the dual partition function in \eref{PF-wtOPgenD-A2B}-\eref{CZ-wtOPDD-A2B} will 
serve as the working definition for the dual $S$-type operation acting on the quiver gauge theory $Y$, decorated by a Wilson 
defect $\CD^\vee$. If the dual theory $Y'$ is Lagrangian, the RHS of \eref{PF-wtOPgenD-A2B} can be rewritten in the standard 
form that makes the gauge group and matter content of the theory manifest. The Wilson defect can then be read off from the 
defect partition function.

\section{$S$-type operation: Partition function analysis for the Abelian quiver}\label{sec:PF-Ab-app}

The partition functions of the theories $(X,Y)$ are given as
\begin{align}
 Z^{(X)}(\vec m, \vec t) =& \int\, ds \, \frac{e^{2\pi i \,(t_1- t_2)\,s}}{\prod^{n}_{i=1}\, \ch{(s-m_i)}}=: \int\, ds \, Z^{(X)}_{\rm int} (s, \vec m, \vec t), \\
Z^{(Y)}(\vec t, \vec m) = &\int\, \prod^{n-1}_{k=1}d\s_k \, \frac{\prod^{n-1}_{k=1}\, e^{2\pi i \,(m_k- m_{k+1})\,\s_k}}{\ch{(\s_1 -t_1)}\,\prod^{n-2}_{k=1}\ch{(\s_k -\s_{k+1})} \,\ch{(\s_{n-1} -t_2)}}\\
=: & \int\, \prod^{n-1}_{k=1}d\s_k \, Z^{(Y)}_{\rm int} (\{\s_k \}, \vec m, \vec t),
\end{align}
where $Z^{(X)}_{\rm int}$ and $Z^{(Y)}_{\rm int}$ are the respective matrix model integrands for $X$ and $Y$, and the parameters $\vec m, \vec t$ are 
unconstrained.
Mirror symmetry of $X$ and $Y$ implies that
\begin{align}
& Z^{(X)}(\vec m, \vec t)  =C_{XY}(\vec m, \vec t)\, Z^{(Y)}(\vec t, -\vec m), \\
& C_{XY}(\vec m, \vec t)=  e^{2\pi i (m_1\,t_1- m_n\,t_2)},
\end{align}
where $C_{XY}$ is a contact term. The partition function of the theory $X'= \CO^\alpha_{\CP}(X)$, where the operation 
$\CO^\alpha_{\CP}$ is specified by \eref{SOp-1a}-\eref{SOp-1b}, is given as
\begin{align}
Z^{\CO^\alpha_{\CP}(X)}= \int\, d{u}^{\alpha}\, \CZ_{\CO^\alpha_{\CP}(X)}(u^\alpha, \vec u, \eta_\alpha, \vec m^f, \vec \mu) \cdot Z^{(X,\CP)}(\vec u, \ldots; \vec t).
\end{align}
The operator $\CZ_{\CO^\alpha_{\CP}(X)}$, constructed from the general prescription of \Secref{Rev-SOps}, is given as:
\begin{align}
\CZ_{\CO^\alpha_{\CP}(X)}= & Z_{\rm FI} ({u}^{\alpha},\eta_{\alpha})\, Z_{\rm 1-loop}^{\rm hyper} ({u}^{\alpha}, \vec m^{\alpha}_F)\, \int \prod^{p}_{j=1} \, d {u_{j}} \,\prod^{p}_{j=1} \delta \Big({u}^{\alpha} - {u}_{j} + {\mu}_{j} \Big) \\
= & \frac{e^{2\pi i \eta_\alpha u^\alpha}}{\prod^{l}_{a=1}\, \ch{(u^\alpha - m^f_a)}}\, \int \prod^{p}_{j=1} \, d {u_{j}} \,\prod^{p}_{j=1} \delta \Big({u}^{\alpha} - {u}_{j} + {\mu}_{j} \Big).
\end{align}
Therefore, we have,
\begin{align}
Z^{(X')}= \int\, ds\, du^\alpha\, \frac{e^{2\pi i \eta_\alpha u^\alpha}\, e^{2\pi i \,(t_1- t_2)\,s}}{\prod^{n-p}_{i=1} \ch{(s-m^{(1)}_i)}\, \prod^{l}_{a=1} \ch{(u^\alpha-m^{(2)}_a)} \, 
\prod^{p}_{j=1} \ch{(s- u^\alpha -m^{\rm bif}_j)}},
\end{align}
where the fundamental masses $\vec{m}^{(1)}, \vec{m}^{(2)}$ and the bifundamental masses $\vec m^{\rm bif}$ are defined as:
\begin{align}
& m^{(1)}_i = m_{i}, \qquad i=1,\ldots, n-p,\\
& m^{(2)}_a= m^f_a, \qquad a=1,\ldots, l,\\
& m^{\rm bif}_j= \mu_j,  \qquad j=1,\ldots, p.
\end{align}

The dual theory can be worked out from the general construction in \eref{PF-wtOPgenD-A2B}-\eref{CZ-wtOPDD-A2B}, 
with the function $ \CZ_{\wt{\CO}^\alpha_{\vec \CP}(Y)}$ being 
given as
\begin{align}
& \CZ_{\wt{\CO}^\alpha_{\vec \CP}(Y)} = \int \, du^\alpha \, \CZ_{\CO^\alpha_{\vec \CP}(X)} \cdot \prod^{p}_{j=1}\, e^{2\pi i (g^j(\{\s_k\}, \CP) + \sum_l b^{jl}t_l)\,u_j},\\
&\prod^{p}_{j=1} e^{2\pi i g^j(\{\s_k\}, \CP) \,u_j}= \Big(\prod^{p-1}_{j=1}\, e^{2\pi i u_j \, (\s_{n-p+j-1} - \s_{n-p+j})}\Big)\,e^{2\pi i u_p\, \s_{n-1}}, \\
& \prod^{p}_{j=1} e^{2\pi i u_j \sum_l b^{jl}t_l} = e^{-2\pi i u_p\, t_2}.
\end{align}
The dual partition function, therefore, can be written as:
\begin{align}\label{Ab-DualPf}
Z^{\wt{\CO}^\alpha_{\vec \CP}(Y)}(\vec{m}'; \vec \eta')= C_{X'Y'}\cdot \int \, & \prod^{n-1}_{k=1}\,  d\s_k\,\prod^{l-1}_{b=1}\, d\tau_b\,
Z^{\rm bif}_{\rm 1-loop}(\tau_1, \s_{n-p}, 0)\, Z^{(\CT_{l-1})}_{\rm int}(\{ \tau_b\}, -\vec m^f, \eta_\alpha -t_2) \nn \\
\times \, &  Z^{(Y)}_{\rm int}(\{ \s_k\}, \vec t, \{-m_1, \ldots, -m_{n-p}, -\mu'_1,\ldots, - \mu'_p \}), 
\end{align}
where the mass parameters $\mu'_j = \mu_j + m^f_1$ for $j=1,\ldots, p$, and the functions $C_{X'Y'}$ and $Z^{(\CT_{l-1})}_{\rm int}$ are given as
\begin{align}
& C_{X'Y'}(\vec m^f, \vec \mu,\vec t)= e^{2\pi i m^f_l(\eta_\alpha -t_2)}\, e^{2\pi i (m_1 t_1 - \mu_p t_2)},\\
& Z^{(\CT_{l-1})}_{\rm int}(\{ \tau_k\}, \vec m^f,  \eta_\alpha -t_2) = 
\frac{\prod^{l-1}_{b=1} e^{2\pi i \tau_b(m^f_b - m^f_{b+1})}}{\prod^{l-2}_{b=1}\ch{(\tau_b -\tau_{b+1})}\, \ch{(\tau_{l-1} - \eta_\alpha +t_2)}}.
\end{align}
From the RHS of \eref{Ab-DualPf}, one can read off the gauge group and the matter content of the dual theory, and the dual partition
function can then be written as:
\begin{align}
Z^{\wt{\CO}^\alpha_{\vec \CP}(Y)}(\vec{m}'; \vec \eta')=:  C_{X'Y'}(\vec m^f, \vec \mu,\vec t) \cdot Z^{(Y')}\Big(\vec{m}'(\vec t, \eta_\alpha); \vec{\eta}'(\vec m, \vec m^f, \vec \mu) \Big),
\end{align}
where $Y'$ is the quiver gauge theory shown in \figref{SimpAbEx1GFI}, and $ C_{X'Y'}$ is a contact term associated with the new duality.
The mirror map can also be read off from the RHS of \eref{Ab-DualPf}, and can be summarized as follows. Note that the quiver $Y'$ can be 
decomposed into two linear quivers -- chain (1)  of $n-1$ gauge nodes and  chain (2)  of $l-1$ gauge nodes, where the $(n-p)$-th  
gauge node of chain (1) is connected to the first gauge node of chain (2) by a single bifundamental 
hypermultiplet. Let us label the FI parameters of the linear chain (1) of $n-1$ nodes as $\eta'^{(1)}_{k} = t'^{(1)}_k - t'^{(1)}_{k+1}$, 
with $k=1,\ldots, n-1$. Also, let us label the FI parameters 
of the linear chain (2) of $l-1$ nodes as  $\eta'^{(2)}_{b} = t'^{(2)}_b - t'^{(2)}_{b+1}$, with $b=1,\ldots, (l-1)$. The mirror maps associated with the 
parameters $\vec{t'}^{(1)}$ and $\vec{t'}^{(2)}$ are then given as
\begin{subequations}
\begin{empheq}[box=\widefbox]{align}
\vec{t'}^{(1)} = & \{m_1, \ldots, m_{n-p}, \mu'_1,\ldots,  \mu'_p \} \nn\\
=: & \{ m^{(1)}_1, \ldots, m^{(1)}_{n-p}, m^{\rm bif}_{1} + m^{(2)}_1, \ldots, m^{\rm bif}_{p} + m^{(2)}_1\} \\
 \vec{t}'^{(2)} = &\{ m^f_1, \ldots, m^f_l\} =: \{ m^{(2)}_1,\ldots, m^{(2)}_l \}.
\end{empheq}
\end{subequations}

Similarly, the hypermultiplet masses of the theory $Y'$ are given in terms of the FI parameters of $X'$. In writing \eref{Ab-DualPf}, we 
chose to set the masses of the bifundamental hypers to zero, while the masses of the fundamental hypers are non-zero. 
The mirror map relating the fundamental masses of $Y'$ to FI parameters of $X'$ is then given as:
\begin{subequations}
\begin{empheq}[box=\widefbox]{align}
& m'^{(1)}_{1} = t_1, \\
&  m'^{(1)}_{n-1} = t_2,\\
& m'^{(2)}_{l-1} = -\eta_\alpha +t_2 =t_3,
\end{empheq}
\end{subequations}
where $m'^{(i)}_{l}$ denotes the mass of the fundamental hyper associated with the $l$-th gauge node in the $i$-th linear chain,
and $\eta_\alpha=t_2 -t_3$.
With this parametrization of masses and FI parameters, the partition functions of $X'$ and $Y'$ are related as
\begin{align}
& Z^{(X')}(\{ \vec m^{(1)},  \vec m^{(2)}, \vec m^{\rm bif} \}, \{ \eta_\alpha, \vec t \}) = C_{X'Y'} \cdot 
\,Z^{(Y')}(\{m'^{(1)}_{1}, m'^{(1)}_{n-1}, m'^{(2)}_{l-1} \}, \{-\vec{t'}^{(1)}, -\vec{t'}^{(2)} \}), \label{MS-Pf-Ab} \\
& C_{X'Y'} = e^{2\pi i (\eta_\alpha -t_2)\, m^{(2)}_l}\, e^{2\pi i (m^{(1)}_1 t_1 - m^{\rm bif}_p t_2)}.
\end{align}

\section{$S$-type operation: Partition function analysis for the non-Abelian quiver}\label{sec:PF-NAb-app}

The partition functions of the theories $(X,Y)$ are given as
\begin{align}
& Z^{(X)}(\vec{m}; \vec{t})=\int \frac{d^2s}{2!} \, \frac{e^{2\pi i \tr{\vec{s}} ({t}_1- {t}_2)}  \sinh^2{\pi(s_1-s_2)}}{\prod^2_{j=1}\prod^4_{i=1} \cosh{\pi(s_j - {m}_i)}}
=: \int \frac{d^2s}{2!} \, Z^{(X)}_{\rm int}(\vec \s, \vec m, \vec t),\\
& Z^{(Y)}(\vec{t}; \vec{m})=\int d \s^{1}\,\Big[ d\vec{\s}^2 \Big]\,d \s^{3} \, \frac{e^{2\pi i \s^{1} (m_1- m_2)} e^{2\pi i \tr \vec{\s}^2 (m_2 -m_3)} e^{2\pi i \s^{3} (m_3- m_4)} \sinh^2{\pi(\s^2_1-\s^2_2)}}{\prod^2_{i=1}\cosh{\pi(\s^1-\s^2_i)}\prod^2_{a=1} \cosh{\pi(\s^2_i - t_a)}\cosh{\pi(\s^3 -\s^2_i)}} \\
& \qquad \qquad =: \int d \s^{1}\,\Big[ d\vec{\s}^2 \Big]\,d \s^{3} \,Z^{(Y)}_{\rm int}(\{\vec \s^\gamma\}, \vec t, \vec m),
\end{align}
where $Z^{(X)}_{\rm int}$ and $Z^{(Y)}_{\rm int}$ are the respective matrix model integrands for $X$ and $Y$, and the parameters $\vec m, \vec t$ are unconstrained. Mirror symmetry of $X$ and $Y$ implies that
\begin{align}
& Z^{(X)}(\vec m ; \vec t) =C_{XY}(\vec m, \vec t) \, Z^{(Y)}(\vec t; - \vec m), \\
& C_{XY}(\vec m, \vec t)= e^{2\pi i t_1(m_1+ m_2)} e^{-2\pi i t_2(m_3+m_4)}.
\end{align}
where $C_{XY}$ is a contact term.\\

The partition function of the theory $X'= \CO_{\vec \CP}(X)$, where the operation $\CO_{\vec \CP}$ is specified in \eref{SOp-2a}-\eref{SOp-2b},
is then given as
\begin{align}
Z^{\CO_{\vec \CP}(X)}= \int\, \prod^3_{i=1}\, d{u}_i\, \CZ_{\CO_{\vec \CP}(X)}(\vec u, \vec \eta, \vec m^f, \vec \mu) \cdot Z^{(X, \vec\CP)}(\vec u, v; \vec t). 
\end{align}
The operator $\CZ_{\CO_{\vec \CP}(X)}$, constructed from the general prescription of \Secref{Rev-SOps}, has the form:
\begin{align}
 \CZ_{\CO_{\vec \CP}(X)}= & \CZ_{\CO^3_{\CP_3} (\CO^2_{\CP_2} \circ \CO^1_{\CP_1}(X))} \cdot \CZ_{\CO^2_{\CP_2} (\CO^1_{\CP_1}(X))} \cdot \CZ_{\CO^1_{\CP_1}(X)} \nn \\
 = & e^{2\pi i \eta_3 u_3}\cdot \Big( \int\, \prod^2_{b=1}\, dm^f_b\, \frac{e^{2\pi i \eta_2 u_2}\,\prod^2_{b=1}\delta(m^f_b -u_2- \mu_b)}{\prod^2_{c=1}\,\ch{(u_2-x^f_c)}}\Big)\cdot\frac{e^{2\pi i \eta_1 u_1}}{\prod^4_{a=1}\, \ch{(u_1-m^f_a)}}.
\end{align}
The partition function of the theory $X'$ can be then written as
\begin{align}
Z^{X'} = \int\,\prod^3_{i=1}\, d{u}_i\,&\frac{d^2s}{2!} \cdot\frac{e^{2\pi i \sum_i \,\eta_i u_i}}{\prod^2_{a=1}\ch{(u_1-m^{(1)}_a)}\, 
\prod^2_{b=1}\ch{(u_1-u_2 -m^{\rm bif}_b)}\, \prod^2_{c=1}\ch{(u_1-m^{(2)}_c)}} \nn \\
& \times \frac{e^{2\pi i \tr{\vec{s}} ({t}_1- {t}_2)}  \sinh^2{\pi(s_1-s_2)}}{\prod^2_{j=1}\prod^3_{i=1} \cosh{\pi(s_j - u_i)}\cdot \ch{(s_j- m^{(0)})}},
\end{align}
where the fundamental masses $m^{(0)}, \vec{m}^{(1)}, \vec{m}^{(2)}$ and the bifundamental masses $\vec m^{\rm bif}$ are given in terms of mass parameters introduced above in the following fashion:
\begin{align}
& m^{(0)}=v, \\
& m^{(1)}_a = m^{f}_{a+2}, \qquad a=1,2,\\
& m^{(2)}_c= x^f_c, \qquad c=1,2,\\
& m^{\rm bif}_b= \mu_b,  \qquad b=1,2.
\end{align}

The dual theory can be read off from the general construction in \eref{PF-wtOPgenD-A2B}-\eref{CZ-wtOPDD-A2B}.
The function $ \CZ_{\wt{\CO}_{\vec \CP}(Y)}$ is given as
\begin{align}
& \CZ_{\wt{\CO}_{\vec \CP}(Y)} = \int \, \prod^3_{i=1}\, d{u}_i \, \CZ_{\CO_{\vec \CP}(X)} \cdot \prod^{3}_{j=1}\, e^{2\pi i (g^j(\{\vec \s^k\}, \vec \CP) + b^{jl}t_l)\,u_j},\\
& \prod^{3}_{j=1}\, e^{2\pi i (g^j(\{\vec \s^k\}, \vec \CP) + b^{jl}t_l)\,u_j} = e^{2\pi i u_1 (-\s^3 + \tr \s^2 -t_2)}\,e^{2\pi i u_2 (\s^3 - t_2)}\, e^{2\pi i u_3 (-\s^1 + t_1)}.
\end{align}
The dual partition function can then be computed in the standard fashion. After some change of variables, the dual partition function can 
be written as
\begin{align}\label{NAb1-DualPf}
&Z^{\wt{\CO}_{\vec \CP}(Y)}(\vec{m}'; \vec \eta')=  C_{X'Y'}\cdot \int \frac{d^2\s}{2!}\,\prod^5_{k=1} d\tau_k\,
\Big( \frac{Z_{\rm FI} (\vec\s, \vec \tau, -\vec \eta')\,\sinh^2{\pi(\s_1-\s_2)}}{\prod^2_{i=1}\, \cosh{\pi(\s_i - t_1 -\eta_3)} \prod^2_{a=1} \cosh{\pi(\s_i - t_a)}\,\ch{(\s_i - \tau_1)}} \Big)\nn \\
& \times \Big( \frac{1}{\ch{(\tr \s - \tau_5 +\eta_2 -t_2)}  \, \prod^3_{k=1}\ch{(\tau_k -\tau_{k+1})}\, \ch{(\tau_4 +\eta_1-t_2)}\,
\ch{(\tau_5-\tau_3)}}\Big),\\
& Z_{\rm FI} (\vec\s, \vec \tau, -\vec \eta') = e^{-2\pi i (m^{(0)} - m^{(2)}_1) \tr \s}\, e^{2\pi i m^{\rm bif}_1\,\tau_1}\, e^{-2\pi i (m^{\rm bif}_1- m^{\rm bif}_2) \tau_2}\,
e^{-2\pi i (m^{\rm bif}_2 - m^{(1)}_1 + m^{(2)}_2 )\tau_3} \nn \\
& \qquad \qquad  \qquad \times e^{-2\pi i (m^{(1)}_1 - m^{(1)}_2)\tau_4}\, e^{-2\pi i (m^{(2)}_1 - m^{(2)}_2)\tau_5},\\
& C_{X'Y'}= e^{2\pi i (2t_1+\eta_3)m^{(0)}}\, e^{2\pi i(\eta_2-t_2)m^{(2)}_1}\, e^{2\pi i(\eta_1-t_2)m^{(1)}_2}.
\end{align}
The RHS of \eref{NAb1-DualPf} can then be identified as the partition function of the theory $Y'$ in \figref{SimpAbEx2GFI}, where 
$\vec \s$ label the matrix integral variables for the $U(2)$ gauge node, while $\{\tau_i\}_{i=1,\ldots, 5}$ label that for the $U(1)$ gauge 
nodes (in the order labelled in the figure). Therefore, we have
\be
Z^{\wt{\CO}_{\vec \CP}(Y)}(\vec{m}'; \vec \eta') =: C_{X'Y'}\cdot Z^{(Y')} (\vec{m}'; -\vec \eta'),
\ee 
where $C_{X'Y'}$ should be identified as the new contact term. The mirror map relating the FI parameters of the theory $Y'$ with the masses 
of the theory $X'$ is given as follows:
\begin{subequations}
\begin{empheq}[box=\widefbox]{align}
& \eta'_0= (m^{(0)} - m^{(2)}_1), \\
&  \eta'_1= m^{\rm bif}_1, \\
& \eta'_2= (m^{\rm bif}_1- m^{\rm bif}_2), \\
& \eta'_3= (m^{\rm bif}_2 - m^{(1)}_1 + m^{(2)}_2 ), \\
& \eta'_4= (m^{(1)}_1 - m^{(1)}_2), \\
& \eta'_5= (m^{(2)}_1 - m^{(2)}_2).
\end{empheq}
\end{subequations}
The mirror map relating the mass parameters of the theory $Y'$ with the FI parameters of the theory $X'$ can also 
be read off from \eref{NAb1-DualPf}.

\bibliography{cpn1-1}
\bibliographystyle{JHEP}

\end{document}